\documentclass[
final
, nomarks
]{dmtcs-episciences}

\usepackage[utf8]{inputenc}
\usepackage{subfigure}

\makeatletter
\newcommand*\bigcdot{\mathpalette\bigcdot@{.5}}
\newcommand*\bigcdot@[2]{\mathbin{\vcenter{\hbox{\scalebox{#2}{$\m@th#1\bullet$}}}}}
\makeatother

\usepackage[all]{xy}
\usepackage[T1]{fontenc}
\usepackage[american]{babel}

\usepackage{graphicx}
\usepackage{xspace}
\usepackage{listings}
\usepackage{stmaryrd}
\usepackage{todonotes}
\usepackage{caption}
\usepackage{amsthm}
\usepackage{amsmath}
\usepackage{amsfonts}
\usepackage{amssymb}
\usepackage{siunitx}
\usepackage{booktabs}

\usepackage{verbatim}
\usepackage{url} 
\usepackage{multirow}
\usepackage{wrapfig}
\usepackage{color}
\usepackage{xcolor}
\usepackage{fancyvrb}
\usepackage{colortbl}
\usepackage{bussproofs}

\usepackage{censor}

\usepackage{times,zlmtt,mathptmx}
\usepackage[scaled=0.9]{helvet}
\usepackage{microtype}

\usepackage{framed}
\usepackage{bbm}
\definecolor{ltblue}{rgb}{0,0.4,0.4}
\definecolor{dkblue}{rgb}{0,0.1,0.6}
\definecolor{dkgreen}{rgb}{0,0.35,0}
\definecolor{dkgreen}{rgb}{0,0.35,0}
\definecolor{dkviolet}{rgb}{0.3,0,0.5}
\definecolor{dkred}{rgb}{0.5,0,0}
\definecolor{orange}{rgb}{0.9,0.5,0.3}
\definecolor{violet}{rgb}{0.7,0,0.7}
\usepackage{listings}

\newcommand{\pure}{{(0)}} 
\newcommand{\acc}{{(1)}} 
\newcommand{\modi}{{(2)}} 
\newcommand{\ppg}{{(1)}} 
\newcommand{\ctc}{{(2)}} 

\newcommand{\rul}[1]{\textrm{(#1)}}

\newcommand{\eqs}{\equiv} 
\newcommand{\eqw}{\sim}

\newcommand{\pair}[1]{\big\langle #1 \big\rangle} 
\newcommand{\lpair}[1]{\langle #1 \rangle_l} 
\newcommand{\rpair}[1]{\langle #1 \rangle_r} 
\newcommand{\pa}{\langle \; \rangle} 
 
\newcommand{\lcopair}[1]{[ #1 ]_l} 
\newcommand{\rcopair}[1]{[ #1 ]_r} 
\newcommand{\copa}{[\;]} 
 
\newcommand{\To}{\Rightarrow}

\newcommand{\downcast}{{\downarrow}} 
\newcommand{\empt}{\mathbb{O}}
\newcommand{\unit}{\mathbbm{1}}

\newcommand{\tagg}{\mathtt{tag}}
\newcommand{\untag}{\mathtt{untag}}
\newcommand{\throw}{\mathtt{throw}}

\newcommand{\try}{\mathtt{try}}
\newcommand{\catch}{\mathtt{catch}}

\newcommand{\lookup}{\mathtt{lookup}}
\newcommand{\update}{\mathtt{update}}

\newcommand{\squad}{\;\;}

\newcommand{\splus}{\!+\!}

\newcommand{\Loc}{\mathit{Loc}}

\newcommand\pca[3]{\{#1\}\kern1pt{#2}\kern1pt\{#3\}}

\usepackage{mathrsfs}

\newcommand{\abr}{\boxed{\mathit{abrupt}}}
\newcommand{\nor}{\boxed{\mathit{normal}}}
 
\newcommand{\Log}{\mathcal{L}} 
\newcommand{\dec}{{(d)}} 
 
\newcommand{\Cz}{\mathscr{C}}

\newcommand{\s}{\mathtt{s}}

\newtheorem{theorem}{Theorem}[section]

\newtheorem{lemma}[theorem]{Lemma}

\theoremstyle{definition}
\newtheorem{definition}[theorem]{Definition}

\theoremstyle{remark}

\newtheorem{remark}[theorem]{Remark}

\theoremstyle{definition}

\newcommand{\coq}{Coq\xspace}

\definecolor{colorsec}{HTML}{345A8A}
\definecolor{colorsubsec}{HTML}{4F81BD}
\definecolor{colorsubsubsec}{HTML}{5388C8}

\hypersetup{
  colorlinks = true,
  citecolor=purple,
  linkcolor=colorsubsec,
}

\definecolor{bg}{rgb}{0.95,0.95,0.95}

\makeatletter

\usepackage[framemethod=tikz]{mdframed}
\usepackage[most]{tcolorbox}

\def\beginlstdelim#1#2#3#4%
{%
  \def\endlstdelim{#2\egroup}%
  {\ttfamily#3#1}\bgroup#4\aftergroup\endlstdelim%
}

\lstdefinelanguage{smt}{
  language=lisp,
  alsoletter=0123456789>=,
  keywords={define-fun,declare-fun,declare-const,set-option,echo,set-info,exit,pop,push,assert},
  classoffset=1,
  morekeywords={Int,Real,Bool},keywordstyle=\color{blue!60!black}\bfseries,
  classoffset=2,
  morekeywords={not,and,or,=>},keywordstyle=\bfseries,
  classoffset=3,
  morekeywords={check-sat,check-sat-assume},keywordstyle=\color{red}\bfseries,
  classoffset=4,
  morekeywords={true,false,0,1,2,3,4,5,6,7,8,9,10},keywordstyle=\color{violet}\bfseries,
  classoffset=0,
  moredelim=**[is][\beginlstdelim{define-fun\ }{\ }{\color{green!50!black}\bfseries}{\color{blue!80!black!50!white}\bfseries}]{define-fun\ }{\ },
  moredelim=**[is][\beginlstdelim{declare-const\ }{\ }{\color{green!50!black}\bfseries}{\color{blue!80!black!50!white}\bfseries}]{declare-const\ }{\ },
  moredelim=**[is][\beginlstdelim{declare-fun\ }{\ }{\color{green!50!black}\bfseries}{\color{blue!80!black!50!white}\bfseries}]{declare-fun\ }{\ },
  mathescape=false
}

\lstdefinelanguage{Coq}{
  alsoletter={<}{:}0123456789,
  morekeywords={Variable,Inductive,CoInductive,Fixpoint,CoFixpoint,%
    Definition,Program, Lemma,Theorem,Corollary,Axiom,Local,Save,Grammar,Syntax,Intro,%
    Trivial,Qed,Intros,Symmetry,Simpl,Rewrite,Apply,Elim,Assumption,%
    Left,Cut,Case,Auto,Unfold,Exact,Right,Hypothesis,Pattern,Destruct,%
    Constructor,Defined,Fix,Record,Proof,Induction,Hints,Exists,%
    Parameter,Parameters,Split,Red,Reflexivity,Transitivity,if,then,else,Opaque,Module,%
    Transparent,Inversion,Absurd,Generalize,Mutual,Cases,of,Analyze,%
    AutoRewrite,Functional,Scheme,params,Refine,using,Discriminate,Try,%
    Require,Load,Import,Scope,Open,Section,End,Ltac,fun,forall,exists,Canonical,Structure,Eval,Notation,as,return,Goal,Class,Module%
  },%
  classoffset=1,
  morekeywords={Type,Prop,bool,nat,Set,let,in,match,with,end,as,<:,Z,farray,bitvector},keywordstyle=\color{blue!60!black}\bfseries,
  classoffset=2,
  morekeywords={Error:,Warning:},keywordstyle=\color{red}\bfseries,
  classoffset=3,
  morekeywords={0,1,2,3,4,5,6,7,8,9,10,11,12,13,14,15,16,16,18,19,20},keywordstyle=\color{violet},
  classoffset=0,
  sensitive, %
  moredelim=**[is][\beginlstdelim{Inductive\ }{\ }{\color{green!50!black}\bfseries}{\color{blue!80!black!50!white}\bfseries}]{Inductive\ }{\ },
  moredelim=**[is][\beginlstdelim{Definition\ }{\ }{\color{green!50!black}\bfseries}{\color{blue!80!black!50!white}\bfseries}]{Definition\ }{\ },
  moredelim=**[is][\beginlstdelim{Lemma\ }{\ }{\color{green!50!black}\bfseries}{\color{blue!80!black!50!white}\bfseries}]{Lemma\ }{\ },
  moredelim=**[is][\beginlstdelim{Axiom\ }{\ }{\color{green!50!black}\bfseries}{\color{blue!80!black!50!white}\bfseries}]{Axiom\ }{\ },
  moredelim=**[is][\beginlstdelim{Theorem\ }{\ }{\color{green!50!black}\bfseries}{\color{blue!80!black!50!white}\bfseries}]{Theorem\ }{\ },
  moredelim=**[is][\beginlstdelim{Class\ }{\ }{\color{green!50!black}\bfseries}{\color{blue!60!black}\bfseries}]{Class\ }{\ },
  moredelim=**[is][\beginlstdelim{Module\ }{\ }{\color{green!50!black}\bfseries}{\color{blue!60!black}\bfseries}]{Module\ }{\ },
  moredelim=**[is][\beginlstdelim{Record\ }{\ }{\color{green!50!black}\bfseries}{\color{blue!60!black}\bfseries}]{Record\ }{\ },
  morecomment=[n]{(*}{*)},%
  morestring=[d]",%
  literate={=>}{{$\Rightarrow$}}1
  {->}{{$\,\to\,$}}1
  {<-}{{$\leftarrow$}}1
  {>->}{{$\rightarrowtail$}}2
  {<->}{{$\leftrightarrow$}}1
  {==>}{{$\Longrightarrow$}}1
  {forall}{{\color{blue!60!black}\bfseries$\forall$}}1
  {exists}{{\color{blue!60!black}\bfseries$\exists$}}1
  {//n}{{$\neq$}}1
  {<=}{{$\overset{?}{\leq}$}}1
  {>=}{{$\overset{?}{\geq}$}}1  
  {>}{{$\overset{?}{>}$}}1
   {<}{{$\overset{?}{<}$}}1
  {:=}{{$\triangleq$}}1
  {\/\\}{{$\wedge$}}1
  {|-}{{$\vdash$}}1
  {\\\/}{{$\vee$}}1
  {'}{'}1
  {⟦}{{$\llbracket$}}1
  {⟧}{{$\rrbracket$}}1
  {-->}{{$\longrightarrow$}}1
}

\lstdefinelanguage{lfsc}{
  language=lisp,
    alsoletter={!}{\%}{@}{\\},
  keywords={check,define,declare,program},
  classoffset=1,
  morekeywords={int,mpz,th_holds,holds,term,sort,type,match,fail,default},keywordstyle=\color{blue!60!black}\bfseries,
  classoffset=2,
  keywords={\%,@,!,\\},keywordstyle=\color{violet}\bfseries,
  classoffset=0,
  moredelim=**[is][\beginlstdelim{define\ }{\ }{\color{green!50!black}\bfseries}{\color{green!50!black!50!white}\bfseries}]{define\ }{\ },
  moredelim=**[is][\beginlstdelim{declare\ }{\ }{\color{green!50!black}\bfseries}{\color{blue!80!black!50!white}\bfseries}]{declare\ }{\ },
  moredelim=**[is][\beginlstdelim{program\ }{\ }{\color{green!50!black}\bfseries}{\color{blue!80!black!50!white}\bfseries}]{program\ }{\ },
  escapechar=\&
}

\lstdefinelanguage{smtcoq}{
  alsoletter=\#0123456789\=,
  classoffset=0,
  keywords={or,and,not,impl,true,false,\=,->},keywordstyle=\color{black}\bfseries,
  classoffset=1,
  morekeywords={0,1,2,3,4,5,6,7,8,9,10,11,12,13,14,15,16,16,18,19,20},keywordstyle=\color{violet}\bfseries,
  classoffset=0,
  sensitive=true,
  moredelim=**[is][\beginlstdelim{:(}{\ }{}{\color{green!50!black}\bfseries}]{:(}{\ },
}

\lstdefinelanguage{ocaml}{
  language=[Objective]caml,
  identifierstyle=\ocidstyle
}

\newcommand*\ocidstyle{%
        \expandafter\id@style\the\lst@token\relax
}
\def\id@style#1#2\relax{%
        \ifcat#1\relax\else
                \ifnum`#1=\uccode`#1%
                        \color{blue!60!black}
                \fi
        \fi
}

\newenvironment{tcb}[2][\small]{%
  \tcblisting{enhanced jigsaw,breakable,lines before break=3,
    listing only,colback=bg,colframe=bg,enlarge
    top by=0mm,top=0pt,bottom=0pt,left=2pt,right=2pt,enhanced,
    before={\vspace{10pt}},
    after={\par\vspace{5pt}\noindent},
    listing options={language=#2,basicstyle={\ttfamily#1\upshape}}%
    }}{\endtcblisting}

\newcommand{\code}[1]{\lstinline!#1!}

\StopCensoring

\makeatother

\usepackage[round]{natbib}

\author{Burak Ekici\affiliationmark{1}
  }
\title[\texttt{IMP} with exceptions over decorated logic]{\texttt{IMP} with exceptions over decorated logic}
\affiliation{
University of Innsbruck,
Department of Computer Science,
Innsbruck, Austria
\email{burak.ekici@uibk.ac.at}
}

\received{2017-4-18}

\revised{2018-2-23, 2018-9-21}

\accepted{2018-10-5}

\begin{document}
\publicationdetails{20}{2018}{2}{11}{3272}
\maketitle


%
%
%
%

\maketitle

\begin{abstract}

In this paper, we facilitate the reasoning about impure programming languages,
by annotating terms with ``decorations'' that describe what
computational (side) effect evaluation of a term may involve.
In a point-free categorical language, called the
``decorated logic'', we formalize the mutable state and
the exception effects first separately, exploiting a
nice duality between them, and then combined.
The combined decorated logic serves as the target language for the denotational semantics of the
\texttt{IMP+Exc} imperative programming language, and allows us to prove equivalences between programs written in \texttt{IMP+Exc}.
The combined logic is encoded in \coq, and this encoding is used to
certify some program equivalence proofs.
%
\end{abstract}

\section{Introduction}
\label{intro}
In programming languages theory, a program is said to have computational effects if,
besides a return value, it has observable interactions with the outside world. For instance,
using/modifying the program state, raising/recovering exceptions, reading/writing
data from/to some file, etc. In order to formally reason about behaviors of a program
with computational effects, one has to take into account these interactions.
One difficulty in such a reasoning is the mismatch between the syntax of operations
with effects and their interpretation. Typically, an operation in an effectful language
with arguments in $X$ that returns a value in $Y$ is not interpreted as a function from $X$ to
$Y$, due to the effects, unless the operation is pure.

The best known \textit{algebraic approach} to formalize
computational effects was initiated by \cite{Moggi:1991}
in his seminal paper. He showed that the effectful
operations of an impure language can be interpreted as arrows
of a Kleisli category for an appropriate monad ($T,\eta,\mu$)
over a base category $\Cz$ with finite products.
For instance, in Moggi's \emph{computational metalanguage},
an operation in an impure language with arguments in
$X$ that returns a value in $Y$
is now interpreted as an arrow from $\llbracket X\rrbracket$ to
$T\llbracket Y\rrbracket$ in $\Cz$
where $\llbracket X\rrbracket$ is the object of \emph{values}
of type $X$ and $T\llbracket Y\rrbracket$
is the object of \emph{computations} that return values of
type $Y$.
The use of monads to formalize effects (such as state, exceptions, input/output
and non-deterministic choice) was popularized by \cite{Wadler:1992},
and implemented in the programming languages Haskell and F$\sharp$.
Using monad transformers, as in \cite{Jaskelioff:2009},
it is usually possible to \lq\lq{}combine\rq\rq{} different
effects formalized by monads.
Moggi's \textit{computational metalanguage} was extended into
the \emph{basic effect calculus} with a notion of \emph{computation type}
by \cite{Filinski:1996} in his effect PCF and by \cite{Levy:1999} in his
call-by-push-value (CBPV). \cite{Simpson:2014}
defined their effect calculus, named \emph{extended effect calculus}
as a canonical calculus incorporating the ideas of Moggi, Filinski and Levy.
Following Moggi, they included a type constructor for computations.
Following Filinski and Levy, they classified
types into value types and computation types.

Being dual to monads, comonads have been
used to formalize context-dependent computations.
Intuitively, an effect which
observes features may arise from a comonad, while an effect which constructs
features may arise from a monad~(\cite{Jacobs:2011}).
\cite{Uustalu:2008}
have structured stream computations,
\cite{Orchard:2010} array computations and \cite{Tzevelekos:2008}
game semantics via the use of comonads.
\cite{Petricek:2013} proposed a unified calculus for tracking
context dependence in functional languages
together  with  a  categorical  semantics  based  on indexed  comonads.
In (\cite{Orchard:2012}), there is a quite nice discussion on
how to choose between a monad or comonad
when either can be used to capture a particular notion of computation.
Also,
\cite{Brookes:1993}
discussed that a computation may
be interpreted by distributive laws of a comonad over a monad
when it is seen as a composition of context-dependence
and effectfulness.
This approach has been applied to clocked causal data-flow computation,
combining causal data-flow and exceptions by~\cite{Uustalu:2005}.

Moggi's approach, using monads in effect modeling,  has been extended to Lawvere theories which
first appeared in \cite{Lawvere:1963}\rq{}s PhD dissertation.
Then, \cite{Linton:1966,Linton:1969} first showed that every Lawvere theory induces a
monad on the category of sets, and then on any category satisfying some
condition called the ``local representability''. 
Therefore, Moggi\rq{}s seminal idea, formalizing computational
effects by monads, made it possible for monadic effects to be formalized
through Lawvere theories. To this extend,
\cite{Plotkin:2002} have shown
that effects
such as the global and the local
state could be formalized by \textit{signatures} of effectful terms and an \textit{equational theory} explaining
the interactions between them.
\cite{Mellies:2010} has refined this \textit{equational theory}
showing that some of the equations modeling the mutable global state can be omitted.
\cite{Hyland:2006,Hyland:2007a}
studied the combination of computational effects in terms of Lawvere theories.

\cite{Plotkin:2009,Plotkin:2013} 
extended Moggi\rq{}s classification
of terms (\emph{values} and \emph{computations})
with a third level called \emph{handlers} for the computational effects
that can be represented by an algebraic theory (\emph{algebraic effects}).
Initially, they introduce an \emph{handler} for the exception
handling, and then account for its generalization to the
other handlers to cope with other algebraic effects
such as stream redirection, explicit non-determinism, CCS,
parameter passing, timeout and rollback~\cite[\S 3]{Plotkin:2013}.
For each algebraic effect, \emph{handling constructs} are used to apply handlers
to effectful computations where
effectful computations can be interpreted as algebraic operations while handling
constructs as homomorphisms from free algebras.
%
This use of handling constructs is inspired from
\cite{Benton:2001}'s work where a single 
construct specialized to handle exceptions is introduced.
Moreover, \cite{Jacobs:2001} formalized the exception effect from the
dual, namely co-algebraic, viewpoint. Exception handling  is also used
to build a Hoare logic for exceptions by~\cite{Schroder:2004}.
%
%
%
%



There is an older formal way of modeling computational effects
called the \emph{effect systems}
by \cite{Lucassen:1988}. They presented
an approach to programming languages for parallel computers.
The key idea was to use an \emph{effect system} to discover expression
scheduling constraints. 
There, every expression comes with three components:
\textit{types} to represent the kinds of the return values,
\textit{effects} to summarize the observable interactions of expressions
and \textit{regions} to highlight the areas of the memory where
expressions may have effects. To this extend, one can simply reason that
if two expressions do not have
overlapping effects, then they can obviously be scheduled in parallel.
The reasoning is done by some inference rules for \textit{types}
and \textit{effects} based on the second order typed $\lambda$-calculus.

\cite{Duval:2010} proposed yet another
paradigm to formalize computational effects by mixing effect systems
and algebraic theories, named \textit{the decorated logic}. The key point of this
paradigm is that
every term
comes with a decoration which exposes its features with respect to
a single computational effect or to several ones keeping
their interpretations close to syntax in reasoning with effects.
In addition, an \textit{equational theory} highlights the
interactions among terms with two sorts of equations:
\textit{weak} equations relate terms with respect only to
their results while \textit{strong} equations relate them
with respect both to their results and effects. 
By and large, decorated logic provides an equational reasoning in between programs written
in imperative languages after being used as a target language for a denotational semantics
of the studied language.

In a decorated logic, a term has three different decorations: pure, accessor and modifier/catcher.
The first two decorations can correspond to Moggi's values and computations, and the third level
can be seen as Plotkin and Pretnar's handlers. An handler operates recursively by its nature, and
handles also the continuation. However, a catcher does not. It returns the continuation unhandled which
should then be handled explicitly. Thus, catchers are non-recursive handlers, so called shallow
handlers introduced by \cite{Kamar13}.
\subsection{On the use of decorated logic}
In this paper, we use Duval's decorated logic to formalize computational effects.
The advantages of using decorated logic in effect formalization is mainly two-folded:
(1) effects of terms are hidden by the decorations, so that it is possible to preserve the
syntax of term signatures. Thereafter, the provided equational reasoning would be
valid for different algebraic models of the same effect.
(2) The equational theory is based on decorated equivalence relations proposing different
reasoning capabilities: one on effects and returned results and the other one only
on returned results.
However, for the time being, it might be inconvenient to use decorated logic to prove
more general properties of algorithms. That is to say,  we can prove equivalences between
programs that admits particular specifications as initializing and describing final values stored
in variables. The total correctness of a theory in a decorated logic,
that guaranties that the theory is not using
too many axioms to become the maximal theory,
 is based on a syntactic completeness property
called relative Hilbert-Post completeness. Section~\ref{sub:tc} 
details mentioned property, and its application to the specific case that this paper covers.
%
\subsection{Organization and contributions}
\begin{figure}[h]   
\renewcommand{\arraystretch}{1}
   $$ \xymatrix@C=8pc@R=1.25pc{
  \texttt{Decorated logic} \ar[rd]|{implementation}
  \ar@(ul,ur)|{equational\ reasoning \ between \ \texttt{IMP+Exc} \ programs}&\\
    & \texttt{Coq}   \ar@(ul,ur)_{certified \ equational\ reasoning} \\
   \texttt{IMP+Exc Programs}
\ar[uu]|{denotational \ semantics}
   \ar@{-->}[ru]
  &
  } 
 $$
 {\vskip -0.5cm}
\renewcommand{\arraystretch}{1}
\caption{The approach} 
\label{thov} 
\end{figure}
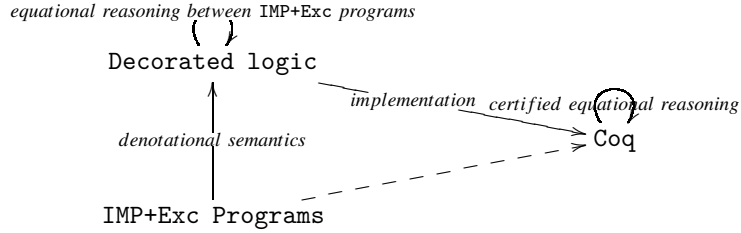

In general terms, in this paper, we extend Moggi's original approach using the classifications of expressions,
provided by the Kleisli category of the monad of exception and the comonad of the state
thanks to the duality between states and exceptions proven by \cite{Dumas:2012b}. The definitions and the results
are presented in terms of equational theories so that one does not need to know the
details about the monad of exceptions nor the comonad of the state.
In more specific terms, this paper designs the decorated logic for the
global state and the exception effects,
and then combines them to serve as a target language for denotational semantics of
imperative programming languages mixing mentioned effects.
It is organized as follows: in Section~\ref{impwexc}, we introduce an imperative programming
language that mixes the state and the exception effects by defining
its small-step operational semantics.
The language we study there is called \texttt{IMP+Exc}
which extends the \texttt{IMP} (or \texttt{while}) with a mechanism to
\emph{raise} and \emph{handle} \emph{exceptions}.
In Section~\ref{dls}, we introduce the decorated logic as a generic framework
extending Moggi's monadic equational logic.
Then, we formally specialize the decorated logic for the state and the exception effects
in Sections~\ref{dlst}~and~\ref{dlexc}, respectively.
In Section~\ref{dlstexc}, we combine these logics.
Finally, Section~\ref{impexcodl} details the use of the combined decorated logic as the
target language for the \texttt{IMP+Exc} denotational semantics.
This provides a rigorous formalism for an equational reasoning between
termination-guaranteed \texttt{IMP+Exc} programs.
I.e., proving two different looking programs
are in fact doing the same job with respect to the state and exception
effects. In Section~\ref{peqp-st-exc}, we presents three proof examples.
Also, we certify such proofs with the
\coq Proof
Assistant. 
See the entire \coq implementation here \footnote{\url{https://github.com/ekiciburak/impex-on-decorated-logic}},
and the approach of the paper in Figure~\ref{thov}.
 {\vskip -0.25cm}
 {\vskip -0.25cm}
This paper builds upon several papers
by \cite{Duval:2010}, \cite{Dumas:2014b}, \cite{DBLP:conf/macis/DumasDEPR15}, \cite{Dumas2014abc}, \cite{Dumas:2012b} and \cite{Reynaud:2014} . The novel points presented here
can be itemized as follows:
\begin{itemize}
\item a combined decorated logic for the states (\cite{Dumas:2014b})
and exceptions (\cite{Dumas2014abc}) effects (this paper explains both logics
again but for the details please refer to the citations),
\item \coq formalization of the combined logic,
\item a denotational semantics for the \texttt{IMP+Exc}  (\texttt{IMP} with exceptions) over  the combined logic,
\item \coq formalization of the \texttt{IMP+Exc} denotational semantics,
\item some equivalence proofs of programs written in \texttt{IMP+Exc} and their verifications in \coq.
\end{itemize}
A preliminary version of this paper has been presented in TFP (Trends in Functional Programming) 2015
but did not appear in the final proceedings.
 Find the mentioned paper here\footnote{\url{ftp://ftp-sop.inria.fr/indes/TFP15/TFP2015_submission_6.pdf}}.

\section{\texttt{IMP} with exceptions}
\label{impwexc}
\texttt{IMP} is a standard Turing complete imperative programming language natively providing
global variables of integer ($\mathtt{Z}$), Boolean ($\mathtt{B}$) and unit ($\mathtt{U}$)
data types, standard integer and Boolean arithmetic
enriched with a set of commands that is made of do-nothing, assignment,
sequence, conditionals and looping operations.  Below, we detail its
syntax
where
\texttt{n}
represents a constant integer term while
\texttt{x}
is an integer global variable.
Note also that abbreviations \texttt{aexp} and \texttt{bexp} respectively denote
arithmetic
and
Boolean
expressions as well as \texttt{cmd} stands for the
commands.
{\vskip -0.5cm}
\begin{figure}[h]
\renewcommand{\arraystretch}{1}
$$\begin{array}{lcl}
\quad \textrm{aexp: } \mathtt{a_1 \ a_2} &::=&
 \mathtt{n}\mid \mathtt{x} \mid \mathtt{a_1 + a_2} 
\mid \mathtt{a_1 - a_2} \mid \mathtt{a_1 \times a_2}\\ 
\quad \textrm{bexp: } \mathtt{b_1 \ b_2} &::=&  \mathtt{true} \mid \mathtt{false} \mid 
\mathtt{a_1 \overset{?}{=} a_2} \mid \mathtt{a_1  \overset{?}{\neq} a_2} \ \mid \
\mathtt{a _1  \overset{?}{>} a_2} \ \mid \ \mathtt{a_1  \overset{?}{<} a_2} \mid 
 \mathtt{b_1\wedge b_2} \mid \mathtt{b_1\vee b_2} \mid \mathtt{\neg b_1} \\
\quad \textrm{cmd: } \mathtt{c_1 \ c_2} &::=& \mathtt{SKIP} \mid \mathtt{x \triangleq a_1}\mid \mathtt{c_1; \ c_2} \mid 
\mathtt{if \ b \ then \ c_1 \ else \ c_2} \mid \mathtt{while \ b \ do \ c_1}
\end{array}$$
{\vskip -0.5cm}
\caption{Standard \texttt{IMP} syntax}
\label{s-imp}
\end{figure}

Neither arithmetic nor Boolean expressions are allowed to modify the state: they are either pure or read-only.
We present, in Figure~\ref{aexp-imp}, the big-step semantics for evaluation of arithmetic
expressions in \texttt{IMP} where we use a big-step transition function $\to_a\colon \mathtt{aexp \times S \to \mathbb{Z}}$.
This function computes an integer value out of an input arithmetic expression and the current program state
(denoted $\mathtt{s}$) which includes contents of variables at a given time.
\begin{figure}[h]
\begin{gather*}
(\mathtt{aconst}) \dfrac{}{(\mathtt{n, \ s}) \to_a \mathtt{n}}\qquad
(\mathtt{var}) \dfrac{}{(\mathtt{x, \ s}) \to_a \mathtt{s(x)}}\qquad
(\mathtt{op-sym}) \dfrac{\mathtt{(a_1, \ s) \to_a \ n_1} \qquad \mathtt{(a_2, \ s) \to_a \ n_2}}
			      {(\mathtt{a_1 \ op \ a_2, s}) \to_a \mathtt{n_1 \  op_\mathbb{Z} \ n_2}}
\end{gather*}
{\vskip -0.45cm}
\caption{Big-step operational semantics for arithmetic expressions}
\label{aexp-imp}
\end{figure}
The symbol \texttt{op} represents the operation symbols ($+$, $-$ or $\times$) given
by the standard syntax in Figure~\ref{aexp-imp},
while $\mathtt{op_\mathbb{Z}\colon \mathtt{\mathbb{Z}} \to \mathtt{\mathbb{Z}} \to \mathtt{\mathbb{Z}}}$
denotes the corresponding binary operations in $\mathtt{\mathbb{Z}}$.
Similarly, in Figure~\ref{bexp-imp}, we present the big-step semantics for evaluation of Boolean
expressions in \texttt{IMP} where we use a big-step transition function $\to_b\colon \mathtt{bexp \times S \to \mathbb{B}}$.
This function simply computes a Boolean value out of an input Boolean expression and the current program state.
\begin{figure}[h]
\begin{gather*}
(\mathtt{true}) \dfrac{}{(\mathtt{true, \ s}) \to_b true }\qquad
(\mathtt{false}) \dfrac{}{(\mathtt{false, \ s}) \to_b false}\\
(\mathtt{op1}) \dfrac{\mathtt{(b_1, \ s) \to_b \ v_1} \qquad \mathtt{(b_2, \ s) \to_b \ v_2}}
			      {(\mathtt{b_1 \ opb \ b_2, s}) \to_b \mathtt{v_1 \  opb_\mathbb{B} \ v_2}}\qquad
(\mathtt{op2}) \dfrac{\mathtt{(b_1, \ s) \to_b \ v_1}}
			      {(\mathtt{\neg \ b_1, s}) \to_b \mathtt{neg \ v_1}}\qquad
\end{gather*}
{\vskip -0.5cm}
\caption{Big-step operational semantics for Boolean expressions}
\label{bexp-imp}
\end{figure}
The constant symbols \texttt{true} and \texttt{false} are Boolean operation symbols given
by the standard syntax in Figure~\ref{s-imp}, while
\emph{true} and \emph{false} are Boolean constructors.  
Similarly,
\texttt{opb} represents the binary operation symbols,
while
$\mathtt{opb_\mathbb{B}\colon \mathbb{B} \to \mathbb{B} \to \mathbb{B}}$ denotes 
the corresponding Boolean operations, and $\mathtt{neg\colon \mathbb{B} \to \mathbb{B}}$ is the Boolean negation.

The small-step operational semantics for evaluation of commands are given in Figure~\ref{ss-cmd}
where we use a small-step transition function $\rightsquigarrow\colon \mathtt{S \times cmd \to S \times cmd}$
which is interpreted as
\emph{at the state~$\mathtt{s}$, one step execution of the
command $\mathtt{c}$ changes the state into $\mathtt{s'}$ and the command $\mathtt{c'}$ is now in further execution}.
\begin{figure}[h]
\begin{gather*}
(\mathtt{sequence})\dfrac{\s, \mathtt{c_1} \rightsquigarrow \mathtt{s\rq{}},  \mathtt{c_1\rq{}} }{\s, (\mathtt{c_1; c_2}) \rightsquigarrow \mathtt{s\rq{}},  (\mathtt{c_1\rq{}; c_2})}\quad
(\mathtt{skip}) \dfrac{}{\s, \ (\mathtt{SKIP}; \mathtt{c}) \rightsquigarrow \s, \mathtt{c}}\\
(\mathtt{assign})\dfrac{\mathtt{(a, \ \s) \to_a n}}{\s, \ (\mathtt{x := a) \rightsquigarrow \s[x \leftarrow n ]}, \mathtt{SKIP}}\\
(\mathtt{cond_1})\dfrac{\mathtt{(b, \ \s) \rightarrow_b}\ true}{\s, \ (\mathtt{if \ b \ then \ c_1 \ else \ c_2}) \rightsquigarrow \s, \mathtt{c_1}}\quad
(\mathtt{cond_2})\dfrac{\mathtt{(b, \ \s) \rightarrow_b}\ false}{\s, \ (\mathtt{if\ b \ then\ c_1 \ else \ c_2}) \rightsquigarrow \s, \mathtt{c_2}}\\
(\mathtt{while_1})\dfrac{\mathtt{(b, \ \s) \rightarrow_b}\ true}{\s, \ (\mathtt{while \ b \ do \ c}) \rightsquigarrow \s, (\mathtt{c; \ while \ b \ do \ c})}\quad
(\mathtt{while_2})\dfrac{\mathtt{(b, \ \s) \rightarrow_b}\ false}{\s, \ (\mathtt{while \ b \ do \ c}) \rightsquigarrow \s, \mathtt{SKIP}}
\end{gather*}
{\vskip -0.45cm}
\caption{Small-step operational semantics for commands}
\label{ss-cmd}
\end{figure}

We need to elucidate that a command $\mathtt{c}$ terminates at a state $\mathtt{s'}$ if 
 $\mathtt{\s, \ c} \rightsquigarrow^* \mathtt{s', \ SKIP}$
 for some state $\mathtt{s'}$, where $\mathtt{\rightsquigarrow^*}$ is the transitive closure of the transition function
 $\mathtt{\rightsquigarrow}$.
Mind also that
$\mathtt{SKIP}$ is allowed to execute at any state $\s$, and $\mathtt{SKIP}$ alone is used to indicate the final step of some command set. 

 \subsection{A mechanism to handle exceptions}
 \label{a-exc-ab}
 Extending the \texttt{IMP} language with a mechanism
 that allows raising exceptions and recovering from them, we enrich the
 command set with
$\mathtt{THROW}$ and $\mathtt{TRY/CATCH}$ blocks.
In addition to the ones in~Figure~\ref{s-imp}, we also consider following commands
in Figure~\ref{imp-acmd}
\begin{figure}[h]
\renewcommand{\arraystretch}{1.7}
$$\begin{array}{lcl}
\quad \textrm{cmd: } \mathtt{c_1 \ c_2} &::=& \ldots \mid {\color{red} \mathtt{THROW}} \ \mathtt{e} \mid {\color{red}\mathtt{TRY}} \ \mathtt{c_1} \ {\color{red}\mathtt{CATCH}} \ \mathtt{e} \Rightarrow \mathtt{c_2}
\end{array}$$
{\vskip -0.5cm}
\caption{Syntax for exceptional commands}
\label{imp-acmd}
\end{figure}
\noindent
where $\mathtt{e}$ is an exception name coming from a finite set $\mathtt{EName}$
which exists by assumption.
There is also a type $\mathtt{EV_e}$ of exceptional values (parameters) for each exception name \texttt{e}.
 The small-step operational semantics for $\mathtt{THROW}$ and $\mathtt{TRY/CATCH}$ commands are shown
 in Figure~\ref{ss-acmd}. 
\begin{figure}[h]
\begin{gather*}
(\mathtt{throw}) \dfrac{\mathtt{e: EName}}{\s, \ (\mathtt{THROW \ e; c}) \rightsquigarrow \mathtt{\s,\ \mathtt{THROW \ e}  }}\squad
(\mathtt{tskip}) \dfrac{}{\mathtt{s},\, \mathtt{TRY \ \mathtt{SKIP} \ CATCH \ e}  \Rightarrow \mathtt{c}\rightsquigarrow \mathtt{s},  \mathtt{SKIP} }\\
(\mathtt{tstep})\dfrac{\mathtt{\s, \mathtt{c_1}\rightsquigarrow \mathtt{\s'}, \mathtt{c_1'}}}{\mathtt{\s}, \mathtt{TRY \ \mathtt{c_1} \ CATCH \ e} \Rightarrow \mathtt{c_2} \rightsquigarrow
 \mathtt{\s',\,TRY \ \mathtt{c_1'} \ CATCH \ e}  \Rightarrow \mathtt{c_2}}\\
(\mathtt{tc_1})\dfrac{\mathtt{e: EName}}{\s, \mathtt{TRY \ (THROW \ e) \ CATCH \ e} \Rightarrow \mathtt{c} \rightsquigarrow \mathtt{s},  \mathtt{c} }\squad
(\mathtt{tc_2})\dfrac{\mathtt{e_1\ e_2: EName \quad e_1 \neq e_2}}{\s, \mathtt{TRY \ (THROW \ e_1) \ CATCH \ e_2} \Rightarrow \mathtt{c} \rightsquigarrow \mathtt{s},  \mathtt{THROW \ e_1} }
\end{gather*}
{\vskip -0.45cm}
\caption{Small-step operational semantics for additional commands}
\label{ss-acmd}
\end{figure}

Exceptional commands are pure in terms of the state effect:
they neither use nor modify the program state.
 However, they introduce another sort of computational effect:
 the exception. In prior, we stated that the command $\mathtt{SKIP}$ alone indicates
the termination of a program. Now, we extend this by saying $\mathtt{THROW \ e}$ is also an end but an
abnormal end.
Intuitively, if an exceptional value of name \texttt{e} is raised in the \texttt{TRY} block and recovered immediately in the \texttt{CATCH},
the program then resumes with the provided continuation.
An exceptional value (of name $\mathtt{e}$) gets propagated if another exceptional value
with different name (say, of name $\mathtt{f}$, s.t. $\mathtt{e \neq f}$)
is being recovered in the \texttt{CATCH}.


In Section~\ref{impexcodl}, we define denotational semantics of the
\texttt{IMP+Exc} language using the decorated logic (generic framework is given
in Section~\ref{dls}) for the state and the exception effects
as the target language. We present this logic in Section~\ref{dlstexc}
as a combination of the logics that we introduce in Sections~\ref{dlst}~and~\ref{dlexc}.

 
 \section{Decorated Logic ($\Log_{dec}$)}
 \label{dls}
 
The decorated logic, as a generic framework, is an extension to monadic equational logic~\cite{Moggi:1991},
that we briefly discuss in Section~\ref{meq},
with the use of decorations on terms and equalities.
It provides a rigorous formalism to do \emph{equational reasoning} between impure programs
written in imperative programming languages with side effects after
being defined as a target language for their denotational semantics.

\subsection{Monadic Equational Logic  ($\Log_{meq}$)}
\label{meq}
The \textit{monadic equational logic} ($\Log_{meq}$)
is the minimal logic that can be interpreted in a category with objects as 
types, arrows as terms and equalities as equations.
I.e., an object $\mathtt{O}$ in the category interprets the
type $\mathtt{X}$ in the logic, just as the usual Leibniz equality,
$\mathtt{f = g}$, interprets the equation $\mathtt{f \cong g}$ in the logic.
The keyword
\lq\lq{}\textit{monadic}\rq\rq{} has little to do with monads.
It rather means that the operations of the logic are
\textit{unary} (or mono-adic). Figure~\ref{syn-leq} presents
the syntax of the logic $\Log_{meq}$.
\begin{figure}[!h]
\renewcommand{\arraystretch}{1.25}
$$\begin{array}{lccl}
\multicolumn{4}{l}{
\textbf{ Grammar for the monadic equational logic: }} \\
\quad \textrm{Types: } & \mathtt{t} &::=& \mathtt{X}\mid \mathtt{Y}\mid \dots \\
\quad \textrm{Terms: } & \mathtt{f,\; g} &::=& \mathtt{id_t}
\mid \texttt{a} \mid \texttt{b} \mid \dots \mid \mathtt{g\circ f}  \\
\quad \textrm{Equations: } & \mathtt{eq} &::=& \mathtt{f} \cong \mathtt{g} \\  
\end{array}$$
{\vskip -0.45cm}
\caption{$\Log_{meq}$: syntax}
\label{syn-leq}
\end{figure}
There, every term has a source and a target type, e.g., $\mathtt{f\colon X\to Y}$. 
Every equation is formed by terms with the same source and target types, e.g., $\mathtt{e: f \cong g}$
where $\mathtt{f,~g\colon~X\to~Y}$.
This syntax is accompanied by the rules shown in Figure~\ref{meq-rls}.
%

\begin{figure}[h]  
{\vskip -0.5cm}
\renewcommand{\arraystretch}{1.5}
$$ \begin{array}{l} 
\textbf{congruence rules} \\ 
\mathtt{\rul{refl} 
  \dfrac{f}{f \cong f}} \quad
\mathtt{\rul{sym} 
  \dfrac{f \cong g}{g \cong f}}  \quad
\mathtt{\rul{trans} 
  \dfrac{f \cong g \squad g \cong h}{f \cong h}}  \quad
\mathtt{\rul{replsubs} 
  \dfrac{f_1 \cong f_2\colon X\to Y \squad g_1 \cong g_2\colon Y\to Z}
    {g_1\circ f_1 \cong g_2\circ f_2}} \\
    \textbf{categorical rules} \\ 
\mathtt{\rul{id}
  \dfrac{X}{id_X\colon X\to X }} \quad 
\mathtt{\rul{comp}
  \dfrac{f\colon X\to Y \quad g\colon Y\to Z}
    {(g\circ f) \colon X\to Z}}  \quad
\mathtt{\rul{ids} 
  \dfrac{f\colon X\to Y}{f\circ id_X \cong f}} \quad 
\mathtt{\rul{idt} 
  \dfrac{f\colon X\to Y}{id_Y\circ f \cong f}}
  \vspace{5pt} \\
\mathtt{\rul{assoc} 
  \dfrac{f\colon X\to Y \squad g\colon Y\to Z \squad h\colon Z\to U}
  {h\circ (g\circ f) \cong (h\circ g)\circ f}}
\end{array}$$
{\vskip -0.5cm}
\renewcommand{\arraystretch}{1}
\caption{$\Log_{meq}$: rules} 
\label{meq-rls} 
\end{figure}

The \textit{congruence rules} say that the relation
 \lq{}$\cong$\rq{} is a congruence meaning that it is an equivalence relation
 (reflexive, symmetric and transitive)
which obeys \textit{replacements} and \textit{substitutions}
of compatible terms with respect to the composition.
The basic categorical rules indicate that there is an identity
morphism $\mathtt{id_X\colon X\to X}$ for each type $\mathtt{X}$,
composition is an associative operation, and composing any term $\mathtt{f}$
with $\mathtt{id}$ is $\mathtt{f}$, up to $\cong$, no matter the composition order.

\subsection{The decorated logic}
\label{ss:dls}
 
The decorated logic extends the monadic equational logic
with a 3-tier effect system for terms
and a 2-tier system for equations made of \lq\lq{}up-to-effects\rq\rq{} (weak)
and \lq\lq{}strong\rq\rq{}~equalities. Figure~\ref{syn-ldec} presents its syntax.
\begin{figure}[h]
\renewcommand{\arraystretch}{1.25}
$$\begin{array}{lccl}
\multicolumn{4}{l}{
\textbf{ Grammar for the decorated logic: }} \\
\quad \textrm{Types: } &\mathtt{t} &::=& \mathtt{X}\mid \mathtt{Y}\mid \dots\\
 \quad \textrm{Decoration for terms: } & \mathtt{(d)} &::=&\pure \mid \acc \mid \modi \\ 
\quad \textrm{Terms: }& \mathtt{f,\; g} &::=&
\mathtt{a^\dec} \mid \mathtt{b^\dec} \mid \dots \mid \mathtt{g\circ f^\dec} \mid    (\mathtt{tpure\ \bigcdot})^\pure   \\
\quad \textrm{Equations: } & \mathtt{eq} &::=& \mathtt{f} \eqs \mathtt{g} \mid \mathtt{f} \eqw \mathtt{g}\\  
\end{array}$$
{\vskip -0.45cm}
\caption{$\Log_{dec}$: syntax}
\label{syn-ldec}
\end{figure}
Each term has a source and a target type as well as a decoration which
describe what computational side effects evaluation of that term may involve,
and used as a superscript $\pure$, $\acc$
or $\modi$: a \textit{pure} term
is decorated with $\pure$, an effect \textit{constructor}
with $\acc$ and an effect \textit{modifier} term comes with the decoration $\modi$.
Each equation is formed by two terms with the same source and
target as well as a decoration which is denoted either
by ``$\eqw$'' (\textit{weak}) or by ``$\eqs$'' (\textit{strong}).
A weak equality between two terms relates them according only to their results, while
a strong equality relates terms  according both to their result and
the side effect evaluations they involve with respect to the effect in question.

The \texttt{tpure} is a special constructor used to introduce decorated pure terms into the logic $\Log_{dec}$.
It inputs a non-decorated pure term from a pure type system
(i.e., \coq's logic) and drops it in with the decoration $\pure$.
For instance, the identity term $\mathtt{id}$
is defined using the \texttt{tpure} constructor, for all types \texttt{X} as follows:
$$\begin{array}{l}
   \xymatrix@R=0.01pc@C=0.5pc{
   \mathtt{id_X^\pure} &\colon& \mathtt{X\to X} &:=& \mathtt{tpure\  (\lambda\, x:\, X.\, x: \, X)}.
   }
\end{array}$$
In Figure~\ref{dl-rls}, we present the inference rules associated to the syntax given in Figure~\ref{meq-rls}.
\begin{remark}
In all of the figures presenting the rules of some decorated logic,
through out the paper, the decorations ``$\mathtt{d_1}$, $\mathtt{d_2}$, $\mathtt{d_3}$, \ldots'' are meant to be in the set $\mathtt{\{0, 1, 2\}}$ unless otherwise stated.
For instance, in the rule (wtos) below decorations $\mathtt{d_1}$ and $\mathtt{d_2}$ cannot take the value $\mathtt{2}$.
\end{remark}
{\vskip -0.25cm}
\begin{figure}[h]  
\renewcommand{\arraystretch}{1.25}
$$ \begin{array}{l} 
\textbf{hierarchy rules} \\ 
\mathtt{\rul{0-to-1}\dfrac{f^\pure}{f^\acc}} \quad
\mathtt{\rul{1-to-2} \dfrac{f^\acc}{f^\modi}} \quad
\mathtt{\rul{stow} \dfrac{f^{(d_1)}\eqs g^{(d_2)}}{f^{(d_1)}\eqw g^{(d_2)}}} \quad
\mathtt{\rul{wtos} \dfrac{f^{(d_1)}\eqw g^{(d_2)}  \squad d_1, d_2 \in \{0, 1\}}{f^{(d_1)}\eqs g^{(d_2)}}}  \\ 
\textbf{congruence rules} \\ 
\mathtt{\rul{refl} 
  \dfrac{f^{(d_1)}}{f^{(d_1)} \eqs f^{(d_1)}}} \qquad
\mathtt{\rul{sym} 
  \dfrac{f^{(d_1)} \eqs g^{(d_2)}}{g^{(d_2)} \eqs f^{(d_1)}}}  \qquad
\mathtt{\rul{trans} 
  \dfrac{f^{(d_1)} \eqs g^{(d_2)} \squad g^{(d_2)} \eqs h^{(d_3)}}{f^{(d_1)} \eqs h^{(d_3)}}}  
      \vspace{5pt} \\
  \mathtt{\rul{wrefl}  
  \dfrac{f^{(d_1)}}{f^{(d_1)} \eqw f^{(d_1)}}} \squad
\mathtt{\rul{wsym} 
  \dfrac{f^{(d_1)} \eqw g^{(d_2)}}{g^{(d_2)} \eqw f^{(d_1)}}} \quad
\mathtt{\rul{wtrans} 
  \dfrac{f^{(d_1)} \eqw g^{(d_2)} \squad g^{(d_2)} \eqw h^{(d_3)}}{f^{(d_1)} \eqw h^{(d_3)}}}  
    \vspace{5pt} \\
\mathtt{\rul{replsubs}  
  \dfrac{f_1^{(d_1)} \eqs f_2^{(d_2)} \colon X\to Y \squad g_1^{(d_3)} \eqs g_2^{(d_4)}\colon Y\to Z}
    {g_1^{(d_3)}\circ f_1^{(d_1)} \eqs g_2^{(d_4)}\circ f_2^{(d_2)}}} \\
\textbf{categorical rules} \\ 
\mathtt{ \rul{comp}  \dfrac{f^{(d_1)}\colon X\to Y \quad g^{(d_1)} \colon Y\to Z}
    {(g\circ f)^{(d_1)} \colon X\to Z}}  \quad
\mathtt{\rul{assoc} 
  \dfrac{f^{(d_1)}\colon X\to Y \squad g^{(d_2)}\colon Y\to Z \squad h^{(d_3)}\colon Z\to U}
  {h^{(d_3)}\circ (g^{(d_2)}\circ f^{(d_1)}) \eqs (h^{(d_3)}\circ g^{(d_2)})\circ f^{(d_1)}}}
      \vspace{5pt} \\
\mathtt{\rul{ids} 
  \dfrac{f^{(d_1)}\colon X\to Y}{f^{(d_1)}\circ id_X^\pure \eqs f^{(d_1)}}} \quad
\mathtt{\rul{idt}
  \dfrac{f^{(d_1)}\colon X\to Y}{id_Y^\pure\circ f^{(d_1)} \eqs f^{(d_1)}}}
        \vspace{5pt} \\
  \rul {tcomp} \dfrac{\mathtt{f^{(p)}: Y \to Z} \squad  \mathtt{g^{(p)}: X \to Y}}
{\mathtt{(tpure \ f)^\pure \circ (tpure\ g)^\pure \eqs (tpure \ (f \ {\color{red}\circ} \ g))^\pure} } 
\end{array}$$
{\vskip -0.25cm}
\renewcommand{\arraystretch}{1}
\caption{$\Log_{dec}$: rules} 
\label{dl-rls} 
\end{figure}
\begin{lemma}
\label{comp}
$\mathtt{\forall f^{(d_1)} \colon X\to Y\\, g^{(d_2)}\colon Y\to Z}$, the annotation $\mathtt{(g \circ f)^{(max(d_1,d_2))}}$
is admissible.
\end{lemma}
\begin{proof}
Trivially follows from case analyses on $\mathtt{d_1}$ and $\mathtt{d_2}$, and the rules (0-to-1), (1-to-2) and (comp). 
\end{proof}

Hierarchically, a \emph{pure} term can be seen as a \emph{constructor} (0-to-1), and
similarly a \emph{constructor} term can be seen as a \emph{modifier} on demand
(1-to-2).

It is obviously free to convert strong equations into weak ones (stow).
However, one has to make sure that the equated terms are not decorated with $\modi$
in order to convert weak equations into strong ones with no further evidence (wtos). 

Both strong and weak equalities are defined to be \emph{equivalence relations} with the assumption that
they are \emph{reflexive}, \emph{transitive} and \emph{symmetric}.
Strong equations form a congruence relation but weak equations do not:
we will see this in detail when we specialize the decorated logic for the
global state and the exception effects in Sections~\ref{dlst}~and~\ref{dlexc},
respectively.

The categorical rules present properties of the term composition:
the decoration of a composition depends on the decoration of its components, always
taking the larger.
I.e., $\mathtt{\forall\, f^\pure\colon X\to Y}$ and $\mathtt{g^\modi\colon Y\to Z}$,
$\mathtt{g \circ f\colon X \to Z}$ takes the decoration $\modi$ (Lemma~\ref{comp}). Composition is an
associative
operation (assoc). The identity term
disappears when to compose on the right (ids), and on the left (idt).
The rule (tcomp)
states that the \texttt{tpure}
constructor preserves the composition of pure terms
up to the strong equality. Meaning that one can first compose pure terms 
outside the decorated environment (in any pure type system) and use the \texttt{tpure}
constructor to translate them into the
$\Log_{dec}$, or translate the terms into the
$\Log_{dec}$ first, and then compose them there.  Notice that
the red colored composition symbol ({\color{red} $\circ$}), in the rule conclusion, stands for the
composition operation for pure terms.
The decoration \texttt{(p)} of terms \texttt{f} and \texttt{g}
is used just to denote the \emph{pure}
terms outside decorated environment, thus it does not take part in
the decorated logic syntax. Similar case applies to the (tcomp) rule
given in Figure~\ref{rls-stexc}.

 \section{The Decorated Logic for the state effect ($\Log_{st}$)}
 \label{dlst}

%
The use and modification of the memory state is the fundamental feature of imperative
languages, and considered as a sort of computational side effect.
In this section, we present a proof
system for the use of the global state which involves access and modify
operations, called the \emph{decorated logic for the state effect} ($\Log_{st}$).
This logic is obtained by extending the generic framework presented in Section~\ref{ss:dls}.
In this case, the decoration $\pure$ is reserved for 
{\em pure} terms, while $\acc$ is for {\em read-only} ({\em accessor})
 and $\modi$ is for {\em read-write} ({\em modifier}) terms.
Two terms are called strongly equal if they return the
same result with the same state manipulation; they are called weakly equal
if they return the same result with different state manipulations.
\begin{figure}[h]
\renewcommand{\arraystretch}{1.25}
$$\begin{array}{lccl}
\multicolumn{4}{l}{
\textbf{Grammar of the decorated logic for the state: }\qquad (\texttt{i} \in \texttt{Loc})} \\
\quad \textrm{Types: } &\mathtt{t,\, s} &::=&
 \mathtt{X}\mid \mathtt{Y} \mid \dots \mid  \mathtt{t\times s}\mid \unit\mid \mathtt{V_{i}} \\
 \quad \textrm{Decorations for terms: } &\mathtt{(d_1), (d_2)} &::=&\pure \mid \acc \mid \modi \\ 
\quad \textrm{Terms: } &\mathtt{f,\;g} &::=& \mathtt{a^\dec} \mid \mathtt{b^\dec}
\mid \dots \mid\mathtt{g\circ f^\dec}\mid  \\ & & & 
\mathtt{\lpair{f^{(d_1)}\colon X \to Y,g^{(d_2)}\colon X \to Z}^{(max(d_1, d_2))}\colon X \to Y  \times  Z} \mid
 \\ & & &
  \lookup_\mathtt{i}^\acc \mid 
  \update_\mathtt{i}^\modi  \mid  (\mathtt{tpure\ \bigcdot})^\pure \\  
\quad \textrm{Equations: } &\mathtt{eq} &::=& \mathtt{f^\dec} \eqs \mathtt{g^\dec} \mid \mathtt{f^\dec} \eqw \mathtt{g^\dec}
\end{array}$$
{\vskip -0.5cm}
\caption{$\Log_{st}$: syntax}
\label{syn-ldecfst}
\end{figure}
 Figure~\ref{syn-ldecfst} shows the grammar of the $\Log_{st}$ where
$\mathbbm{1}$ is the singleton type while $\mathtt{V_i}$ is the
type of values that can be stored in any location \texttt{i}. We assume that
there is a set of locations called $\mathtt{Loc}$. Given types $\mathtt{X}$ and $\mathtt{Y}$, we have
$\mathtt{X\times Y}$ representing type products.

Terms are
closed under composition ($\circ$) and pairing ($\mathtt{\lpair{\_, \_}}$). I.e., for all terms
$\mathtt{f\colon X\to Y}$ and $\mathtt{g\colon Y\to Z}$, we have $\mathtt{g \circ f\colon X\to Z}$. Similarly,
for all
$\mathtt{f\colon X\to Y}$ and $\mathtt{g\colon X\to Z}$, there is  $\mathtt{\lpair{f, g}\colon X\to Y\times Z}$.
Notice that the pair subscript `$\mathtt{l}$'
denotes the left pairs. One can define in a symmetric way the
right pairs for terms $\mathtt{f\colon X\to Y}$ and $\mathtt{g\colon X\to Z}$ as
$\mathtt{\rpair{f,g}:= permut \circ \lpair{g,f}}$ where $\mathtt{permut := \lpair{\pi_2,\pi_1}}$.
In the same way, one can respectively obtain left and right products of terms $\mathtt{f\colon X_1\to Y_1}$ and
$\mathtt{g\colon X_2\to Y_2}$ as $\mathtt{f\times_l g:= \lpair{f \circ \pi_1, g\circ \pi_2}}$ and
$\mathtt{f\times_r g:= \rpair{f \circ \pi_1, g\circ \pi_2}}$.
The term pairs/products are used to impose
some order of term evaluation since the evaluation result depends on the order
that the mutable state is accessed/modified. I.e.,
the product of two terms can be intuitively interpreted as they run on the global state in parallel, while
sequential products, put forward in~\cite[\S 2.3]{Dumas:2014b}, enforce terms to use the state in sequence.
%
The decoration of a pair/product depends on the decoration of its components, always
taking the larger.
I.e., $\mathtt{\forall\, f^\pure\colon X\to Y}$ and $\mathtt{g^\modi\colon X\to Z}$,
the term
$\mathtt{\lpair{f,g}\colon X\to Y\times Z}$ takes the decoration $\modi$.
Note that in $\Log_{st}$, we do not necessarily stick to the sequential products,
even pairs/products of modifiers (intuitively parallel execution of modifiers)
are allowed to be constructed.
However, they cannot be used
in the provided equational reasoning, since they may lead to conflicts on the returned result
due to possible hazardous parallel modifications of the global state.
We can have equational reasoning only when the left component is at most an accessor.
This restriction is given by the rules (w$\_$lpair$\_$eq)  and (s$\_$lpair$\_$eq) in Figure~\ref{rls-st}.
In the Coq implementation of this logic, as detailed in Section~\ref{lst-coq}, we
only allow the construction of pairs/products of modifiers under contradictory assumptions. See the
constructor $\mathtt{is\_pair}$ of the inductive type \texttt{is}.

The interface terms are $\lookup_\mathtt{i}
\colon \unit \times \mathtt{S} \to \mathtt{V_i}$ and $\update_\mathtt{i}\colon \mathtt{V_i}  \times \mathtt{S} \to \unit \times \mathtt{S}$
where $\mathtt{S}$ denotes the distinguished object of states which never appears in the decorated setting.
The use of decorations provides a new schema where term signatures are constructed
without any occurrence of the state object. For instance,
 $\lookup^\acc_\mathtt{i}
\colon \unit \to \mathtt{V_i}$ is an accessor while
 $\update_\mathtt{i}^\modi\colon \mathtt{V_i} \to \unit$ is a modifier. 
This way, we keep signatures close to their syntax and
compose compatible terms as usual.
The term $\lookup$ reads the value stored in a given location while $\update$
modifies it. We can call them \emph{the unique sources of impurity}, since only 
the terms including  $\lookup$ or $\update$ are impure; meaning those do not include them are pure
with respect to the state effect.

The identity term $\mathtt{id}$,
the canonical pair projections $\mathtt{\pi_1}$, $\mathtt{\pi_2}$,
the empty pair $\mathtt{\pa}$
and \texttt{constants} are
translated from a pure type system with type
products using the \texttt{tpure} constructor, for all types \texttt{X} and \texttt{Y}, as follows:
$$\begin{array}{l}
   \xymatrix@R=0.01pc@C=0.5pc{
   \mathtt{id_X^\pure} &\colon& \mathtt{X\to X} &:=& \mathtt{tpure\  (\lambda\, x:\, X.\, x: \, X)}\\
    \mathtt{\pi_1^\pure} &\colon& \mathtt{X\times Y\to X} &:=& \mathtt{tpure \ fst}\\
    \mathtt{\pi_2^\pure} &\colon& \mathtt{X\times Y\to Y} &:=& \mathtt{tpure \ snd}\\
    \mathtt{\pa_X^\pure} &\colon& \mathtt{X \to \unit} &:=& \mathtt{tpure\  (\lambda\ x:\ X.\ void: \ \unit)}\\
    \mathtt{constant_x^\pure} &\colon& \mathtt{\unit \to X} &:=& \mathtt{tpure \ (\lambda\ \_.\ x: X)}\\
   }
\end{array}$$
where \texttt{fst} and \texttt{snd} are constructors of product types.
%

%

The intended model of the above grammar is built with respect to the set of states $\mathtt{S}$
where a pure term $\mathtt{p^\pure: X\to Y}$ is interpreted as a function $\mathtt{p: X\to Y}$,
an accessor $\mathtt{a^\acc: X\to Y}$ as a function $\mathtt{a: X\times S \to Y}$, and a
modifier $\mathtt{m^\ctc: X\to Y}$ as a function $\mathtt{m: X\times S\to Y\times S}$. 
The complete and detailed category theoretical model is given in~\cite[\S 5.1]{Ekici:2015t}.
\begin{figure}[h]
\renewcommand{\arraystretch}{1.25}
\begin{tabular}{l} 
%
%
\textbf{Rules of the decorated logic for the state: }\\ 
(pwrepl)$\mathtt{\dfrac{f_1^{(d_1)}\eqw f_2^{(d_2)}\colon X\to Y \squad g^\pure\colon Y\to Z} {g^\pure\circ f_1^{(d_1)} \eqw g^\pure\circ f_2^{(d_2)} }}$  
\squad
(wsubs)$\mathtt{\dfrac{g^{(d_3)}\colon X\to Y \squad f_1^{(d_1)}\eqw f_2^{(d_2)}\colon Y\to Z} {f_1^{(d_1)} \circ g^{(d_3)} \eqw f_2^{(d_2)}\circ g^{(d_3)} }}$ \squad
\vspace{5pt} \\
(replsubs) 
  $\mathtt{\dfrac{f_1^{(d_1)} \eqs f_2^{(d_2)} \colon X\to Y \squad g_1^{(d_3)} \eqs g_2^{(d_4)}\colon Y\to Z}
    {g_1^{(d_3)}\circ f_1^{(d_1)} \eqs g_2^{(d_4)}\circ f_2^{(d_2)}}}$ \squad
(w$\_$unit) 
  $\dfrac{\mathtt{f^{(d_1)}\colon X\to \unit}}{\mathtt{f^{(d_1)} \eqw \pa_X^\pure}}$ 
  \vspace{5pt} \\
(ax$_1$) 
  $\dfrac{}{\lookup_\mathtt{i}^\acc \circ \update_\mathtt{i}^\modi \eqw \mathtt{id_{V_i}}^\pure} $ \squad
(ax$_2$)
  $\dfrac{\mathtt{\forall i, j\in Loc, \ i \neq j}}
  {\lookup_\mathtt{i}^\acc \circ \update_\mathtt{j}^\modi \eqw \lookup_\mathtt{i}^\acc \circ
  \pa_{V_j}^\pure} $ 
  \vspace{5pt} \\
(effect) 
  $\dfrac{\mathtt{f_1^{(d_1)},f_2^{(d_2)}\colon X\to Y \;\;
  f_1^{(d_1)}\eqw f_2^{(d_2)} \;\; \pa_Y^\pure\circ f_1^{(d_1)} \eqs \pa_Y^\pure\circ f_2^{(d_2)}}}{\mathtt{f_1^{(d_1)}\eqs f_2^{(d_2)}}}$
  \vspace{5pt} \\
(local$\_$global) 
  $\dfrac{\mathtt{f_1^{(d_1)},f_2^{(d_2)}\colon X\to \unit \quad \forall\ \mathtt{i\in Loc}, \, \lookup_\mathtt{i}^\acc\circ f_1^{(d_1)} \eqw \lookup_\mathtt{i}^\acc\circ f_2^{(d_2)}}}
    {\mathtt{f_1^{(d_1)}\eqs f_2^{(d_2)}}}$
    \vspace{5pt} \\
(w$\_$lpair$\_$eq)  
  $\dfrac{\mathtt{f_1^{(d_1)}: X \to Y \squad f_2^{(d_2)}: X \to Z
  \squad d_1\in\{0,1\}}}{\mathtt{\pi_1^\pure \circ \lpair{f_1,f_2}^{(max(d_1,d_2))} \sim f_1^{(d_1)}}}$ 
  \vspace{5pt} \\
(s$\_$lpair$\_$eq) 
  $\dfrac{\mathtt{f_1^{(d_1)}: X \to Y \squad f_2^{(d_2)}: X \to Z\squad d_1\in\{0,1\}}}{\mathtt{\pi_2^\pure \circ \lpair{f_1,f_2}^{(max(d_1,d_2))} \equiv f_2^{(d_2)}}}$
\end{tabular}
{\vskip -0.15cm}
\caption{$\Log_{st}$: rules}
\label{rls-st} 
\end{figure}
The syntax given in  Figure~\ref{syn-ldecfst} is enriched with two sets of rules
presented in Figures~\ref{rls-st} and~\ref{dl-rls}.
%
%
%
Weak equalities do not form a congruence:
the term replacement cannot be done unless the replaced term is pure. I.e.,
given an equation $\mathtt{f_1^{(d_1)} \eqw f_2^{(d_2)}\colon X\to Y}$ and a
term $\mathtt{g\colon Y\to Z}$,  it is possible to get the
equation $\mathtt{g \circ f_1 \eqw g \circ f_2}$ only when the term $\mathtt{g}$ is pure.
At this stage, we have no information about the modifications that
$\mathtt{f_1}$ and $\mathtt{f_2}$ make on the memory state. Therefore, the post
executed impure term $\mathtt{g}$ would destroy this result equality, for instance
by reading the location $\mathtt{i}$ on which $\mathtt{f_1}$ and $\mathtt{f_2}$ has performed different
modifications (pwrepl).
However,
the term substitution can be done regardless of the term decoration.
I.e., given the equation $\mathtt{f_1^{(d_1)} \eqw f_2^{(d_2)}\colon Y\to Z}$ and a
term $\mathtt{g^{(d_3)}\colon X\to Y}$, it is possible to get the
equation $\mathtt{f_1 \circ g \eqw f_2 \circ g}$ independent from the decoration of the term $\mathtt{g}$.
We already now that $\mathtt{f_1}$ and $\mathtt{f_2}$ return the same result, executing any
term $\mathtt{g}$ in advance would not end them returning different results  (wsubs).
Strong equalities form a congruence by allowing both term substitutions and
replacements independent from the term decorations (replsubs).
%

Any term $\mathtt{f\colon X \to \unit}$ with no result returned ``\texttt{void}''
(the unique inhabitant of $\unit$ type) 
has an obvious result equality with the canonical empty pair 
$\mathtt{\pa_X}$ (w$\_$unit).
%
%

The fundamental equations are given with the rules (ax$_1$) and (ax$_2$).
The former states that
by updating the location \texttt{i} with a value \texttt{v}
and then observing the same location, one gets the value \texttt{v}.
This outputs the same value with the identity term $\mathtt{id_{V_i}}$, if it takes \texttt{v}
as an argument. However, notice that these two ways of getting the value \texttt{v} have
different state manipulations which makes them \emph{weakly equal}.
The latter, (ax$_2$), 
is to assume that updating the location \texttt{j} with a value
$\mathtt{v}$ and then reading the content of a different location \texttt{i} would return the same value
with first throwing out the value $\mathtt{v}$ then observing the content of the location \texttt{i}. They
have different manipulations on the state so that they are \emph{weakly equal}. 

Two modifiers $\mathtt{f_1^\modi,f_2^\modi:X\to Y}$ modify the state in the same way if 
and only if $\mathtt{\pa_Y\circ f_1 \eqs \pa_Y\circ f_2:X\to \unit}$, where
$\mathtt{\pa_Y:Y\to\unit}$ throws out the returned value.
So that $\mathtt{f_1^\modi,f_2^\modi:X\to Y}$ are \emph{strongly equal}  if and only if 
$\mathtt{f_1\eqw f_2}$ and $\mathtt{\pa_Y\circ f_1 \eqs \pa_Y\circ f_2}$ (effect).
Notice that this rule is valid also for the other decorations of terms 
$\mathtt{f_1}$ and $\mathtt{f_2}$.
%

Locally, the strong equality between two modifiers
$\mathtt{f_1^\modi,f_2^\modi:X\to \unit}$ can also be expressed as a pair of weak equations:  
$\mathtt{f_1\eqw f_2}$ and 
$\mathtt{\forall \mathtt{i:}Loc,
\lookup_\mathtt{i} \circ f_1 \eqw \lookup_\mathtt{i} \circ f_2}$.
The latter intuitively means that $\mathtt{f_1}$ and $\mathtt{f_2}$ leaves the memory with the same values
stored in all (finitely many) locations after being executed.
Given that both return ``\texttt{void}''
there is no explicitly need to check if $\mathtt{f_1\eqw f_2}$.
It suffices to see whether $\mathtt{\forall \mathtt{i:}Loc, \lookup_\mathtt{i} \circ f_1 \eqw \lookup_\mathtt{i} \circ f_2}$
to end up with $\mathtt{f_1\eqs f_2}$ (local$\_$global).
The rule is valid also for the other decorations of terms 
$\mathtt{f_1}$ and $\mathtt{f_2}$.

With (w$\_$lpair$\_$eq) and (w$\_$rpair$\_$eq) term pairs are characterized: 
the (left) pair structure 
$\mathtt{\lpair{f_1,f_2}}$ cannot be used when $\mathtt{f_1}$ and $\mathtt{f_2}$, both are modifiers,
since it may lead to a conflict on the returned result. 
However, it can be used only when $\mathtt{f_1}$ is an accessor. 
We state by (w$\_$lpair$\_$eq) that
$\mathtt{\lpair{f_1,f_2}^{(max(d_1,d_2))}}$ has only result equality with $\mathtt{f_1^{(d_1)}}$ and
by (w$\_$rpair$\_$eq) that it has
both result and effect equality with $\mathtt{f_2^{(d_2)}}$.


These rules are designed to be sound with respect to a categorical model
detailed in~\cite[\S 5.2, \S 5.3, \S 5.4, \S 5.5]{Ekici:2015t}. However, their syntactic completeness is not immediate.
\cite{DBLP:conf/macis/DumasDEPR15}
defines a new syntactic completeness property,
subsuming a consistency check, called the relative Hilbert-Post completeness.
In \cite[\S 5.4]{Ekici:2015t}, it is proven that this set of rules is complete with due respect. 

\subsection{Decorated properties of the memory state}
\label{dpotms}
In~\cite[\S 3]{Plotkin:2002},
an equational representation of the mutable state has been introduced.
The decorated version of such representation is given as follows:
\begin{itemize}
\item[(1)$_d$] 
Annihilation lookup-update.
\textit{Reading the content of
a location $\mathtt{i}$ and then updating it with the
obtained value is just like doing nothing.}
 $\mathtt{ \forall\, i\in Loc ,\;
  \update^\modi_i \circ \lookup^\acc_i \eqs id^\pure_\unit 
  : \unit \to \unit }$.
\item[(2)$_d$] 
Interaction lookup-lookup.
\textit{Reading twice the same
location $\mathtt{i}$ is the same as reading it once.}
 $\mathtt{ \forall\, i\in Loc ,\;
  \lookup^\acc_i\circ \pa^\pure_{V_i} \circ \lookup^\acc_i \eqs \lookup^\acc_i 
  : \unit \to V_i }$.
\item[(3)$_d$] 
Interaction update-update. 
\textit{Storing value the values  $\mathtt{x}$ and  $\mathtt{y}$
in a row to the same location~ $\mathtt{i}$ is just like storing  $\mathtt{y}$ in it.} 
 $\mathtt{\forall\, i\in Loc ,\;\; 
  \update^\modi_i \circ \pi^\pure_2 \circ (\update_i^\modi\times_r id^\pure_{V_i}) \eqs}\,
  \mathtt{\update^\modi_i} \circ \mathtt{\pi^\pure_2 : V_i\times V_i \to \unit }$.  
\item[(4)$_d$] 
Interaction update-lookup. 
\textit{Storing the value  $\mathtt{x}$ in a location  $\mathtt{i}$ and then reading the content of
 $\mathtt{i}$, one gets the value  $\mathtt{x}$.} 
 $\mathtt{\forall\, i\in\Loc ,\;  
  \lookup^\acc_i\circ \update^\modi_i \eqw id^\pure_{V_i} 
  : V_i \to V_i}$.
\item[(5)$_d$] 
Commutation lookup-lookup.
\textit{The order of reading
two different locations  $\mathtt{i}$ and  $\mathtt{j}$ does not matter.} 
  $\mathtt{ \forall\, i\ne j\in Loc ,\; 
  (id_{V_i}^\pure \times_r \lookup_j^\acc) \circ \pi_1^{-1\pure} 
  \circ \lookup_i^\acc \eqs
  permut_{j,i}^\pure \circ  (id_{V_j}^\pure \times_r \lookup_i^\acc)\, \circ} \\ 
  \mathtt{ \pi_1^{-1\pure} \circ \lookup_j^\acc 
  : \unit \to V_i\times V_j}$ where $\mathtt{\pi_1^{-1\pure} := \lpair{id, \pa}^\pure}$.
\item[(6)$_d$] 
Commutation update-update.
\textit{The order of storing
in two different locations  $\mathtt{i}$ and  $\mathtt{j}$ does not matter.}  
 $\mathtt{ \forall\, i\ne j\in Loc ,\;
   \update^\modi_j \circ \pi^\pure_2 \circ (\update^\modi_i \times_r id^\pure_{V_j}) \eqs}\,
   \mathtt{\update^\modi_i} \circ \mathtt{\pi^\pure_1} \circ \\ 
   \mathtt{(id^\pure_{V_i} \times_l \update^\modi_j):}
   \mathtt{V_i\times V_j \to \unit }$. 
\item[(7)$_d$] 
Commutation update-lookup.
\textit{The order of storing
in a location  $\mathtt{i}$ and reading another location~ $\mathtt{j}$ does not
matter.} 
  $\mathtt{\forall\, i\ne j\in Loc ,\; 
  \lookup^\acc_j \circ \update^\modi_i \eqs 
  \pi^\pure_2 \circ (\update^\modi_i \times_r id^\pure_{V_j})} 
    \circ \\ \mathtt{(id^\pure_{V_i } \times_l \lookup^\acc_j)}  \circ 
    \mathtt{\pi^{-1\pure}_1 
  : V_i \to V_j}$.
\item[(8)$_d$] Commutation lookup-constant.
\textit{Just after storing a constant  $\mathtt{c}$ in a location  $\mathtt{i}$,
observing the content of  $\mathtt{i}$ is the same as regenerating the
constant  $\mathtt{c}$.} 
 $\mathtt{\forall\, i\in Loc ,\ \forall\, c\in V_i;\,
  \lookup^\acc_i \circ \update^\modi_i \circ} \\ \mathtt{constant\, c^\pure} \eqs
  \mathtt{constant\, c^\pure \circ \update^\modi_i \circ constant\, c^\pure
  : \unit \to V_i}$.
\end{itemize} 

These are the archetype properties that we have proved within the scope of the
 logic $\Log_{st}$. To see these
 proofs,
 check out author's PhD thesis
 \cite[\S 5.3]{Ekici:2015t}.
 Besides, we have implemented the $\Log_{st}$ in \coq to certify mentioned proofs. Section~\ref{lst-coq}
 details this implementation.

%

 \subsection{$\Log_{st}$ in \coq}
 \label{lst-coq}
In this section, we aim to highlight some crucial points of
the $\Log_{st}$ implementation in \coq. It mainly consists of four steps:
(1) implementing the terms, (2) assigning the decorations over terms,
(3) stating the rules, and (4) proving properties of the memory state
referred in Section~\ref{dpotms}.

We represent the set of
memory locations by a Coq parameter $\mathtt{Loc: Type}$. Since memory locations may contain
different types of values, we also assume an arrow type $\mathtt{Val: Loc \to Type}$
that is the type of values contained in
each location. This fixes a type for every location.
Note that the system thus does not support reasoning about \emph{strong updates}.


\begin{tcb}[\scriptsize]{Coq}
Parameters (Loc: Type)  (Val: Loc -> Type).
\end{tcb}
We define the terms of $\Log_{st}$ using an inductive predicate called
$\mathtt{term}$. It establishes a new \coq \texttt{Type} out of two input \texttt{Type}s. 
The type $\mathtt{term \ Y \ X}$ is dependent. It depends on the \texttt{Type} instances
\texttt{X} and \texttt{Y}, and represents the arrow type $\mathtt{X \to Y}$ in the
decorated framework. As opposed to a flat grammar with a typing predicate, we prefer
a dependently typed implementation for higher readability.

\begin{tcb}[\scriptsize]{Coq}
 Inductive term: Type -> Type -> Type :=
  | tpure:  forall {X Y: Type}, (X -> Y) -> term Y X
  | comp:   forall {X Y Z: Type}, term X Y -> term Y Z -> term X Z
  | pair: forall {X Y Z: Type}, term X Z -> term Y Z -> term (X*Y) Z
  | lookup: forall i:Loc, term (Val i) unit    
  | update: forall i:Loc, term unit (Val i). 
 Infix "o" := comp (at level 70).
\end{tcb}
The constructor $\mathtt{tpure}$ takes a Coq side (pure) function and translates it 
into the decorated environment. 
The \texttt{comp} constructor deals with the composition of two compatible terms. I.e., given a pair
of terms $\mathtt{f : term \ X \ Y}$ and $\mathtt{g: term \ Y \ Z}$, then the composition $\mathtt{f \circ g}$ would be an
instance of
the type $\mathtt{term \ X \ Z}$. For the sake of conciseness, infix `$\circ$' is used to denote the term composition.
Similarly, the (left) \texttt{pair} constructor is to constitute pairs of compatible terms. I.e., given
$\mathtt{f: term \ Y \ X}$ and $\mathtt{g: term \ Z \ X}$, we have pair $\mathtt{\lpair{f, g}: term \ (Y \times Z) \ X}$.
Instead of the symbol $\mathtt{\lpair{\_,\_}}$, we use the keyword \texttt{pair} in the implementation.
The terms $\mathtt{\lookup}$ and $\mathtt{\update}$ come as no surprise; just that the singleton type $\unit$
and the type of values $\mathtt{V_i}$ are
respectively called \texttt{unit} and $\mathtt{Val \ i}$ in the code.
The terms
such as the identity, the pair projections, the empty pair and the constant function can
be derived from the native Coq functions with the use of \texttt{tpure} constructor as follows:
\begin{tcb}[\scriptsize]{Coq}
 Definition id  {X: Type}     : term X X      := tpure id.
 Definition pi1 {X Y: Type}   : term X (X*Y)  := tpure fst. 
 Definition pi2 {X Y: Type}   : term Y (X*Y)  := tpure snd.
 Definition forget {X}        : term unit X   := tpure (fun _ => tt). 
 Definition constant {X: Type} (v: X): term X unit := tpure (fun _ => v).
\end{tcb}
Remark that
\texttt{id} is overloaded: defined one (on the left) is the identity of the decorated logic while the other one
is the identity of \coq's logic.
The pair projections are named \texttt{pi1} and \texttt{pi2} while the unique mapping
$\mathtt{\pa_{X}}$ from any type $\mathtt{X}$ to $\unit$ is named \texttt{forget} in the implementation.

The decorations are enumerated under the new type called \texttt{kind}: \texttt{pure} (0), \texttt{ro} (1) and \texttt{rw} (2) and
inductively assigned to \texttt{term}s via the predicate called \texttt{is}. This predicate 
builds a proposition out of a \texttt{term} and a 
decoration. I.e., $\mathtt{\forall i: Loc}$, \texttt{is ro (lookup i)} is a \texttt{Prop} instance, ensuring that  ``\texttt{lookup i}'' is an accessor.

Notice that on the paper, we always mention the decoration of a term as a superscript. However, with such a Coq implementation,
we do not need to additionally carry that information with a term. Instead, we inject it inside the rules as predicates, and check
if a rule is applicable or not via this information. See Remark~\ref{rmrules}. 
\begin{tcb}[\scriptsize]{Coq}
 Inductive kind := pure | ro | rw.
 Inductive is: kind -> forall X Y, term X Y -> Prop :=
  | is_tpure: forall X Y (f: X -> Y), is pure (@tpure X Y f)
  | is_comp: forall k X Y Z (f: term X Y) (g: term Y Z), is k f -> is k g -> is k (f o g)
  | is_pair: forall k X Y Z (f: term X Z) (g: term Y Z), is ro f -> is k f -> is k g -> is k (pair f g)
  | is_lookup: forall i, is ro (lookup i)   
  | is_update: forall i, is rw (update i)
  | is_pure_ro: forall X Y (f: term X Y), is pure f -> is ro f
  | is_ro_rw: forall X Y  (f: term X Y), is ro f -> is rw f.
\end{tcb}
Any term that is built by the \texttt{tpure} constructor is pure ($\mathtt{is\_tpure}$).
The decoration of any term composition
depends on its components and always takes the upper decoration
($\small\mathtt{pure}$ \textless \ $\mathtt{ro}$ \textless \ $\mathtt{rw}$).
E.g., given a modifier term and a read-only term, their composition
will be a modifier, as well.
This trivially follows from ($\mathtt{is\_comp}$),
($\mathtt{is\_pure\_ro}$) and ($\mathtt{is\_ro\_rw}$): see Lemma~\ref{comp},
and the corresponding Coq proof
here 
\footnote{\url{https://github.com/ekiciburak/decorated-logic-for-states-effect/blob/master/Decorations.v\#L76-L79}}.
The decoration of a (left) pair of terms also depends on its
components always taking the upper with the restriction that
the first component can at most be an accessor.
This is also trivial given ($\mathtt{is\_pair}$),
($\mathtt{is\_pure\_ro}$) and ($\mathtt{is\_ro\_rw}$). See the 
Coq proof of this fact here 
\footnote{\url{https://github.com/ekiciburak/decorated-logic-for-states-effect/blob/master/Decorations.v\#L81-L84}}.
We declare that the term \texttt{lookup} is an accessor ($\mathtt{is\_lookup}$), and 
the term \texttt{update} is a modifier ($\mathtt{is\_update}$).
The last two constructors ($\mathtt{is\_pure\_ro}$) and ($\mathtt{is\_ro\_rw}$)
define the decoration hierarchies.

It is easy to derive that any \texttt{tpure} built term is pure. I.e., the purity proof of
the first pair projection:
\begin{tcb}[\scriptsize]{Coq}
Lemma is_pi1 X Y: is pure (@pi1 X Y).
Proof. apply is_tpure. Qed.
\end{tcb}
We now state the rules up to weak
and strong equalities by defining them in a mutually inductive way: mutuality
here is used to enable the constructors including both weak and strong equalities.
We use the notation $\mathtt{==}$ and $\mathtt{\eqw}$ to denote strong and
weak equalities, respectively.
\begin{tcb}[\scriptsize]{Coq}
 Definition idem X Y (x y: term X Y) := x = y.
 Inductive strong: forall X Y, relation (term X Y) :=
 | refl X Y: Reflexive (@strong X Y)
 | sym: forall X Y, Symmetric (@strong X Y)
 | trans: forall X Y, Transitive (@strong X Y)
 | replsubs: forall X Y Z, Proper (@strong X Y ==> @strong Y Z ==> @strong X Z) comp
 | ids: forall X Y (f: term X Y), f o id == f
 | idt: forall X Y (f: term X Y), id o f == f
 | assoc: forall X Y Z T (f: term X Y) (g: term Y Z) (h: term Z T), f o (g o h) == (f o g) o h
 | wtos: forall X Y (f g: term X Y), is ro f -> is ro g -> f ~ g -> f == g
 | s_lpair_eq: forall X Y' Y (f1: term Y X) (f2: term Y' X), is ro f1 -> pi2 o pair f1 f2 == f2
 | effect: forall X Y (f g: term Y X), forget o f == forget o g -> f ~ g -> f == g
 | local_global: forall X (f g: term unit X), (forall i: Loc, lookup i o f ~ lookup i o g) -> f == g
 | tcomp: forall X Y Z (f: Z -> Y) (g: Y -> X), tpure (compose g f) == tpure g o tpure f
 with weak: forall X Y, relation (term X Y) :=
 | wsym: forall X Y, Symmetric (@weak X Y)
 | wtrans: forall X Y, Transitive (@weak X Y)
 | pwrepl: forall A B C (g: term C B), (is pure g) ->  Proper (@weak B A ==> @weak C A) (comp g)
 | wsubs: forall A B C, Proper (@weak C B ==> @idem B A ==> @weak C A) comp
 | stow: forall  X Y (f g: term X Y), f == g -> f ~ g
 | w_lpair_eq: forall X Y' Y (f1: term Y X) (f2: term Y' X), is ro f1 -> pi1 o pair f1 f2 ~ f1
 | w_unit: forall X (f g: term unit X), f ~ g    
 | ax1: forall i, lookup i o update i ~ id
 | ax2: forall i j, i //n j -> lookup j o update i ~ lookup j o forget
   where "x == y" := (strong x y) and "x ~ y" := (weak x y).
\end{tcb}
\noindent
The rule \texttt{tcomp}
states that the \texttt{tpure}
constructor preserves the composition of pure
terms up to the strong equality: one can first compose pure terms on \coq side
(using higher order function \texttt{compose})
and then apply \texttt{tpure} constructor to translate them into decorated settings or
can translate the terms first and then compose them in decorated settings.
\begin{remark}
\label{rmrules}
In a decorated logic, it is crucial to verify the decorations of the terms in applying/rewriting a rule. If the rule is
applicable for all decorations, then it is not necessary to check the decorations of
terms which appear in that rule.
Otherwise put, decoration checks are necessary only when the rule premise has restrictions over term decorations. I.e.,
see the constructor $\mathtt{w\_lpair\_eq}$ above.
We apply the same strategy for the logics presented
in Sections~\ref{dlexc} and~\ref{dlstexc} when implementing them in \coq.
\end{remark}
This framework allows us to express and prove, in \coq, the decorated versions of the properties mentioned in
Section~\ref{dpotms}. E.g.,  the statement \texttt{commutation update-update} looks like:
\begin{tcb}[\scriptsize]{Coq}
(**  Commutation update update **)
Theorem CUU: forall i j: Loc, i//nj ->  update j o (pi2 o (rprod (update i) (@id (Val j)))) == 
					update i o (pi1 o (lprod (@id (Val i)) (update j))). 
\end{tcb}
\noindent
where
\begin{tcb}[\scriptsize]{Coq}
Definition permut {X Y}: term (X*Y) (Y*X) := pair pi2 pi1.
Definition rpair {X Y Z} (f: term Y X) (g: term Z X): term (Y*Z) X := permut o pair g f.
Definition lprod {X Y X' Y'} (f: term X X') (g: term Y Y'): term (X*Y) (X'*Y') := pair (f o pi1) (g o pi2).
Definition rprod {X Y X' Y'} (f: term X X') (g: term Y Y') := permut o pair (g o pi2) (f o pi1).
\end{tcb}
\noindent
The full \coq proofs of such properties can be found here \footnote{\url{https://github.com/ekiciburak/decorated-logic-for-states-effect/blob/master/Proofs.v}}, and the entire implementation there
\footnote{\url{https://github.com/ekiciburak/decorated-logic-for-states-effect}}.
%

 \section{The Decorated Logic for the exception effect ($\Log_{exc}$)}
 \label{dlexc}

Exception handling is provided by most modern programming languages to
deal with anomalous or exceptional events which require special processing.
In this section, we present a proof
system for exceptions, which involves raising and handling operations,
called the \emph{decorated logic for the exception effect} ($\Log_{exc}$).
This logic is obtained by extending the generic framework presented in Section~\ref{ss:dls}.
In this context,
the decoration $\pure$ is reserved for 
{\em pure} terms, while $\ppg$ is for {\em propagators}
and $\modi$ is for {\em catchers}.
%
A fundamental feature of the exceptions mechanism is the distinction 
between \textit{ordinary} (\textit{non-exceptional}) values
and \textit{exceptions} (or \emph{exceptional values}). 
%
Two terms are called strongly equal if they behave the same on ordinary and exceptional 
values; they are called weakly equal
if they behave the same on ordinary values but differently on exceptional ones.

It has been shown by~\cite{Dumas:2012b} that the core part of this
proof system is dual to one for the state ($\Log_{st}$). Based on this nice duality,
we build the logic $\Log_{exc}$, and detail it in the following.
{\vskip -0.65cm}
\begin{figure}[h]
\renewcommand{\arraystretch}{1.25}
$$\begin{array}{lccl}
\multicolumn{4}{l}{
\textbf{ Grammar of the decorated logic for the exception: }\qquad (\texttt{e} \in \texttt{EName})} \\
\quad \textrm{Types: } &\mathtt{t,\, s} &::=&
 \mathtt{X}\mid \mathtt{Y} \mid \dots \mid  \mathtt{t\splus s}\mid \empt\mid \mathtt{EV_e}\\
 \quad \textrm{Decoration for terms: } &\mathtt{(d_1), (d_2)} &::=&\pure \mid \acc \mid \modi \\ 
\quad \textrm{Terms: } &\mathtt{f,\;g} &::=&  \mathtt{a}^\mathtt{(d)} \mid \mathtt{b}^\mathtt{(d)}\mid \dots
\mid \mathtt{g\circ f}^\mathtt{(d)}\mid \\ & & & 
\mathtt{\lcopair{f^{(d_1)}\colon X\to Y\mid g^{(d_2)}\colon Z\to Y} }^\mathtt{(max (d_1, d_2))}\mathtt{\colon X + Z \to Y} \mid \\
 & & & \tagg _e^\acc \mid \untag_e^\modi \mid (\mathtt{\downcast \ f})^\acc \mid  (\mathtt{tpure\ \bigcdot})^\pure \\  
\quad \textrm{Equations: } &\mathtt{eq} &::=& \mathtt{f^\dec} \eqs \mathtt{g^\dec} \mid \mathtt{f^\dec} \eqw \mathtt{g^\dec}  
\end{array}$$
{\vskip -0.45cm}
\caption{$\Log_{exc}$: syntax}
\label{syn-exc}
\end{figure} 
{\vskip -0.25cm}

Figure~\ref{syn-exc} shows the grammar of $\Log_{exc}$ where
$\empt$ is the empty (uninhabited) type while $\mathtt{EV_e}$ is the
type of parameters for each exception name \texttt{e}. We assume that
there is a finite set of exception names called \texttt{EName}. Given types $\mathtt{X}$ and $\mathtt{Y}$, we have
$\mathtt{X \splus Y}$ denoting co-product (disjoint union or sum) types.
Terms are
closed under composition ($\mathtt{\circ}$) and co-pairing ($\mathtt{\lcopair{\_\mid \_}}$). I.e., for all terms
$\mathtt{f\colon X\to Y}$ and $\mathtt{g\colon Y\to Z}$, we have $\mathtt{g \circ f\colon X\to Z}$. Similarly,
for all
$\mathtt{f\colon X\to Y}$ and $\mathtt{g\colon Z\to Y}$, there is  $\mathtt{\lcopair{f\mid g}\colon X \splus Z\to Y}$.
Notice that the co-pair subscript `$\mathtt{l}$' denotes
the left co-pairs.
One can define in a symmetric way the
right co-pairs for terms $\mathtt{f\colon X\to Y}$ and $\mathtt{g\colon Z\to Y}$ as
$\mathtt{\rcopair{f\mid g}:= \lcopair{g,f} \circ permut}$ where $\mathtt{permut := \lcopair{in_2 \mid in_1}}$.
Similarly, one can respectively obtain left and right co-products (sums) of terms
$\mathtt{f\colon X_1\to Y_1}$ and
$\mathtt{g\colon X_2\to Y_2}$ as $\mathtt{f\splus_l g:= \lcopair{in_1 \circ f \mid in_2 \circ g}}$ and
$\mathtt{f\splus_r g:= \rcopair{in_1 \circ f \mid in_2 \circ g}}$.
%
The decoration of a co-pair (co-product) depends on the decoration of its components, always
taking the larger.
I.e.,
$\mathtt{\forall\, f^\pure\colon X\to Z}$ and $\mathtt{g^\modi\colon Y\to Z}$,
$\mathtt{\lcopair{f \mid g}\colon X \splus Y \to Z}$ takes the decoration $\modi$.
Being dual to the pairs in  $\Log_{st}$ (which impose an evaluation order),
co-pairs in $\Log_{exc}$ are used to have  \emph{case distinction} among terms.
Co-pairs of catchers are allowed to be constructed in the logic $\Log_{exc}$.
However, they cannot be used
in the provided equational reasoning as they lead to ambiguous case distinctions over
input exceptional arguments for the component terms.
I.e., it is not obvious
to which input argument the recovery would apply when both are exceptional.
The intended equational reasoning can be done only when the left term is at most a
propagator.
The restriction is given by the rules (w$\_$lcopair$\_$eq)  and (s$\_$lcopair$\_$eq)   in Figure~\ref{rls-exc}.

The interface terms are $\mathtt{\tagg_\mathtt{e}
\colon EV_e \to \empt \splus E}$ and $\mathtt{\untag_\mathtt{e}\colon \empt \splus E \to EV_e \splus E}$
where $\mathtt{E}$ denotes the distinguished object of exceptions which never appears in the decorated setting.
The use of decorations provides a new schema where term signatures are constructed
without any occurrence of it. 
For instance,
 $\mathtt{\tagg^\acc_\mathtt{e}
\colon EV_e \to \empt}$ is a thrower while
 $\mathtt{\untag_\mathtt{e}^\modi\colon \empt \to EV_e}$ is a catcher. 
This way, we keep signatures close to their syntax and
compose compatible terms as usual.
The term $\mathtt{\tagg_\mathtt{e}}$ encapsulates  an  ordinary
value  with  an  exception  of  name \texttt{e}
while the term $\untag_\mathtt{e}$ recovers the value from the exceptional case.

 The `$\mathtt{\downarrow}$' symbol denotes the
\texttt{downcast}
term that takes as input a
term and prevents it from catching exceptions. It is used when to define the
$\mathtt{try/catch}$ block in this setting. See Definition~\ref{def:tc}.

The identity term $\mathtt{id}$,
the canonical co-pair inclusions $\mathtt{in_1}$ and $\mathtt{in_2}$, and
the empty co-pair $\mathtt{\copa_X}$ (used to convert the type of
input exceptional value into the given type; \texttt{X} in this case)
are
translated from a pure type system with
sum types using the \texttt{tpure} constructor, for all types \texttt{X} and \texttt{Y}, as follows:
{\vskip -0.25cm}
$$\begin{array}{l}
   \xymatrix@R=0.01pc@C=0.5pc{
   \mathtt{id_X^\pure} &\colon& \mathtt{X\to X} &:=& \mathtt{tpure\  (\lambda\, x:\, X.\, x: \, X)}\\
    \mathtt{in_1^\pure} &\colon& \mathtt{X\to X \splus Y} &:=& \mathtt{tpure \ inl}\\
    \mathtt{in_2^\pure} &\colon& \mathtt{Y\to X \splus Y} &:=& \mathtt{tpure \ inr}\\
    \mathtt{\copa_X^\pure} &\colon& \mathtt{\empt \to X} &:=& \mathtt{tpure \ (\lambda \ \_:\empt. \ x: X)}
   }
\end{array}$$
where \texttt{inl} and \texttt{inr} are constructors of sum types, and in the definition of $\mathtt{\copa_X}$, $\mathtt{X}$ is
assumed to be inhabited.

The intended model of the grammar of the logic $\Log_{exc}$ is built with respect to the set of exceptions $\mathtt{E}$
where a pure term $\mathtt{p^\pure: X\to Y}$ is interpreted as a function $\mathtt{p: X\to Y}$,
a propagator $\mathtt{pp^\acc: X\to Y}$ as a function $\mathtt{pp: X\to Y \splus E}$, and a
catcher $\mathtt{c^\ctc: X\to Y}$ as a function $\mathtt{c: X\splus E\to Y\splus E}$.
The complete and detailed category theoretical model is given in~\cite[\S 6.1]{Ekici:2015t}.



\begin{definition}
\label{def:throw}
For each type $\mathtt{Y}$ and exception name $\mathtt{e}$, the propagator $\mathtt{\throw_{Y,e}^\ppg}$ is
defined as:
$$\mathtt{\throw_{Y,e}^\ppg := \copa_Y^\pure \circ \tagg_e^\ppg \colon EV_e\to Y}$$
 \end{definition}
 \noindent
Intuitively, raising an exception of name $\mathtt{e}$
is first tagging the given ordinary value with $\mathtt{e}$
and then coercing the empty type into $\mathtt{Y}$ for the continuation issues. 
\begin{definition}
\label{def:tc}
For each propagators $\mathtt{f^\ppg\colon X\to Y}$, $\mathtt{g^\ppg\colon EV_e\to Y}$ and each
exception name $\mathtt{e}$, the propagator $\mathtt{\try(f)\catch(e\To g)^\ppg}$ is defined in three steps, 
as follows: 
\renewcommand{\arraystretch}{1.25}
$$ \begin{array}{lcll}
\mathtt{\mathtt{Catch}(e\Rightarrow g)^\modi} & := &
  \mathtt{\lcopair{\; id_Y^\pure \;|\; g^\ppg \circ\untag_e^\ctc  \;} 
}& \colon  \mathtt{Y\splus\empt\to Y}\\ 
\mathtt{\mathtt{Try}(f)\mathtt{Catch}(e\Rightarrow g)^\ctc} & := & 
 \mathtt{ \mathtt{Catch}(e\Rightarrow g)^\modi \circ in_{1_{Y, \empt}}^\pure
   \circ f^\ppg 
 } &\colon\mathtt{X\to Y} \\
\mathtt{\try(f)\catch(e\To g)^\ppg} & := &
 \mathtt{ \downcast\big(\mathtt{Try}(f)\mathtt{Catch}(e\Rightarrow g)^\ctc\big) 
  } &\colon\mathtt{X\to Y} \\
\end{array}$$
\renewcommand{\arraystretch}{0.75}
\end{definition}
\noindent
To handle an exception, 
the intermediate expressions $\mathtt{Catch (e\To g)}$ 
and $\mathtt{\mathtt{Try}(f)\mathtt{Catch}(e\To g)}$ are private catchers 
and the expression $\mathtt{\try(f)\catch(e\To g)}$ is a public propagator: 
the downcast operator intuitively used to prevent $\mathtt{\try(f)\catch(e\To g)}$
from catching exceptions 
with name $\mathtt{e}$ which might have been raised before its execution.
Below we depict the $\mathtt{\try(f)\catch(e\To g)}$ definition
as a diagram:
$$
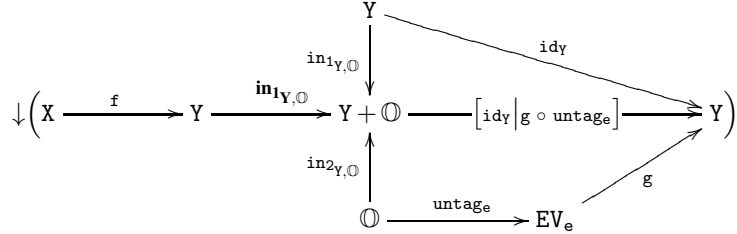
\xymatrix@R=1.75pc@C=3.75pc{
& & \mathtt{Y}  \ar[d]_{\mathtt{in_{1_{Y,\empt}}}} \ar[rrd]^{\mathtt{id_{Y}}}&  \\
\downcast\Big(\mathtt{X} \ar[r]^{\mathtt{f}} & \mathtt{Y} \ar[r]^{\mathtt{\mathbf{in_{1_{Y,\empt}}}}} & \mathtt{Y + \empt} \ar[rr]|{\big[\mathtt{id_Y} \big| \mathtt{g} \ \circ \ \mathtt{untag_e} \big]} & & \mathtt{Y}\Big)    \\
& & \empt \ar[u]^{\mathtt{in_{2_{Y,\empt}}}} \ar[r]^{\mathtt{untag_e}} & \mathtt{EV_e} \ar[ru]_{\mathtt{g}}&   \\ 
}
$$
This, inside the downcast, intuitively tells us that if the term \texttt{f} throws an exception,
then within the \texttt{Catch} block using the case distinction, provided by the copair,
the exception is handled via the \texttt{untag} (unhandled exception gets propagated) and the continuation
is the execution of the term \texttt{g}. If \texttt{f} does not throw any exception then no handling is performed, we
have \texttt{id} term in execution. Note also that the term $\mathtt{in_{1_{Y,\empt}}}$ used in the
Definition~\ref{def:tc} is the horizontal one in the above diagram, and implicitly means that \texttt{Y}
and $\mathtt{Y + \empt}$ isomorphic objects. The inclusions $\mathtt{in_{1_{Y,\empt}}}$ (the vertical in the above diagram)
and $\mathtt{in_{2_{Y,\empt}}}$ play a role in the equational reasoning,
given in Figure~\ref{rls-exc}, that we provide on the top of $\Log_{exc}$ syntax.

The definition of $\mathtt{\try(f)\catch(e\To g)}$ corresponds to 
the Java mechanism for exceptions as in ~\cite[\S 14]{Gosling:2005}
and in~(\cite{Jacobs:2001}) with
the following control flow
(where \texttt{exc?} means ``\emph{is this value an exception?}''):
an \emph{abrupt} termination returns an uncaught exception 
and a \emph{normal} termination returns an ordinary value.
{\vskip -0.75cm}
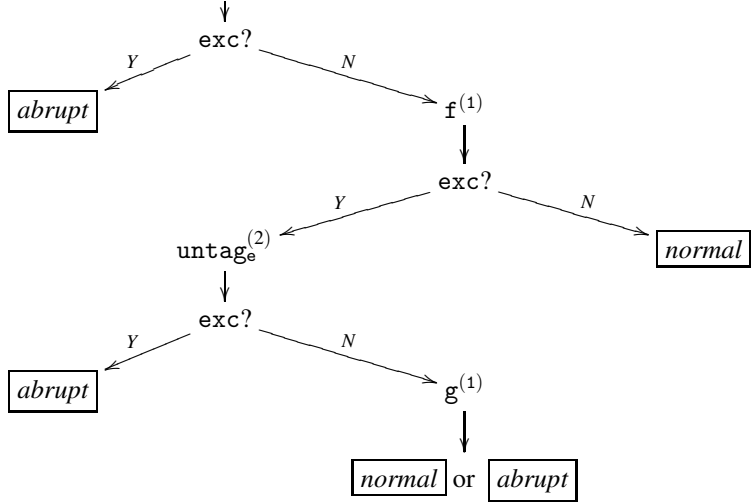
$$ \xymatrix@C=2pc@R=0.8pc{
& \ar[d] && \\
& \mathtt{exc?} \ar[ld]_{Y}\ar[rd]^{N} && \\
\abr && \mathtt{f^\ppg} \ar[d] & \\
&& \mathtt{exc?} \ar[ld]_{Y}\ar[rd]^{N} & \\
& \untag_\mathtt{e}^\modi \ar[d] && \nor \\
& \mathtt{exc?} \ar[ld]_{Y}\ar[rd]^{N} && \\ 
\abr && \mathtt{g^\ppg} \ar[d] & \\
&& \txt{\nor \mbox{ or } \abr} \\
} $$
\begin{remark}
The decorated terms $\mathtt{throw^\ppg}$ and $\mathtt{throw/catch^\ppg}$ stated in
Definitions~\ref{def:throw}~and~\ref{def:tc} will serve, in Section~\ref{impexcodl} (see
the translator function \texttt{dCmd}), as
interpretations of the \texttt{IMP+Exc} commands \texttt{THROW} and \texttt{TRY/CATCH}. 
\end{remark}
\begin{figure}[h]
\renewcommand{\arraystretch}{1.25}
\begin{tabular}{l} 
\textbf{Rules of the decorated logic for the exception: }\\
(pwsubs)$\mathtt{\dfrac{g^\pure\colon X\to Y \squad f_1^{(d_1)}\eqw f_2^{(d_2)}\colon Y\to Z} {f_1^{(d_1)} \circ g^\pure \eqw f_2^{(d_2)}\circ g^\pure}}$ \squad
(wrepl)$\mathtt{\dfrac{f_1^{(d_1)}\eqw f_2^{(d_2)}\colon X\to Y \squad g^{(d_3)}\colon Y\to Z} {g^{(d_3)}\circ f_1^{(d_1)} \eqw g^{(d_3)}\circ f_2^{(d_2)} }}$ 
    \vspace{5pt} \\
(replsubs)  
  $\mathtt{\dfrac{f_1^{(d_1)} \eqs f_2^{(d_2)}\colon X\to Y \squad g_1^{(d_3)} \eqs g_2^{(d_4)}\colon Y\to Z}
    {g_1^{(d_3)}\circ f_1^{(d_1)} \eqs g_2^{(d_4)}\circ f_2^{(d_2)}}}$ 
\squad

(w$\_$empty)  
  $\dfrac{\mathtt{f^{(d_1)}\colon \empt\to X}}{\mathtt{f^{(d_1)} \eqw \copa_X^\pure}}$ 
            \vspace{5pt} \\
  (w$\_$downcast)  
  $\dfrac{\mathtt{f^\ctc\colon Y\to X}}{\mathtt{(\mathtt{\downarrow}\ f)^\acc \eqw f^\modi}}$
  \vspace{5pt} \\
(eax$_1$)  
  $\dfrac{}{\untag^\modi_\mathtt{e} \circ \tagg^\acc_\mathtt{e} \eqw \mathtt{id_{EV_e}^\pure}} $\squad
(eax$_2$)
  $\dfrac{\mathtt{\forall e_1, e_2 \in EName,\ e_1 \neq e_2}}
  {\untag^\modi_\mathtt{e_1} \circ \tagg^\acc _\mathtt{e_2} \eqw \copa^\pure_{\mathtt{EV_{e_1}}} \circ \tagg^\acc_\mathtt{e_2} } $ 
  \vspace{5pt} \\
(eeffect) 
  $\mathtt{\dfrac{f_1^{(d_1)},f_2^{(d_2)}\colon Y\to X \;\;
  f_1^{(d_1)}\eqw f_2^{(d_2)} \;\; f_1^{(d_1)} \circ \copa_Y^\pure \eqs f_2^{(d_2)} \circ \copa_Y^\pure}{f_1^{(d_1)}\eqs f_2^{(d_2)}}}$
  \vspace{5pt} \\
(elocal$\_$global) 
  $\mathtt{\dfrac{ f_1^{(d_1)},f_2^{(d_2)}\colon \empt\to X \quad \forall\ \mathtt{e \in EName}, \ f_1^{(d_1)} \circ
  \tagg_\mathtt{e}^\ppg
   \eqw f_2^{(d_2)} \circ
  \tagg _\mathtt{e}^\ppg}
    {f_1^{(d_1)}\eqs f_2^{(d_2)}}}$ \vspace{5pt} \\
(w$\_$lcopair$\_$eq)  
  $\mathtt{\dfrac{f_1^{(d_1)}: X \to Y \squad f_2^{(d_2)}: Z \to Y \squad d_1 \in \{0,1\}}{[f_1\mid f_2]^{(max(d_1,d_2))} \circ \mathtt{in_1}^\pure \sim f_1^{(d_1)}}}$  
    \vspace{5pt} \\
(s$\_$lcopair$\_$eq)   
  $\mathtt{\dfrac{f_1^{(d_1)}: X \to Y \squad f_2^{(d_2)}: Z \to Y \squad d_1 \in \{0,1\}}{[f_1\mid f_2]^{(max(d_1,d_2))} \circ \mathtt{in_2}^\pure \equiv f_2^{(d_2)}}}$ 
\end{tabular}
\vspace{-0.15cm}
\caption{$\Log_{exc}$: rules} 
\label{rls-exc}
\end{figure}
The syntax given in  Figure~\ref{syn-exc} is enriched with two sets of rules
presented in Figures~\ref{rls-exc} and~\ref{dl-rls}.
%
%
%
Weak equalities do not form a congruence:
the term substitution cannot be done unless the substituted
term is pure. I.e.,
given the equation $\mathtt{f_1^{(d_1)} \eqw f_2^{(d_2)}\colon Y\to Z}$ and a
term $\mathtt{g\colon X\to Y}$,  it is possible to get the
equation $\mathtt{f_1 \circ g \eqw f_2 \circ g}$ only when the term $\mathtt{g}$ is pure.
At this stage, we have no information about the behaviors of
$\mathtt{f_1}$ and $\mathtt{f_2}$ on exceptional values. Therefore, the pre-executed term $\mathtt{g}$
would destroy this result equality unless being pure, for instance,
by throwing an exception of name $\mathtt{e}$ for which $\mathtt{f_1}$ and $\mathtt{f_2}$ perform different
behaviors: say one is propagating, while the other is recovering from it (pwsubs).
However, the term replacement can be done regardless of the term decoration.
I.e., given the equation $\mathtt{f_1^{(d_1)} \eqw f_2^{(d_2)}\colon X\to Y}$ and a
term $\mathtt{g^{(d_3)}\colon Y\to Z}$,  it is possible to get the
equation $\mathtt{g \circ f_1 \eqw g \circ f_2}$ independent from the decoration of the term $\mathtt{g}$.
Since $\mathtt{f_1}$ and $\mathtt{f_2}$ behave the same on ordinary values,
executing any
term $\mathtt{g}$ after $\mathtt{f_1}$ and $\mathtt{f_2}$ would not end them
behave different on ordinary values  (wrepl).
Strong equalities form a congruence by allowing both term substitutions and
replacements regardless of the term decorations (replsubs).


Any term $\mathtt{f\colon \empt \to X}$ with no input parameter 
has an equivalence on ordinary values with the empty co-pair 
$\mathtt{\copa_X}$ (w$\_$empty).
The rule (w$\_$downcast) states that the term
($\mathtt{\downcast \ f}$) behaves as $\mathtt{f}$, if the argument is ordinary.
%
The fundamental equations are given with the rules (eax$_1$) and (eax$_2$).
The former states that
encapsulating an ordinary value with an exception of name \texttt{e}
followed by an immediate recovery would be equivalent to ``doing nothing''
in terms of ordinary values.
Clearly, this is only a weak equation since its
sides behave different on exceptional values: left hand side may recover but right hand side
definitely propagates.
The latter, (eax$_2$),
is to assume that encapsulating an ordinary value $\mathtt{v}$ with an exception of name  $\mathtt{e_2}$
and then trying to recover it from a different exception of name  $\mathtt{e_1}$
would just lead $\mathtt{e_2}$ to be propagated.
Similarly, if the ordinary value $\mathtt{v}$ is encapsulated with $\mathtt{e_2}$ with no recovery attempt
afterwards would again lead $\mathtt{e_2}$ to be propagated.
These two operations behave the same
on ordinary values but different on exceptional ones. For instance,
left hand side recovers the input value (encapsulated with the exception name $\mathtt{e_1}$) while
right hand side propagates it. 

Two catchers $\mathtt{f_1^\modi,f_2^\modi:X\to Y}$ behave the same on exceptional values
if and only if $\mathtt{f_1 \circ \copa_X \eqs f_2 \circ \copa_X}$, where
$\mathtt{\copa_X\colon \empt \to X}$ throws out exceptional values.
So that $\mathtt{f_1^\modi,f_2^\modi:X\to Y}$ are \emph{strongly equal}  if and only if 
$\mathtt{f_1\eqw f_2}$ and $\mathtt{f_1 \circ \copa_X \eqs f_2 \circ \copa_X}$ (eeffect).
The rule is valid also for the other decorations of terms 
$\mathtt{f_1}$ and $\mathtt{f_2}$.
 
Strong equality between two catchers
$\mathtt{f_1^\modi,f_2^\modi: \empt\to X}$ can also be expressed as a pair of weak equations:
$\mathtt{f_1 \eqw f_2}$ and $\mathtt{\forall \mathtt{e: ENname}, f_1 \circ \tagg_\mathtt{e} 
\eqw f_2\circ \tagg_\mathtt{e}}$.
The latter intuitively means that $\mathtt{f_1}$ and $\mathtt{f_2}$ behaves the same on
all (finitely many) exceptional values when executed.
Given that both behave the same on ordinary arguments (due to (w$\_$empty)),
there is no explicitly need to check if $\mathtt{f_1\eqw f_2}$.
It suffices to see whether $\mathtt{\forall \mathtt{e: EName}, f_1 \circ \tagg_\mathtt{e} 
\eqw f_2\circ \tagg_\mathtt{e}}$
to end up with $\mathtt{f_1\eqs f_2}$ (elocal$\_$global).
This rule is valid also for the other decorations of terms 
$\mathtt{f_1}$ and $\mathtt{f_2}$.

With (w$\_$lcopair$\_$eq) and (w$\_$rcopair$\_$eq), term co-pairs (sums) are characterized: 
the (left) co-pair structure 
$\mathtt{\lcopair{f_1\mid f_2}}$ cannot be used when $\mathtt{f_1}$ and $\mathtt{f_2}$, both are catchers,
since it may lead to a conflict on exceptional values.  When $\mathtt{f_1}$ is a propagator,
with (w-copair-eq), we assume that
ordinary values of type $\mathtt{X}$ are treated by 
$\mathtt{\lcopair{f_1\mid f_2}^{(max(d_1,d_2))}}$ as they would be by $\mathtt{f_1^{(d_1)}}$ and
with (s-copair-eq) that
ordinary values of type $\mathtt{Z}$ and exceptional values are treated by
$\mathtt{\lcopair{f_1\mid f_2}^{(max(d_1, d_2))}}$ as they would be~by~$\mathtt{f_2^{(d_2)}}$. 

Similar to the ones of the logic $\Log_{st}$, the rules of the logic $\Log_{exc}$ also
designed to be sound with respect to a categorical model which is detailed in~\cite[\S 6.2, \S 6.3, \S 6.4, \S 6.5]{Ekici:2015t}.
In~(\cite{DBLP:conf/macis/DumasDEPR15}), we prove that this set of rules is 
complete with respect to the notion of relative Hilbert-Post completeness.

\subsection{Decorated properties of the exception effect}
\label{dpotee}
Similar to the one for the state effect presented in Section~\ref{dpotms},
we propose an equational representation of the exception effect with the following decorated equations:
\begin{itemize}
\item[(1)$_d$] 
Annihilation tag-untag.
\textit{Untagging an exception of name
 $\mathtt{e}$ and then raising it again is just like doing nothing.}
 $\mathtt{ \forall\, e\in EName,\;
 \tagg_e^\ppg \circ \untag_e^\ctc \eqs id^\pure_\empt 
  : \empt \to \empt}$.
\item[(2)$_d$]
Commutation untag-untag.
\textit{Untagging two distinct exception names
can be done in any~order.}\\
 $ \mathtt{\forall\, e\ne r \in EName,\;
 (\untag_e +_r id_{EV_r})^\ctc \circ in_2^\pure \circ \untag_r^\ctc \eqs} \\
 \mathtt{(id_{EV_e} +_l \untag_r)^\ctc \circ in_1^\pure \circ \untag_e^\ctc\colon
  \empt \to EV_e + EV_r}$.
\item[(3)$_d$]
Propagator-propagates.
\textit{A propagator term always propagates the exception.}\\
 $\mathtt{\forall\, e \in EName,\; a^\ppg\colon X\to Y, \;\; 
 a^\ppg \circ \copa_X^\pure \circ \tagg_e^\ppg \eqs
 \copa_Y^\pure \circ \tagg_e^\ppg\colon EV_e\to Y}$.
\item[(4)$_d$]
Recovery.
\textit{The parameter used for throwing an exception may be recovered.}\\
$\mathtt{\Big(\forall\, f^\ppg,\,g^\ppg\colon X\to \empt, \;\;
\copa_Y^\pure \circ f^\ppg \eqs \copa_Y^\pure \circ g^\ppg \implies f^\ppg \eqs g^\ppg\Big) \implies}\\
\mathtt{\Big(\forall\, e \in EName,\; u_1^\pure,u_2^\pure:X\to EV_e,}
\mathtt{\big(\throw_e^\ppg \circ u_1^\pure
 \eqs \throw_e^\ppg \circ u_2^\pure\big) \implies u_1^\pure \eqs u_2^\pure\colon X\to  EV_e \Big)}.$


\item[(5)$_d$]
Try. 
\textit{The strong equation is compatible with $\try$/$\catch$.}\\
$\mathtt{\forall\, e \in EName,\;
a_1^\ppg,\,a_2^\ppg\colon X\to Y,\, b^\ppg: EV_e\to Y,\;
 a_1^\ppg \eqs a_2^\ppg \implies} \\
\mathtt{\try(a_1)\catch(e\To b)^\ppg \eqs \try(a_2)\catch(e\To b)^\ppg\colon X \to Y}.$
 
\item[(6)$_d$]
Try$_0$.
\textit{Pure code inside $\try$ never triggers 
  the code inside $\catch$. }\\
$\mathtt{\forall\, e \in EName,\;
u^\pure\colon X\to Y,\, b^\ppg: EV_e\to Y,\,} 
\mathtt{\try(u)\catch(e\To b)^\ppg \eqs u^\pure\colon X\to Y}.$ 
 
\item[(7)$_d$]
Try$_1$.
\textit{The code inside $\catch$ is executed
as soon as an exception is thrown inside $\try$.}\\
$\mathtt{\forall\, e \in EName,\;
u^\pure\colon X\to EV_e,\, b^\ppg: EV_e\to Y,} \,
\mathtt{\try(\throw_e \circ u)\catch(e\To b)^\ppg  \eqs b^\ppg \circ u^\pure\colon X\to Y}.$ 
 
 \item[(8)$_d$]
Try$_2$.
\textit{An exception gets propagated, if the exception name is not pattern matched in}
\texttt{catch}.\\
$\mathtt{ \forall\;(e \neq f)\in EName,\;  
u^\pure\colon X\to EV_f,\, b^\ppg: EV_e\to Y,} \\
\mathtt{\try(\throw_f \circ u)\catch(e\To b)^\ppg\eqs \throw_f^\ppg \circ u^\pure\colon X\to Y}. $
 
 
\end{itemize}
These are the archetype properties that we have proved within the scope of the
 $\Log_{exc}$. To see these
 proofs,
 check out~\cite[\S 6.7]{Ekici:2015t}.
 Besides, we have implemented the $\Log_{exc}$ in \coq to certify mentioned proofs.
 Section~\ref{lexc-coq} briefly discusses this implementation. Notice that the premise of the
 property (4)$_d$ is a very specific mono requirement. It intuitively says that if there is a strong equality
 between two propagators (i.e., $\mathtt{f^\ppg}$ and $\mathtt{g^\ppg}$)
 after removing the exceptional values they may propagate, then they are
 strongly equal. In the absence of this requirement the property is not valid.

 \subsection{$\Log_{exc}$ in \coq}
 \label{lexc-coq}
Coq implementation of  $\Log_{exc}$ follows the same approach with the one for $\Log_{st}$
as summarized in Section~\ref{lst-coq}.
We represent the set of
exception names by a Coq parameter $\mathtt{EName: Type}$. An arrow type $\mathtt{EVal: EName \to Type}$ is assumed
as the type of values (parameters) for each exception name. We then inductively define terms
and assign decorations
over them.
There, we
respectively use keywords \texttt{epure}, \texttt{ppg} and 
\texttt{ctc} instead of (0), (1) and (2). 
The rules up to weak
and strong equalities are stated in a mutually inductive way to allow
constructors including both types of equalities, similar to the approach presented
in Section~\ref{lst-coq}. 
We choose not to replay the entire \coq encoding here, but at least give Coq formalizations of
Definitions~\ref{def:throw}~and~\ref{def:tc}: 

\begin{tcb}[\scriptsize]{Coq}
 Definition throw (X: Type) (e: EName)  := (@empty X) o tag e.
 Definition try_catch (X Y: Type) (e: EName) (f: term Y X) (g: term Y (Val e)) := 
 				downcast (copair (@id Y) (g o untag e) o in1 o f).
\end{tcb}
The encodings of other terms are contained in 
this file \footnote{\url{https://github.com/ekiciburak/decorated-logics-for-exceptions-effect/blob/master/Terms.v}}.

We can conclude that such a framework allows us to express and prove, in \coq, the decorated versions of the properties mentioned in
Section~\ref{dpotee}. E.g.,  the statement \texttt{propagator-propagates} looks like:
\begin{tcb}[\scriptsize]{Coq}
(**  Propagator propagates **)
 Lemma PPT: forall X Y (e: EName) (a: term Y X), is ppg a ->  a o ((@empty X) o tag e) == (@empty Y) o tag e.
\end{tcb}
The full \coq proofs of such properties can be found here
\footnote{\url{https://github.com/ekiciburak/decorated-logics-for-exceptions-effect/blob/master/Proofs.v}},
and the entire implementation there\footnote{\url{https://github.com/ekiciburak/decorated-logics-for-exceptions-effect}}.

\section{Combining $\Log_{st}$ and $\Log_{exc}$}
\label{dlstexc}
In order to formally cope with different computational effects,
one needs to compose the related formal models. For instance,
using monad transformers (\cite{Jaskelioff:2009}),
it is usually possible to combine
effects formalized by monads, as encoded in Haskell. Handler compositions allow
combining effects modeled by algebraic handlers,
as implemented in Eff by \cite{Bauer:2015,Bauer:2014,Pretnar:2013}
and in Idris by~\cite{Brady:2013}.
%
To combine effects formalized in decorated settings, we just need to compose the related logics.
In this section,
we formally study the combination of the state and the exception effects using the
logics $\Log_{st}$ and $\Log_{exc}$. We call the newly born logic
\emph{the decorated logic for the state and the exception}, and denote it $\Log_{st+exc}$.
To start with, we give the
syntax of $\Log_{st+exc}$ below in Figure~\ref{syn-ldecfstexc}.
{\vskip -0.7cm}
\begin{figure}[h]
\renewcommand{\arraystretch}{1.25}
$$\begin{array}{lccl}
\multicolumn{4}{l}{
\textbf{Grammar of the decorated logic for the state and the exception: }\quad (\texttt{i} \in \texttt{Loc})\quad (\texttt{e} \in \texttt{EName})} \\
\quad \textrm{Types: } &\mathtt{t,\, s} &::=&
 \mathtt{X}\mid \mathtt{Y} \mid \dots \mid  \mathtt{t\times s} \mid  \mathtt{t\splus s} \mid \unit   \mid \empt
\mid \mathtt{V_{i}} 
\mid \mathtt{EV_e} \\
 \quad \textrm{Decoration for terms: }& \mathtt{(d_1, d_2), (d_3, d_4)} &::=&(0,0) \mid (0,1) \mid (0,2)
	\mid (1,0) 
	\mid (1,1) \mid
	 \\ & & & 
	 (1,2) \mid 
	(2,0) \mid (2,1) \mid (2,2)  \\ 
	
\quad \textrm{Terms: } &\mathtt{f,\;g} &::=&\mathtt{a^{(d_1,d_2)}} \mid \mathtt{b^{(d_1,d_2)}}
\mid \dots \mid\mathtt{g\circ f^{(d_1,d_2)}} \mid
\\ & & &
  \mathtt{\lpair{f^{(d_1, d_2)},g^{(d_3,d_4)}}^{(max(d_1,d_3),max(d_2, d_4))}}  \mid  
\\ & & &
\mathtt{\lcopair{f^{(d_1, d_2)}\mid g^{(d_3,d_4)}}^{(max(d_1,d_3),max(d_2,d_4))}}\mid 
 \\ & & &
  \lookup_\mathtt{i}^{(1,0)} \mid 
  \update_\mathtt{i}^{(2,0)} \mid
   \tagg_\mathtt{e}^{(0,1)} \mid \untag_\mathtt{e}^{(0,2)} \mid 
        \\ & & &
    (\mathtt{\downarrow\ f})^{(0,1)} \mid
       (\mathtt{tpure\ \bigcdot})^{(0,0)} \\
\quad \textrm{Equations: } &\mathtt{eq} &::=& \mathtt{f^{(d_1,d_2)}} \eqs\eqs \mathtt{g^{(d_1,d_2)}} \mid \mathtt{f^{(d_1,d_2)}} \eqs\eqw \mathtt{g^{(d_1,d_2)}}   \mid
   \\ & & &
 \mathtt{f^{(d_1,d_2)}} \eqw\eqs \mathtt{g^{(d_1,d_2)}}  \mid \mathtt{f^{(d_1,d_2)}} \eqw\eqw \mathtt{g^{(d_1,d_2)}}
\end{array}$$
{\vskip -0.5cm}
\caption{$\Log_{st+exc}$: syntax}
\label{syn-ldecfstexc}
\end{figure}
{\vskip -0.25cm}
The decorations are paired off to cover all possible combinations: 
the decoration symbol on the left is given in terms of the state effect while the one on the right 
is of the exception. I.e., $\mathtt{f^{(1,2)}}$ says that $\mathtt{f}$ may \textit{access} to 
the state alongside \textit{catching} exceptions.
The decoration of a (co)-pair/(co)-product or a composition depends on the decorations of its components,
always taking the larger. I.e., $\mathtt{\forall\, f^{(1,2)}\colon X\to Y}$ and $\mathtt{g^{(2,1)}\colon Y\to Z}$,
$\mathtt{g \circ f\colon X \to Z}$ takes the decoration $(2,2)$.
The pairs/products of compatible terms $\mathtt{f_1^{(2,2)}}$,  $\mathtt{g_1^{(2,2)}}$, and
similarly the co-pair/co-products of compatible terms $\mathtt{f_2^{(2,2)}}$,  $\mathtt{g_2^{(2,2)}}$
can be constructed within the scope of $\Log_{st+exc}$ but cannot be used
in the provided equational reasoning. This is because,
$\mathtt{f_1^{(2,2)}}$ and $\mathtt{g_1^{(2,2)}}$,
as two modifiers, may lead to conflicts on the
returned results over any type of (exceptional or ordinary) arguments
due to the possible hazardous parallel modifications of the global state,
while $\mathtt{f_2^{(2,2)}}$ and $\mathtt{g_2^{(2,2)}}$,
as two catchers, may yield in ambiguous
case distinctions over input exceptional arguments. I.e., it is not obvious
to which input argument the recovery would apply when both are exceptional.
The rules (w$\_$lpair$\_$eq), (s$\_$lpair$\_$eq), (w$\_$lcopair$\_$eq) and
(s$\_$lcopair$\_$eq), in Figure~\ref{rls-stexc}, enforce these restrictions.

The types of  $\Log_{st+exc}$ is the union of the types of  $\Log_{st}$ and  $\Log_{exc}$.
Similarly, the terms of  $\Log_{st+exc}$ is the union of the terms of  $\Log_{st}$ and  $\Log_{exc}$.
The interface terms for the
state effect are pure with respect to the exception and vice versa: $\lookup^{(1,0)}$,
$\update^{(2,0)}$, $\tagg^{(0,1)}$ and $\untag^{(0,2)}$.
As in Sections~\ref{dlst}~and~\ref{dlexc}, we use the special \texttt{tpure} constructor to translate pure terms
such as
the identity $\mathtt{id}$,
the canonical pair projections $\mathtt{\pi_1}$, $\mathtt{\pi_2}$,
the empty pair $\mathtt{\pa}$,
the canonical co-pair inclusions $\mathtt{in_1}$, $\mathtt{in_2}$,
the empty co-pair $\mathtt{\copa}$
and \texttt{constants}
from a pure type system with product and
sum types using the \texttt{tpure} constructor, for all types \texttt{X} and \texttt{Y}, as:
$$\begin{array}{l}
   \xymatrix@R=0.01pc@C=0.5pc{
   \mathtt{id_X^{(0,0)}} &\colon& \mathtt{X\to X} &:=& \mathtt{tpure\  (\lambda\, x:\, X.\, x: \, X)}\\
    \mathtt{\pi_1^{(0,0)}} &\colon& \mathtt{X\times Y\to X} &:=& \mathtt{tpure \ fst} \\
   }
\end{array}$$
$$\begin{array}{l}
   \xymatrix@R=0.01pc@C=0.5pc{
           \mathtt{\pi_2^{(0,0)}} &\colon& \mathtt{X\times Y\to Y} &:=& \mathtt{tpure \ snd}\\
              \mathtt{\pa_X^{(0,0)}} &\colon& \mathtt{X \to \unit} &:=& \mathtt{tpure \ (\lambda\ x:\ X.\ void: \ \unit)}\\
               \mathtt{in_1^{(0,0)}} &\colon& \mathtt{X\to X \splus Y} &:=& \mathtt{tpure \ inl}\\
    \mathtt{in_2^{(0,0)}} &\colon& \mathtt{Y\to X \splus Y} &:=& \mathtt{tpure \ inr}\\
    \mathtt{\copa_X^{(0,0)}} &\colon& \mathtt{\empt \to X} &:=& \mathtt{tpure \ (\lambda \ \_:\empt. \ x: X)}\\
    \mathtt{constant_x^{(0,0)}} &\colon& \mathtt{\unit \to X} &:=& \mathtt{tpure \ (\lambda\ \_.\ x: X)}
   }
\end{array}$$
where \texttt{fst} and \texttt{snd} are constructors of  product types while
\texttt{inl} and \texttt{inr} are of sum types, and in the definition of $\mathtt{\copa_X}$, $\mathtt{X}$ is
assumed to be inhabited.

The rule combinations need a bit of reformulation
as we summarize below:
\begin{itemize}
\item 
The decoration symbol $\mathtt{\pure}$
freely converts into $\mathtt{\acc}$ and $\mathtt{\modi}$, while the symbol $\mathtt{\acc}$ just into $\mathtt{\modi}$ when the
other symbol is fixed. I.e., $\mathtt{f^{(0,2)}}$ freely
converts into  $\mathtt{f^{(1,2)}}$. See all cases below:
\begin{itemize}
\item
$\mathtt{\dfrac{f^{(0,d)}}{f^{(1,d)}}}$, \squad
$\mathtt{\dfrac{f^{(1,d)}}{f^{(2,d)}}}$,\squad
$\mathtt{\dfrac{f^{(d,0)}}{f^{(d,1)}}}$, \squad
$\mathtt{\dfrac{ f^{(d,1)}}{f^{(d,2)}}}$ \ for $\mathtt{d \in \{0, 1, 2\}}$
\end{itemize}
\item We have all possible combinations of  equality sorts: $\eqs\eqs$, $\eqs\eqw$, $\eqw\eqs$
and $\eqw\eqw$. The first equality symbol relates terms with respect to the state effect. I.e.,
$\mathtt{f \eqs\eqw g}$ means that $\mathtt{f}$ and $\mathtt{g}$ are strongly equal
with respect to the state,
while being weakly equal
with respect to the exception.  Below we present the conversion rules between these four sorts.
The burden here is that a strong equality symbol can always be freely converted into a weak one independent of
according to which effect it relates terms. But, to convert a weak equality symbol into a strong one, we need to
make sure that the related terms are decorated either with $\mathtt{\pure}$ or $\mathtt{\acc}$ with respect to the
effect they are weakly related. 
\begin{itemize}
\item
{($\eqs\eqs$-to-$\eqs\eqw$)}$\mathtt{\dfrac{f^{(2,2)} \eqs\eqs g^{(2,2)}}{f^{(2,2)} \eqs\eqw g^{(2,2)}}}$, \squad
{($\eqs\eqs$-to-$\eqw\eqs$)}$\mathtt{\dfrac{f^{(2,2)} \eqs\eqs g^{(2,2)}}{f^{(2,2)} \eqw\eqs g^{(2,2)}}}$
 
 \item
{($\eqs\eqw$-to-$\eqw\eqw$)}$\mathtt{\dfrac{f^{(2,2)} \eqs\eqw g^{(2,2)}}{f^{(2,2)} \eqw\eqw g^{(2,2)}}}$, \squad
{($\eqw\eqs$-to-$\eqw\eqw$)}$\mathtt{\dfrac{f^{(2,2)} \eqw\eqs g^{(2,2)}}{f^{(2,2)} \eqw\eqw g^{(2,2)}}}$
 
   \item
{($\eqw\eqs$-to-$\eqs\eqs$)}$\mathtt{\dfrac{f^{(1,2)} \eqw\eqs g^{(1,2)}}{f^{(1,2)} \eqs\eqs g^{(1,2)}}}$,
 \squad
{($\eqs\eqw$-to-$\eqs\eqs$)}$\mathtt{\dfrac{f^{(2,1)} \eqs\eqw g^{(2,1)}}{f^{(2,1)} \eqs\eqs g^{(2,1)}}}$

  \item
{($\eqw\eqw$-to-$\eqs\eqw$)}$\mathtt{\dfrac{f^{(1,2)} \eqw\eqw g^{(1,2)}}{f^{(1,2)} \eqs\eqw g^{(1,2)}}}$, \squad
{($\eqw\eqw$-to-$\eqw\eqs$)}$\mathtt{\dfrac{f^{(2,1)} \eqw\eqw g^{(2,1)}}{f^{(2,1)} \eqw\eqs g^{(2,1)}}}$
 

\end{itemize}
\item The rules of the logic $\Log_{st+exc}$ are presented in Figure~\ref{rls-stexc}
as a union of the ones given in Figures~\ref{rls-st}~and~\ref{rls-exc} in terms of
new equality sorts and refined term decorations. 
There, we replay the whole rule bodies, and implicitly assume that
all equality sorts are equivalence relations respecting the
properties \emph{reflexivity}, \emph{symmetry}, and \emph{transitivity}.
\end{itemize}
\begin{figure}\small
\renewcommand{\arraystretch}{2}
\begin{tabular}{l} 
\textbf{Rules of the decorated logic for the state and the exception: }\\ 
  (assoc)  
    $\mathtt{\dfrac{f^{(d_1,d_2)}\colon X\to Y \squad g^{(d_3,d_4)}\colon Y\to Z \squad h^{(d_5,d_6)}\colon Z\to T}
  {h^{(d_5,d_6)}\circ (g^{(d_3,d_4)}\circ f^{(d_1,d_2)}) \eqs\eqs (h^{(d_5,d_6)}\circ g^{(d_3,d_4)})\circ f^{(d_1,d_2)}} }$
            \vspace{5pt} \\
(ids)  
    $\mathtt{\dfrac{f^{(d_1,d_2)}\colon X\to Y}{f^{(d_1,d_2)}\circ id_X^{(0,0)} \eqs\eqs f^{(d_1,d_2)}}}$ \squad
 (idt) 
    $\mathtt{\dfrac{f^{(d_1,d_2)}\colon X\to Y}{id_Y^{(0,0)}\circ f^{(d_1,d_2)} \eqs\eqs f^{(d_1,d_2)}}}$ 
              \vspace{5pt} \\
 (pwrepl)$\mathtt{\dfrac{f_1^{(d_1,d_2)}\eqw\eqs f_2^{(d_3,d_4)}\colon X\to Y \squad g^{(0,d_5)}\colon Y\to Z}
 {g^{(0,d_5)}\circ f_1^{(d_1,d_2)} \eqw\eqs g^{(0,d_5)}\circ f_2^{(d_3,d_4)} }}$     
 \squad \
 (wsubs)$\mathtt{\dfrac{g^{(d_5,d_6)}\colon X\to Y \squad f_1^{(d_1,d_2)}\eqw\eqs f_2^{(d_3,d_4)}\colon Y\to Z}
 {f_1^{(d_1,d_2)} \circ g^{(d_5,d_6)} \eqw\eqs f_2^{(d_3,d_4)}\circ g^{(d_5,d_6)} }}$ 
           \vspace{5pt} \\
(pwsubs)$\mathtt{\dfrac{g^{(d_5,0)}\colon X\to Y \squad f_1^{(d_1,d_2)}\eqs\eqw f_2^{(d_3,d_4)}\colon Y\to Z}
{f_1^{(d_1,d_2)} \circ g^{(d_5,0)} \eqs\eqw f_2^{(d_3,d_4)}\circ g^{(d_5,0)}}}$     
\squad
 (wrepl)$\mathtt{\dfrac{f_1^{(d_1,d_2)}\eqs\eqw f_2^{(d_3,d_4)}\colon X\to Y \squad g^{(d_5,d_6)}\colon Y\to Z}
{g^{(d_5,d_6)}\circ f_1^{(d_1,d_2)} \eqs\eqw g^{(d_5,d_6)}\circ f_2^{(d_3,d_4)} }}$ 
          \vspace{5pt} \\
(replsubs) 
  $\mathtt{\dfrac{f_1^{(d_1,d_2)} \eqs\eqs f_2^{(d_3,d_4)}\colon X\to Y \squad g_1^{(d_5,d_6)} \eqs\eqs g_2^{(d_7,d_8)}\colon Y\to Z}
    {g_1^{(d_5,d_6)}\circ f_1^{(d_1,d_2)} \eqs\eqs g_2^{(d_7,d_8)}\circ f_2^{(d_3,d_4)}}}$ 
    \vspace{5pt} \\
 (w$\_$unit) 
  $\dfrac{\mathtt{f^{(d_1,d_2)}\colon X\to \unit}}{\mathtt{f^{(d_1,d_2)} \eqw\eqs \pa_X^{(0,0)}}}$ \squad
   (w$\_$empty)  
  $\dfrac{\mathtt{f^{(d_1,d_2)}\colon \empt\to X}}{\mathtt{f^{(d_1,d_2)} \eqs\eqw \copa_X^{(0,0)}}}$ \squad
  (w$\_$downcast)  
  $\dfrac{\mathtt{f^{(d_1,2)}\colon Y\to X}}{\mathtt{(\mathtt{\downarrow}\ f)^{(d_1,1)} \eqs\eqw f^{(d_1,2)}}}$ 
  \vspace{5pt} \\
 (ax$_1$) 
  $\dfrac{}{\lookup_\mathtt{i}^{(1,0)} \circ \update_\mathtt{i}^{(2,0)} \eqw\eqs \mathtt{id_{V_i}^{(0,0)}}} $ 
\squad (ax$_2$)
  $\dfrac{\mathtt{\forall i, j\in Loc, \ i \neq j}}
  {\lookup_\mathtt{i}^{(1,0)} \circ \update_\mathtt{j}^{(2,0)} \eqw\eqs \lookup_\mathtt{i}^{(1,0)} \circ
  \pa_{V_i}^{(0,0)}} $ 
  \vspace{5pt} \\
   (eax$_1$) 
  $\dfrac{}{\untag^{(0,2)}_\mathtt{e} \circ \tagg^{(0,1)}_\mathtt{e} \eqs\eqw \mathtt{id_{EV_e}^{(0,0)}}} $
\quad \quad \ (eax$_2$)
  $\dfrac{\mathtt{\forall e_1, e_2 \in EName,\ e_1 \neq e_2}}
  {\untag^{(0,2)}_\mathtt{e_1} \circ \tagg^{(0,1)} _\mathtt{e_2} \eqs\eqw \copa^{(0,0)}_{\mathtt{EV_e}} \circ \tagg^{(0,1)}_\mathtt{e_2} } $
  \vspace{5pt} \\
 (effect) 
  $\dfrac{\mathtt{f_1^{(d_1,d_2)},f_2^{(d_3,d_4)}\colon X\to Y \;\;
  f_1^{(d_1,d_2)}\eqw\eqs f_2^{(d_3,d_4)} \;\; \pa_Y^{(0,0)}\circ f_1^{(d_1,d_2)} \eqs\eqs \pa^{(0,0)}_Y\circ f_2^{(d_3,d_4)}}}
  {\mathtt{f_1^{(d_1,d_2)}\eqs\eqs f_2^{(d_3,d_4)}}}$ 
  \vspace{5pt} \\
   (eeffect) 
  $\mathtt{\dfrac{f_1^{(d_1,d_2)},f_2^{(d_3,d_4)}\colon Y\to X \;\;
  f_1^{(d_1,d_2)}\eqs\eqw f_2^{(d_3,d_4)} \;\; f_1^{(d_1,d_2)} \circ \copa^{(0,0)}_Y \eqs\eqs f_2^{(d_3,d_4)} \circ \copa^{(0,0)}_Y}
  {f_1^{(d_1,d_2)} \eqs\eqs f_2^{(d_3,d_4)}}}$
  \vspace{5pt} \\ 
 (local$\_$global) 
  $\dfrac{\mathtt{f_1^{(d_1,d_2)},f_2^{(d_3,d_4)}\colon X\to \unit \quad \forall\ \mathtt{i\in Loc}, \, \lookup^{(1,0)}_\mathtt{i}\circ f_1^{(d_1,d_2)} \eqw\eqs \lookup^{(1,0)}_\mathtt{i}\circ f_2^{(d_3,d_4)}}}
    {\mathtt{f_1^{(d_1,d_2)}\eqs\eqs f_2^{(d_3,d_4)}}}$ 
    \vspace{5pt} \\
 (elocal$\_$global) 
  $\mathtt{\dfrac{ f_1^{(d_1,d_2)},f_2^{(d_3,d_4)}\colon \empt\to X \quad \forall\ \mathtt{e \in EName}, \ f_1^{(d_1,d_2)} \circ
  \tagg^{(0,1)}_\mathtt{e}
   \eqs\eqw f_2^{(d_3,d_4)} \circ
  \tagg^{(0,1)}_\mathtt{e}}
    {f_1^{(d_1,d_2)} \eqs\eqs f_2^{(d_3,d_4)}}}$
    \vspace{5pt} \\
 (w$\_$lpair$\_$eq) 
  $\dfrac{\mathtt{f_1^{(d_1,d_2)}: X \to Y \squad f_2^{(d_3,d_4)}: X \to Z \squad d_1 \in \{0,1\}}}
  {\mathtt{\pi_1^{(0,0)} \circ \lpair{f_1,f_2}^{((max(d_1,d_3),max(d_2,d_4))} \eqw\eqs f_1^{(d_1,d_2)}}}$ 
      \vspace{5pt} \\
 (s$\_$lpair$\_$eq) 
  $\dfrac{\mathtt{f_1^{(d_1,d_2)}: X \to Y \squad f_2^{(d_3,d_4)}: X \to Z \squad d_1 \in \{0,1\} }}
  {\mathtt{\pi_2^{(0,0)} \circ \lpair{f_1,f_2}^{((max(d_1,d_3),max(d_2,d_4))} \eqs\eqs f_2^{(d_3,d_4)}}}$
  \vspace{5pt} \\
 (w$\_$lcopair$\_$eq)  
  $\mathtt{\dfrac{f_1^{(d_1,d_2)}: X \to Y \squad f_2^{(d_3,d_4)}: Z \to Y \squad d_2 \in \{0,1\}}
  {[f_1\mid f_2]^{((max(d_1,d_3),max(d_2,d_4))}  \circ \mathtt{in_1}^{(0,0)} \eqs\eqw f_1^{(d_1,d_2)}}}$ 
    \vspace{5pt} \\
 (s$\_$lcopair$\_$eq)   
  $\mathtt{\dfrac{f_1^{(d_1,d_2)}: X \to Y \squad f_2^{(d_3,d_4)}: Z \to Y \squad d_2 \in \{0,1\}}
  {[f_1\mid f_2]^{((max(d_1,d_3),max(d_2,d_4))}  \circ \mathtt{in_2}^{(0,0)} \eqs\eqs f_2^{(d_3,d_4)}}}$ 
  \vspace{5pt} \\
 (tcomp) $\dfrac{\mathtt{f^{(p,p)}: Y \to Z} \squad  \mathtt{g^{(p,p)}: X \to Y}}
{\mathtt{(tpure \ f)^{(0,0)} \circ (tpure\ g)^{(0,0)} \eqs\eqs (tpure \ (f \,{\color{red}\circ}\, g))^{(0,0)}} }$
\end{tabular}
\vspace{-0.15cm}
\caption{$\Log_{st+exc}$: rules}
\label{rls-stexc} 
\end{figure}
We plan it as a future work to come up with a more general and systematic way to combine effects formalized within decorated logics.

\subsection{Decorated properties of the state and exception effects}
\label{dpotsee}
The properties given in Sections~\ref{dpotms}~and~\ref{dpotee} are now stated with the
refined term decorations, and related with the equation sort $\eqs\eqs$.
I.e., the statements \texttt{propagator-propagates}
and \texttt{update-update}
look like:
\begin{align*}
\nonumber
 &  \forall\, \mathtt{e \in \mathtt{EName,\; a^{(0,1)}\colon X\to Y, \;\; 
 a^{(0,1)} \circ \copa_X^{(0,0)} \circ \tagg_e^{(0,1)}} \eqs\eqs
  \mathtt{\copa_Y^{(0,0)} \circ \tagg_e^{(0,1)}\colon EV_e\to Y.}}\\
& \mathtt{\forall\, i\ne j\in Loc ,\;
   \update^{(2,0)}_j \circ \pi^{(0,0)}_2 \circ (\update^{(2,0)}_i \times_r id^{(0,0)}_{V_j})}  \eqs \eqs \\
  & \ \ \ \ \ \ \ \ \ \ \ \
  \ \ \ \ \ \ \ \ \ \ \ \
  \ \ \ \ \ \ \ \ \ \ \ 
  \ \ \ \ \ \ \ \ \ \ \ \
  \ \ \ \ \ \ \ \ \ \ \ \ \
  \mathtt{ \update^{(2,0)}_i \circ \pi^{(0,0)}_1 \circ (id^{(0,0)}_{V_i} \times_l \update^{(2,0)}_j) 
    : V_i\times V_j \to \unit.} 
  \end{align*}
These are the archetype properties that we can prove within the scope of the
 $\Log_{st+exc}$. Although it is doable,
 we prefer not to prove them for this generic framework (skipped since it would take substantial amount of time);
%
instead, we first specialize them
in a way to serve as  a target language for a denotational semantics of \texttt{IMP+Exc}, and then
prove them for the specialized version. Also, we
encode the specialized version in \coq and certify related proofs. Section~\ref{impexcodl} gives
the related details.

\section{\texttt{IMP+Exc} over the combined decorated logic $\Log_{st+exc}$}
\label{impexcodl}
Now, it comes to define a denotational semantics for the \texttt{IMP+Exc}
language, with the 
combined decorated logic for the state and the exception ($\Log_{st+exc}$) as the target language.
Recall that by doing this, we aim to prove some (strong) equalities between
terminating programs written in \texttt{IMP+Exc} with respect to the state and
the exception effects.

In \texttt{IMP+Exc},
the values that can be stored in any location (variable) \texttt{i} are just 
integers.
So that any occurrence of $\mathtt{(V_i)}$ in term signatures
of $\Log_{st+exc}$ is
replaced by
$\mathbb{Z}$. I.e., $\lookup^{(1,0)}\colon \unit \to \mathbb{Z}$ and
$\update^{(2,0)}\colon \mathbb{Z} \to \unit$.
We now define a denotational semantics of \texttt{IMP+Exc}
expressions over combined decorated settings using two translator functions $\mathtt{daExp}$
and  $\mathtt{dbExp}$. The former takes an arithmetic expression as input
and outputs a decorated term of
type $\mathtt{term}\  \mathbb{Z}\  \mathtt{\unit}$, while the latter
takes a Boolean expression and returns a decorated term of type
$\mathtt{term}\  \mathbb{B}\  \mathtt{\unit}$:
$$\begin{array}{lll}
   \xymatrix@R=0.001pc@C=0.025pc{
\mathtt{daExp \ n} &\Rightarrow& \mathtt{(constant\ n)}^{(0,0)} \\
\mathtt{daExp \ x}  &\Rightarrow& \mathtt{(lookup_x)}^{(1,0)}\\
\mathtt{daExp}\ (\mathtt{a_1\splus  a_2})  &\Rightarrow& \mathtt{(tpure\ add)}^{(0,0)} \circ  \mathtt{\lpair{\mathtt{daExp \ a_1},  \mathtt{daExp \ a_2}}^{(d,0)}}\\
\mathtt{daExp}\ (\mathtt{a_1\times  a_2})  &\Rightarrow& \mathtt{(tpure\ mlt)}^{(0,0)} \circ  \mathtt{\lpair{\mathtt{daExp \ a_1},  \mathtt{daExp \ a_2}}^{(d,0)}}\\
\mathtt{daExp}\ (\mathtt{a_1 -  a_2})  &\Rightarrow& \mathtt{(tpure\ subt)}^{(0,0)} \circ  \mathtt{\lpair{\mathtt{daExp \ a_1},  \mathtt{daExp \ a_2}}^{(d,0)}}\\
\mathtt{dbExp \ b} &\Rightarrow& \mathtt{(constant\ b)}^{(0,0)} \\
\mathtt{dbExp}\ (\mathtt{a_1 \overset{?}{=}  a_2})  &\Rightarrow& \mathtt{(tpure\ chkeq)}^{(0,0)} \circ  \mathtt{\lpair{\mathtt{daExp \ a_1},  \mathtt{daExp \ a_2}}^{(d,0)}}\\
   \mathtt{dbExp}\ (\mathtt{a_1 \overset{?}{\neq}  a_2})  &\Rightarrow& \mathtt{(tpure\ chkneq)}^{(0,0)} \circ  \mathtt{\lpair{\mathtt{daExp \ a_1},  \mathtt{daExp \ a_2}}^{(d,0)}}\\
         \mathtt{dbExp}\ (\mathtt{a_1 \overset{?}{>} a_2})  &\Rightarrow& \mathtt{(tpure\ chkgt)}^{(0,0)} \circ  \mathtt{\lpair{\mathtt{daExp \ a_1},  \mathtt{daExp \ a_2}}^{(d,0)}}\\
\mathtt{dbExp}\ (\mathtt{a_1  \overset{?}{<}  a_2})  &\Rightarrow& \mathtt{(tpure\ chklt)}^{(0,0)} \circ  \mathtt{\lpair{\mathtt{daExp \ a_1},  \mathtt{daExp \ a_2}}^{(d,0)} }\\
}
\end{array}$$
$$\begin{array}{lll}
   \xymatrix@R=0.001pc@C=0.025pc{
   \mathtt{dbExp}\ (\mathtt{b_1 \wedge b_2})  &\Rightarrow& \mathtt{(tpure\ andB)}^{(0,0)} \circ  \mathtt{\lpair{\mathtt{dbExp \ b_1},  \mathtt{dbExp \ b_2}}^{(d,0)}}\\
\mathtt{dbExp}\ (\mathtt{b_1 \vee b_2})  &\Rightarrow& \mathtt{(tpure\ orB)}^{(0,0)} \circ  \mathtt{\lpair{\mathtt{dbExp \ b_1},  \mathtt{dbExp \ b_2}}^{(d,0)}}\\
\mathtt{dbExp}\ (\mathtt{\neg b})  &\Rightarrow& \mathtt{(tpure\ notB)}^{(0,0)} \circ  \mathtt{dbExp \ b}^{(\mathtt{d},0)} 
}
\end{array}$$
\noindent
In ``$\mathtt{dbExp \,b}$'' ($6^{th}$ line above on the left), \texttt{b} can be either of the Boolean expressions \texttt{true} and \texttt{false}.
The constructor
\texttt{tpure} is applied to given unary and binary functions. For instance $\mathtt{add\colon (\mathbb{Z} \times \mathbb{Z}) \to \mathbb{Z}}$
takes an instance of an integer tuple and returns their sum. To see the definition of these functions in a \coq implementation,
please check out this file \footnote{\url{https://github.com/ekiciburak/impex-on-decorated-logic/blob/master/Functions.v}}.
\begin{remark}
\label{rmk71}
An expression in in \texttt{IMP+Exc} can have memory access right (i.e., a variable \texttt{x}) but can never
throw or catch exceptions. To calculate the decoration \texttt{d} of 
an arithmetic expression pair, i.e., $\mathtt{\lpair{daExp\,\, a_1, daExp\,\, a_2}^{(d, 0)}}$,
we use the following strategy:
$$\begin{array}{l}
   \xymatrix@R=0.01pc@C=0.5pc{
\mathtt{d := let \,\,f^{(d_1,0)} = daExp(a_1)\,\, in \,\,
                           let \,\,g^{(d_2, 0)} = daExp(a_2)\,\, in \,\,
                           max (d_1, d_2).}
   }
\end{array}$$
The same strategy follows for Boolean expressions, too.

\end{remark}

We have some additional rules to make use of some pure algebraic 
operations in the combined decorated setting presented in Figure~\ref{ar-imps}
where the
pure term $\mathtt{lpi\,\,b\,\,f\colon \unit \to \unit}$, within the rule (imp-li), is used to
bridge successive loop iterations
as long as the loop conditional evaluates into decorated logic's true (\texttt{constant}\ \emph{true}).
Also, the pure term $\mathtt{pbl\colon \mathbb{B}\to \unit \splus \unit}$
forms a bridge between the usual Boolean data type and its correspondence
in the decorated settings which is the type $\unit \splus \unit$.
$$\begin{array}{l}
   \xymatrix@R=0.01pc@C=0.5pc{
\mathtt{lpi\ (b: term\ \unit \ (\unit+\unit)) \ (f: term\ \unit\ \unit)} &:=& \mathtt{tpure \ (\lambda x:\unit.x).} \\
   \mathtt{pbl} &:=& \mathtt{tpure \ (bool\_to\_two).}\\
}\\
\text{where}\
   \xymatrix@R=0.01pc@C=0.5pc{
   \mathtt{bool\_to\_two\ (b: bool)} &:=& \mathtt{(if\ b\ then\ (inl\ void)\ else\ (inr\ void)).}
   }\\
\text{such that}\
   \xymatrix@R=0.01pc@C=0.5pc{
   \mathtt{void: \unit} \text{ is the unique constructor of the type } \mathtt{\unit ,} \ \text{and}\\
    \mathtt{inl,\ inr: \unit \to (\unit+\unit)}  \text{ are the canonical inclusions.} 
   }
\end{array}$$
The rule $\mathtt{(imp_6)}$ (functional extensionality), in Figure~\ref{ar-imps},
is to say that if two pure functions on Coq side are point-wise equal, then they are strongly equal in the
decorated setting. Here we take them strongly equal since strong and weak equalities are indistinguishable when
the related terms are pure.
The idea is to be able to use Coq's Leibniz equality as the strong equality in the decorated
setting. 

Note also that in
$\mathtt{(imp_2)}$ and $\mathtt{(imp_4)}$ by replacing \emph{false}
into \emph{true}
we get $\mathtt{(imp_3)}$ and $\mathtt{(imp_5)}$ that are
not explicitly stated in Figure~\ref{ar-imps}.
\begin{figure}[h]
$$\begin{array}{l}
   \xymatrix@R=0.01pc@C=0.5pc{
\mathtt{(imp_1)\squad \dfrac{\forall p,\ q: \mathbb{Z}, \ (f:  \mathbb{Z} \times  \mathbb{Z} \to  \mathbb{Z})}
{tpure \ f \circ \lpair{constant \ p, constant \ q} \eqs\eqs (constant \ f(p, q))}} \\
\mathtt{(imp_2)\squad \dfrac{\forall p,\ q: \mathbb{Z}, \ (f:  \mathbb{Z} \times  \mathbb{Z} \to  \mathbb{B})
\squad f(p,q) = \emph{false}}
{tpure \ f \circ \lpair{constant \ p, constant \ q} \eqs\eqs \mathtt{constant \ \emph{false}}}}\\
\mathtt{(imp_4)\squad \dfrac{\forall p,\ q: \mathbb{B}, \ (f:  \mathbb{B} \times  \mathbb{B} \to  \mathbb{B})
\squad f(p,q) = \emph{false}}
{tpure \ f \circ \lpair{constant \ p, constant \ q} \eqs\eqs \mathtt{constant\  \emph{false}}}}\\
   \mathtt{(imp-li)}\squad \dfrac{\mathtt{\forall (b: term\ \unit \ (\unit+\unit)) \ (f: term\ \unit\ \unit)}}
{\mathtt{lpi \ b\ f \eqs\eqs \big[(lpi \ b\ f) \circ f \big| id \big]_\mathtt{l} \circ b}} \\
\mathtt{(imp_6)} \dfrac{\mathtt{ (\forall x,\ f\ x \ = \ g\ x)}}
{\mathtt{tpure \ f \eqs tpure\ g}}\\
}
\end{array}$$
{\vskip-0.5cm}
\caption{Additional rules on pure terms: \texttt{IMP+Exc} specific}
\label{ar-imps}
\end{figure}

\begin{lemma}
\label{pblf}
$\mathtt{pbl^{(0,0)} \circ (constant}\ false\mathtt{)}^{(0,0)} \eqs\eqs \mathtt{in_2}.$
\end{lemma}
\proof
unfolding all term definitions, we have
$\mathtt{tpure}$
$\mathtt{(\lambda b: bool.}$
$\mathtt{if}$
$\mathtt{b}$
$\mathtt{then}$
$\mathtt{(inl \ void)}$
$\mathtt{else \ (inr \ void))}$
$\circ$ $\mathtt{tpure}$
$\mathtt{(\lambda \ \_ : void.true)}$
$\mathtt{\eqs\eqs}$
$\mathtt{tpure \ inl}$.
Now, we obtain
$\mathtt{\forall x: \unit, \, inl\, void = inl \,x}$
by first rewriting \texttt{tcomp} from left to right, and
then applying $\mathtt{imp_6}$ which is trivial since Leibniz equality `$\mathtt{=}$' is reflexive. \qed
\begin{lemma}
\label{pblt}
$\mathtt{pbl^{(0,0)} \circ (constant}\ true\mathtt{)}^{(0,0)} \eqs\eqs \mathtt{in_1}.$
\end{lemma}
\proof It follows the same steps with the proof of Lemma~\ref{pblf} \qed
\begin{remark}
See this file
\footnote{\url{https://github.com/ekiciburak/impex-on-decorated-logic/blob/master/Derived_co_Pairs.v\#L122-L133}} for the \coq
certified proofs of the Lemmas~\ref{pblf}~and~\ref{pblt}.
\end{remark}
The fact that \texttt{IMP+Exc} commands are of type $\mathtt{\unit \to \unit}$,
 in $\throw_\mathtt{e}^{(0,1)} := \copa_\mathtt{Y}^{(0,0)} \circ \tagg_\mathtt{e}^{(0,1)}\colon \mathtt{EV_e} \to \mathtt{Y}$,
we replace $\mathtt{EV_e}$ and $\mathtt{Y}$ with $\unit$. This means that
we stick to a single exceptional value (parameter),
for each exception name $\mathtt{e \in EName}$.

Below, we recursively define the \texttt{IMP+Exc} commands within $\Log_{st+exc}$ using a
translator function $\mathtt{dCmd}$ which establishes a decorated term
of type $\mathtt{term}\ \mathtt{\unit} \  \mathtt{\unit}$ out of an input command:
$$\begin{array}{c}
   \xymatrix@R=0.0001pc@C=0.25pc{
\mathtt{dCmd \ ( SKIP)} &\Rightarrow& \mathtt{(id \ \unit)^{(0,0)}} \\
\mathtt{dCmd \ (x \triangleq a)}  &\Rightarrow& \mathtt{(update_x)^{(2,0)}} \circ \ \mathtt{(daExp \ a)}^{(\mathtt{d_1,0})}\\
\mathtt{dCmd \ (c_1; c_2)} &\Rightarrow& \mathtt{(dCmd \ c_2)^{(d_1,d_2)}} \circ \mathtt{(dCmd \ c_1)^{(k_1,k_2)}} \\
\mathtt{dCmd \ (if \ b \ then \ c_1 \ else \ c_2)} &\Rightarrow& \mathtt{\Big[dCmd \ c_1\Bigm| dCmd \ c_2\Big]_\mathtt{l}^{(d_1,d_2)}} \circ \ \mathtt{pbl}^{(0,0)} \circ \mathtt{(dbExp \ b)^{(d_3,0)}} \\
\mathtt{dCmd \ (while \ b\ do \ c)} &\Rightarrow& \mathtt{\Big[(lpi \ (\mathtt{pbl} \circ (dbExp \ b)) \ (dCmd \ c)) \circ (dCmd \ c)\Bigm| id\Big]_\mathtt{l}^{(d_1,d_2)}} \\
& &\circ 
\ \mathtt{pbl}^{(0,0)} \circ \mathtt{(dbExp \ b)^{(d_3,0)}} \\
\mathtt{dCmd \ (THROW \ e)}  &\Rightarrow& \mathtt{throw\ e}^{(0,1)}\\
\mathtt{dCmd \ (TRY \ c_1 \ CATCH \ e \Rightarrow c_2)} &\Rightarrow& \mathtt{\try\ (dCmd\ c_1)\ \catch\ (e\To (dCmd \ c_2))^{(d_1,d_2)}}
	}
\end{array}$$
\begin{remark}
To calculate the decorations $\mathtt{d_1}$ and $\mathtt{d_2}$ (or $\mathtt{k_1}$ and $\mathtt{k_2}$), we use
the following strategy:
$$\begin{array}{l}
   \xymatrix@R=0.01pc@C=0.5pc{
\mathtt{d_1 := let \,\,f^{(d'_1,d'_2)} = dCmd(c_1)\,\, in \,\,
                           let \,\,g^{(d'_3, d'_4)} = dCmd(c_2)\,\, in \,\,
                           max (d'_1, d'_3).
}
\\
\mathtt{d_2 := let \,\,f^{(d'_1,d'_2)} = dCmd(c_1)\,\, in \,\,
                           let \,\,g^{(d'_3, d'_4)} = dCmd(c_2)\,\, in \,\,
                           max (d'_2, d'_4).
} }
\end{array}$$
For the strategy to calculate $\mathtt{d_3}$, see Remark~\ref{rmk71}.
Also, recall Definition~\ref{def:tc}:
translation of any \texttt{IMP+Exc} command cannot be a public catcher, even the one
for \texttt{TRY/CATCH}. Thus, \texttt{dCmd} function outputs terms
at most wih decoration $\acc$ with respect to the exception effect.
\end{remark}

In Figure~\ref{cw-d}, the diagram on the left schematizes the command $\mathtt{if\ b\ then\ c_1 \ else \ c_2}$:
if the Boolean expression \texttt{dbExp b} evaluates into (\texttt{constant} \emph{true}) then by Lemma~\ref{pblt},
we have the command $\mathtt{c_1}$ in execution,
$\mathtt{c_2}$ otherwise by Lemma~\ref{pblf}.
As for the loops, it is well know that as long as the looping
condition evaluates into (\texttt{constant} \emph{true}),
loop body gets executed. 
This is depicted in~Figure~\ref{cw-d} (the diagram on the right),
as the arrow $\mathtt{lpi \ b \ c}$ is each time replaced by the 
whole diagram itself. The rule \texttt{(imp-li)} allows us to do so. 
If the looping condition evaluates into (\texttt{constant} \emph{false}),
using  Lemma~\ref{pblf},
we then have the term $\mathtt{id_\unit}$ in execution forcing the loop to terminate.
Recall that the case distinction in the diagrams are provided by the term inclusions.

\begin{figure}[h]
$$\begin{array}{ll}
\xymatrix@R=2pc@C=2.5pc{
 & &\unit  \ar[d]_{\mathtt{in_1}} \ar[rd]^{\mathtt{c_1}}&   \\ 
\unit\ar[r]^{\mathtt{dbExp \ b}} & \mathbb{B} \ar[r]^{\mathtt{pbl}} & \unit + \unit  \ar[r]|
{\big[\mathtt{c_1} \big| \mathtt{c_2}\big]} & \unit     
\\
& &\unit  \ar[u]^{\mathtt{in_2}} \ar[ru]_{\mathtt{c_2}}&  
} & \quad
\xymatrix@R=2pc@C=2.5pc{
&  & \unit \ar[d]_{\mathtt{in_1}} \ar[r]^{\texttt{c}} & \unit \ar@/^-2.5pc/@[blue][llld]_{\mathtt{lpi\ b\ c
}}&   \\ 
\unit\ar[r]^{\mathtt{dbExp \ b}} & \mathbb{B} \ar[r]^{\mathtt{pbl}} & \unit + \unit  \ar[rr]|
{\big[\texttt{(lpi b c)} \ \circ \ \mathtt{c} \big| \mathtt{id_{\unit}}\big]} && \unit     
\\
& & \unit  \ar[u]^{\mathtt{in_2}} \ar[rru]_{\mathtt{id_{\unit}}}&  
} 
\end{array}$$
{\vskip-0.5cm}
\caption{$\mathtt{(cond\ b\ c_1\ c_2)}$ and $\mathtt{(while\ b\ do\ c)}$ in $\Log_{st+exc}$}
\label{cw-d}
\end{figure}
Figure~\ref{tt-d} respectively
visualizes the formal behaviors of $\mathtt{THROW}$ and $\mathtt{TRY/CATCH}$ commands where
the basis is the core decorated terms for the exception effect.
They are formulated as in Definitions~\ref{def:throw}~and~\ref{def:tc}
with a single difference in their signatures: domains and co-domains are set
to $\mathtt{\unit}$.

\begin{figure}[h]
$$\begin{array}{ll}
\xymatrix@R=2pc@C=3.0pc{ \\
\unit \ar[r]^{\mathtt{tag_e}} & \empt \ar[r]^{[]_{\unit}} & \unit
} & \quad
\xymatrix@R=1.75pc@C=3.25pc{
& & \unit  \ar[d]_{\mathtt{in_1}} \ar[rrd]^{\mathtt{id_{\unit}}}&  \\
\unit \ar[r]^{\mathtt{c_1}}&\mathbb{\unit} \ar[r]^{\mathtt{in_1}} & \unit + \empt  \ar[rr]|{\big[\mathtt{id_\unit} \big| \mathtt{c_2} \ \circ \ \mathtt{untag_e} \big]} & & \unit    \\
& & \empt \ar[u]^{\mathtt{in_2}} \ar[r]^{\mathtt{untag_e}} & \mathtt{\unit} \ar[ru]_{\mathtt{c_2}}&   \\ 
} 
\end{array}$$
{\vskip-0.5cm}
\caption{$\mathtt{(THROW\ e)}$ and $\mathtt{(TRY \ c_1 \ CATCH \ e \Rightarrow c_2)}$ in $\Log_{st+exc}$}
\label{tt-d}
\end{figure} 

We now encode the \texttt{IMP+Exc} denotational semantics, with the $\Log_{st+exc}$ as the target language, in \coq.
Arithmetic and
Boolean expressions are inductively forming new \coq \texttt{Type}s, called  \texttt{aExp} and
\texttt{bExp} respectively.
As for the type constructors, we use the syntactic operators given as parts of
\texttt{aexp} and \texttt{bexp} in Figure~\ref{s-imp}. The difference lies in the naming:
notations are translated into plain text.
It is easy to match them one another as they are given in the
same order. Notice also that the implementation of
the constant Boolean expressions \texttt{true} and \texttt{false}
are subsumed by the constructor \texttt{bconst}.


\begin{tcb}[\scriptsize]{Coq}
 Inductive aExp : Type :=
    | aconst: Z    -> aExp
    | var   : Loc  -> aExp
    | plus  : aExp -> aExp -> aExp
    | subtr : aExp -> aExp -> aExp
    | mult  : aExp -> aExp -> aExp.

 Inductive bExp : Type :=
    | bconst: bool -> bExp
    | eq    : aExp -> aExp -> bExp
    | neq   : aExp -> aExp -> bExp
    | gt    : aExp -> aExp -> bExp
    | lt    : aExp -> aExp -> bExp
    | ge    : aExp -> aExp -> bExp
    | le    : aExp -> aExp -> bExp
    | and   : bExp -> bExp -> bExp
    | or    : bExp -> bExp -> bExp
    | neg   : bExp -> bExp.
\end{tcb}
We interpret the functions \texttt{daExp} and  \texttt{dbExp}  
in \coq using following fixpoints:
\begin{tcb}[\scriptsize]{Coq}
 Fixpoint daExp (e: aExp): term Z unit :=
  match e with
    | aconst n      => constant n
    | var x         => lookup x
    | plus a1 a2    => tpure add o pair (daExp a1) (daExp a2)
    | subtr a1 a2   => tpure subt o pair (daExp a1) (daExp a2)
    | mult a1 a2    => tpure mlt o pair (daExp a1) (daExp a2)
  end.

 Fixpoint dbExp (e: bExp): term bool unit :=
  match e with
    | bconst n      => constant n
    | eq a1 a2      => tpure chkeq o pair (daExp a1) (daExp a2)
    | neq a1 a2     => tpure chkneq o pair (daExp a1) (daExp a2)
    | gt a1 a2      => tpure chkgt o pair (daExp a1) (daExp a2)
    | lt a1 a2      => tpure chklt o pair (daExp a1) (daExp a2)
    | ge a1 a2      => tpure chkge o pair (daExp a1) (daExp a2)
    | le a1 a2      => tpure chkle o pair (daExp a1) (daExp a2)
    | and b1 b2     => tpure andB o pair (dbExp b1) (dbExp b2)
    | or b1 b2      => tpure orB o pair (dbExp b1) (dbExp b2)
    | neg b         => tpure notB o (dbExp b)
  end.
\end{tcb}
We follow a similar idea to  implement commands. We inductively define
a \coq type \texttt{Cmd} of \texttt{IMP+Exc} commands whose constructors are the
members of \texttt{IMP+Exc} command set as presented in Figures~\ref{s-imp}~and~\ref{imp-acmd}.
Notice that some commands are encoded with different names.
I.e., the assignment command `$\mathtt{\triangleq}$' is called \texttt{assign},
the sequencing command `\texttt{;}' is called \texttt{sequence} 
while ``$\mathtt{if \ then \ else}$'' block
is named \texttt{cond} in the
implementation. It is easy to match them one another since they are presented in the
same order.

\begin{tcb}[\scriptsize]{Coq}
 Inductive Cmd  : Type :=
    | skip      : Cmd
    | sequence  : Cmd   -> Cmd  -> Cmd
    | assign    : Loc   -> aExp -> Cmd 
    | cond      : bExp  -> Cmd  -> Cmd -> Cmd
    | while     : bExp  -> Cmd  -> Cmd
    | THROW     : EName -> Cmd
    | TRY_CATCH : EName -> Cmd -> Cmd -> Cmd.
\end{tcb}
We now interpret the \texttt{dCmd} function 
in \coq using the below fixpoint:

\begin{tcb}[\scriptsize]{Coq}
 Fixpoint dCmd (c: Cmd): (term unit unit) :=
  match c with
    | skip              => (@id unit)
    | sequence c0 c1    => (dCmd c1) o (dCmd c0)
    | assign j e0       => (update j) o (daExp e0)
    | cond b c2 c3      => copair (dCmd c2) (dCmd c3) o (pbl o (dbExp b))
    | while b c4        => (copair (lpi (pbl o (dbExp b)) (dCmd c4) o (dCmd c4)) (@id unit)) o 
                           (pbl o (dbExp b))
    | THROW e           => (throw unit e)
    | TRY_CATCH e c1 c2 => (try_catch e (dCmd c1) (dCmd c2))
  end.
\end{tcb}
Now, we retain sufficient material to state and prove equivalences between
programs written in
\texttt{IMP+Exc}. Also, the discussed \coq implementation allows us to certfy them in \coq.

\subsection{Program equivalence proofs}
\label{peqp-st-exc}
\noindent
In this section, we finally prove equivalences of several programs written in \texttt{IMP+Exc}, using the
denotational semantics characterized within the scope of the
logic $\Log_{st+exc}$.
Note that for the sake of simplicity, we will use $\mathtt{u_x}$, $\mathtt{l_x}$, $\mathtt{(t \ op)}$ and $\mathtt{(c \ p)}$ instead of $\mathtt{(\update\ x)}^{(2,0)}$, $\mathtt{(\lookup\ x)}^{(1,0)}$,
$\mathtt{(tpure \ op)}^{(0,0)}$ and $\mathtt{(constant \ p)}^{(0,0)}$, respectively.
\begin{remark}
Recall that the use of term products is to impose some order of term evaluation
on the mutable state.
\texttt{IMP+Exc} specific properties of the mutable state
are slightly different than their generic versions
(mentioned in Section~\ref{dpotms}) due to the fact that the language does not allow parallel term evaluations,
meaning that every term is evaluated in the given sequence.
Therefore, we no more need to use term products in property statements.
%
The properties we use through out the following proofs are re-stated in Figure~\ref{IMP-pp-st}.
The full certified \coq proofs of these properties can be found here
\footnote{\url{https://github.com/ekiciburak/impex-on-decorated-logic/blob/master/Proofs.v}}.
\begin{figure}[h]
\footnotesize
\begin{enumerate}
\item 
\texttt{interaction update-update}
 $ \forall\, \mathtt{x}\in \texttt{Loc}\ \mathtt{p,q: \mathbb{Z}},\ \mathtt{u_x\circ (c\ p) \circ 
 u_x \circ (c\ q)}$ $\mathtt{\eqs\eqs u_x\circ (c\ p)}
 $  

\item 
\texttt{commutation update-update}
 $ \forall\, \mathtt{x \neq y}\in \texttt{Loc}\ \mathtt{p,q: \mathbb{Z}},\ \mathtt{u_x\circ (c\ p) \circ 
 u_y \circ (c\ q)}$ $\mathtt{\eqs\eqs u_y\circ (c\ q) \circ u_x\circ (c\ p)}
 $  

\item 
\texttt{commutation-lookup-constant-update}
  $ \forall\, \mathtt{x}\in \mathtt{Loc},\mathtt{p, q}\in \mathbb{Z} , \;
  \mathtt{\lpair{l_x, (c\ q)}} \circ \mathtt{u_x} \circ \mathtt{(c\ p)} \eqs\eqs \mathtt{\lpair{(c\ p), (c\ q)}}
  \circ \mathtt{u_x} \circ \mathtt{(c\ p)} 
  $
\end{enumerate}  
{\vskip -0.4cm}
\caption{Primitive properties of the state: \texttt{IMP+Exc} specific}
\label{IMP-pp-st}
\end{figure}

%
\end{remark}

\begin{remark}
Below, we state three lemmata using the \texttt{IMP+Exc} notation introduced in Figures~\ref{s-imp}
and~\ref{imp-acmd}.
However, we introduce a new set of
notations for the \coq encoding to increase the readability score: browse
this set of notations here
\footnote{\url{https://github.com/ekiciburak/impex-on-decorated-logic/blob/master/IMPEX_to_COQ.v\#L185-L205}}
where, i.e., the \texttt{assign} command is denoted by `$::=$' while
the \texttt{sequence} command by `$;;$'. These notations do not appear through out the paper, but
might be of help in reading the lemma statements in the \coq encoding.
Notice also that they are not so pretty,
due to the fact that \coq internally reserves prettier notations for other issues.
\end{remark}

\noindent
Another point here to notice is that the proofs in the following might be long and hard to follow.
If you find it so, please try reading the \coq codes. They are written in parallel with
the ones on the paper. Starting from Lemma~\ref{l77}, we give overall explanations
about the way we compute the proof using our semantics before diving it into the detailed rule applications.
Note also that proofs are chosen to be presented in a way that the sides of the equations are simplified until
obtaining a trivial equation to solve. 

\begin{remark}
All the statements we prove below are strong equations. The reason is that \texttt{IMP+Exc} (or \texttt{IMP})
language does not have a \texttt{return} command. Thus, one cannot compare the values that two programs return.
When we use some combined decorated logic as a target language for the semantics of another language with
the \texttt{return} command (i.e., the \texttt{C} language), then it would make sense to prove sentences with
weak equations. Recall also that any strong equation can be seen as a weak equation.
\end{remark}

\begin{lemma}
For all exceptionally pure commands $\mathtt{f, \, g}$ $\mathtt{(doesNotThrowTC\, (f) = true}$,
$\mathtt{doesNotThrowTC}$ $\mathtt{(g) = true})$
and $\mathtt{b}$ $\mathtt{\in}$ $\mathtt{ \{ true, false\}}$, if program pieces $\mathtt{prog1}$
	and $\mathtt{prog2}$ are given as
in the following listings,
then $\mathtt{dCmd \ (prog1)}$ $\mathtt{\eqs\eqs}$  $\mathtt{dCmd \ (prog2)}$.

\noindent\begin{minipage}{.40\textwidth}
\begin{lstlisting}[caption=prog1,frame=tlrb]{Name}
/* prog1 */
if b then f else g;
\end{lstlisting}
\end{minipage}
\qquad\qquad
\begin{minipage}{.40\textwidth}
\begin{lstlisting}[caption=prog2,frame=tlrb]{Name}
/* prog2 */
if b then (if b then f else g)
else g;
	
\end{lstlisting}
\end{minipage}

\end{lemma}
\noindent
Note that
the function $\mathtt{doesNotThrowTC: cmd \rightarrow Bool}$ takes any command,
recursively checks whether the input involves either \texttt{THROW} or \texttt{TRY/CATCH}, 
and returns \texttt{true} if that is the case; \texttt{false} otherwise.
Browse this function, in a Coq implementation, here
\footnote{\url{https://github.com/ekiciburak/impex-on-decorated-logic/blob/master/IMPEX_to_COQ.v\#L148}}.

\proof We sketch the diagrams of both programs below:
$$\begin{array}{ll}
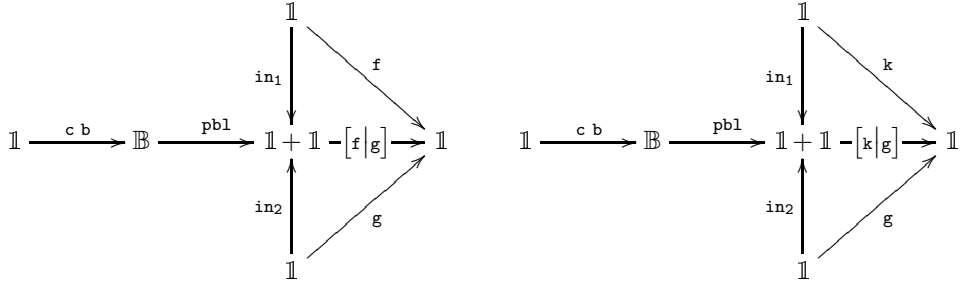
\xymatrix@R=3pc@C=3pc{
 & & \unit  \ar[d]_{\mathtt{in_1}} \ar[rd]^{\mathtt{f}}&   \\ 
\unit\ar[r]^{\mathtt{c \ b}} &\mathbb{B} \ar[r]^{\mathtt{pbl}} & \unit + \unit  \ar[r]|
{\big[\mathtt{f} \big| \mathtt{g}\big]} & \unit     
\\
& & \unit  \ar[u]^{\mathtt{in_2}} \ar[ru]_{\mathtt{g}}&  
} & \quad
\xymatrix@R=3pc@C=3pc{
& & \unit  \ar[d]_{\mathtt{in_1}} \ar[rd]^{\mathtt{k}}&   \\ 
\unit\ar[r]^{\mathtt{c \ b}} &\mathbb{B} \ar[r]^{\mathtt{pbl}} & \unit + \unit  \ar[r]|
{\big[\mathtt{k} \big| \mathtt{g}\big]} & \unit     
\\
& &\unit  \ar[u]^{\mathtt{in_2}} \ar[ru]_{\mathtt{g}}&  
} 
\end{array}$$
\noindent where $\mathtt{k}$ $=$ $\mathtt{(if\ b\ then\ f\ else\ g) }$. The
statement we would like to prove is
\begin{equation}
\mathtt{\big[f\big|g\big]_l \circ pbl \circ c\ b} \eqs\eqs \mathtt{\big[k\big|g\big]_l \circ pbl \circ c\ b}.
\end{equation}
Using the rules
of the logic $\Log_{st+exc}$, in the below given order, the idea is to simplify both
sides of the statement into the same shape with respect to the
equality sort $\eqs\eqs$.
The proof proceeds by a case analysis on \texttt{b}.  

%
If $\texttt{b = false}$, 
by unfolding the definitions of $\mathtt{pbl}$ and $(\mathtt{c} \ false)$, we have
\begin{equation}
 \mathtt{\big[}   \mathtt{f}   \mathtt{\big|}   \mathtt{g}   \mathtt{\big]_l}   \circ   \mathtt{t \ (bool\_to\_two)}   \circ   \mathtt{t \ (\lambda x:unit.false)}   \eqs\eqs   \mathtt{\big[}   \mathtt{k}   \mathtt{\big|}   \mathtt{g}   \mathtt{\big]_l}   \circ   \mathtt{t \ (bool\_to\_two)}   \circ   \mathtt{t\ (\lambda x:unit.false)} .
 \end{equation}
We rewrite (tcomp) on both sides, and get
\begin{equation}
 \mathtt{\big[}   \mathtt{f}   \mathtt{\big|}   \mathtt{g}   \mathtt{\big]_l}   \circ   \mathtt{t \ (\lambda x:unit.bool\_to\_two\ false)}   \eqs\eqs   \mathtt{\big[}   \mathtt{k}   \mathtt{\big|}   \mathtt{g}   \mathtt{\big]_l}   \circ   \mathtt{t \ }   \mathtt{(\lambda x:unit.bool\_two} \   \mathtt{false)} .
 \end{equation}
Now, we cut
\begin{equation}
\label{eq:cut}
\mathtt{t \ }\mathtt{(\lambda x:unit.bool\_to\_two\ false)} \eqs\eqs \mathtt{in_2}
\end{equation}
and rewrite it back in the goal.
So that we obtain
\begin{equation}
 \mathtt{\big[}   \mathtt{f}   \mathtt{\big|}   \mathtt{g}   \mathtt{\big]_l}   \circ   \mathtt{in_2}   \eqs\eqs   \mathtt{\big[}   \mathtt{k}   \mathtt{\big|}   \mathtt{g}   \mathtt{\big]_l}   \circ   \mathtt{in_2} .
 \end{equation}
Then, we use (s$\_$lcopair$\_$eq), and finally have $\mathtt{g \eqs\eqs g}$
which is trivial since $\eqs\eqs$ is reflexive. 
It remains to show that the cut statement in Equation~\ref{eq:cut}
holds.
By simplifying  $\mathtt{t \ }$ $\mathtt{(\lambda x: unit.}$ $\mathtt{bool}$ $\mathtt{\_to}$ $\mathtt{\_two}$ $\mathtt{false)}$ and unfolding
$\mathtt{in_2}$, we have
\begin{equation}
 \mathtt{t \ }   \mathtt{(\lambda x: unit.}   \mathtt{inr}   \ \mathtt{x)}   \eqs\eqs   \mathtt{t \ }   \mathtt{(inr)} .
 \end{equation}
Now,
we apply (imp$_6$) and get
\begin{equation}
\mathtt{\forall x:unit, inr \ x =} \mathtt{inr \ x}
\end{equation}
which is
trivial since the Leibniz equality `$=$' is reflexive.

If $\texttt{b = true}$, by following above procedure with \texttt{true} (instead of \texttt{false})
we first handle
\begin{equation}
\mathtt{\big[f\big|g\big]_l \circ in_1} \eqs\eqs \mathtt{\big[k\big|g\big]_l \circ in_1}
\end{equation}
and then freely convert $\eqs\eqs$ into $\eqs\eqw$.
There, rewriting the rule  (w$\_$lcopair$\_$eq) yields $\mathtt{f}$ $\eqs\eqw$ $\mathtt{k}$.
We unfold \texttt{k} with $\mathtt{b = true}$ and get
\begin{equation}
 \mathtt{f}   \eqs\eqw   \mathtt{\big[}   \mathtt{f}   \mathtt{\big|}   \mathtt{g}   \mathtt{\big]_l}   \circ   \mathtt{in_1} .
 \end{equation}
Now by rewriting  (w$\_$lcopair$\_$eq), we have $\mathtt{f \eqs\eqw f}$.
This
is again trivial, since the equality sort
$\eqs\eqw$ is reflexive.\qed

\begin{lemma}
\label{l77}
For all $\mathtt{x: Loc}$, if program pieces $\mathtt{prog3}$ and $\mathtt{prog4}$ are given as in the following listings, then
$\mathtt{dCmd \ (prog3)}$ $\eqs\eqs$ $\mathtt{dCmd \ (prog4)}$.

\noindent\begin{minipage}{.40\textwidth}
\begin{lstlisting}[caption=prog3,frame=tlrb]{Name}
/* prog3 */
x := 2;
while (x < 11)
	do (x := x + 4);
\end{lstlisting}
\end{minipage}
\qquad\qquad
\begin{minipage}{.40\textwidth}
\begin{lstlisting}[caption=prog4,frame=tlrb]{Name}
/* prog4 */
x := 14;
\end{lstlisting}
\end{minipage}

\end{lemma}

\proof 
In the proof structure we intend to reduce 
$\mathtt{prog3}$,
first dealing with the pre-loop assignments and the looping pre-condition. Since it evaluates into \textit{true}, in the second step we identify things related to the first loop iteration. The third step primarily studies the second and then the third loop iteration after which the looping pre-condition switches to \textit{false}. Finally, we explain the program termination and show that $\mathtt{prog3}$ does exactly the same state manipulation
with $\mathtt{prog4}$. Note also that we do not need to check the results they returned, since all \texttt{IMP+Exc} commands, thus programs,
return $\mathtt{void\colon U}$.

Below is the sketch of $\mathtt{prog3}$:
$$\begin{array}{ll}
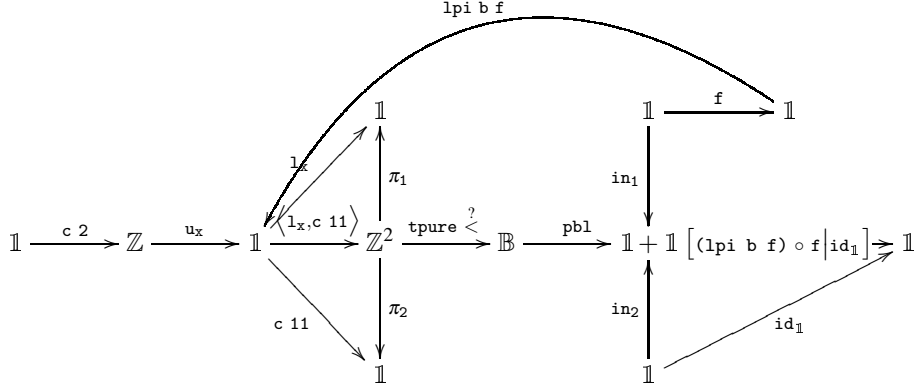
\xymatrix@R=3pc@C=2.75pc{
& & & \unit &  & \unit \ar[d]_{\mathtt{in_1}} \ar[r]^{\texttt{f}} & \unit \ar@/^-5pc/@[blue][lllld]_{\mathtt{lpi\ b\ f}}&   \\ 
\unit\ar[r]^{\mathtt{c\ 2}} & \mathbb{Z} \ar[r]^{\mathtt{u_x}}& \unit\ar[r]^{\mathtt{\pair{l_x, c\ 11}}}  \ar[ru]^{\mathtt{l_x}} \ar[rd]_{\mathtt{c\ 11}}& \mathbb{Z}^2 \ar[r]^{\mathtt{tpure \ \overset{?}{<}}} \ar[u]_{\mathtt{\pi_1}} 
\ar[d]^{\mathtt{\pi_2}}&\mathbb{B} \ar[r]^{\mathtt{pbl}}  & \unit + \unit  \ar[rr]|
{\big[\texttt{(lpi b f)} \ \circ \ \mathtt{f} \big| \mathtt{id_{\unit}}\big]}&  & \unit      
\\
& & & \unit& & \unit  \ar[u]^{\mathtt{in_2}} \ar[rru]_{\mathtt{id_{\unit}}}&  
} 
\end{array}$$
\noindent
where $\mathtt{f = (x \triangleq x+4)}$ and $\mathtt{b = (x \overset{?}{<} 11)}$. Using the rules
of the logic $\Log_{st+exc}$, we simplify this diagram
into the one given below with respect to the equality sort $\eqs\eqs$:
$$\begin{array}{ll}
\xymatrix@R=2pc@C=3pc{
\unit \ar[r]^{\mathtt{c \ 14}} & \mathbb{Z} \ar[r]^{\mathtt{u_x}} & \unit
} 
\end{array}$$
which is actually \texttt{prog4} when sketched. 

 
 \begin{enumerate}
\item 
Initially, we have
\begin{equation}
\big[\mathtt{(lpi }\
\mathtt{b}\ \mathtt{f)} \circ \mathtt{f} \mathtt{\big|} \mathtt{id_{\unit}} \mathtt{\big]} 
\circ \mathtt{pbl} \circ \mathtt{(t} \ \mathtt{\overset{?}{<})} \circ \mathtt{\pair{l_x, (c\ 11)}} \circ \mathtt{u_x}
\circ \mathtt{(c\ 2)} \eqs\eqs \mathtt{u_x} \circ \mathtt{(c\ 14) }.
\end{equation}
%
Let us simplify it as far as possible. By rewriting $\mathtt{commutation-lookup-constant-update}$
(see Figure~\ref{IMP-pp-st}), we obtain
\begin{equation}
 \mathtt{\big[}   \mathtt{(lpi \ b \ f)}   \circ   \mathtt{f}   \mathtt{\big|}   \mathtt{id_{\unit}}   \mathtt{\big]}   \circ   \mathtt{pbl}  
 \circ   \mathtt{(t} \  \mathtt{\overset{?}{<})} 
 \mathtt{\circ \pair{(c\ 2), (c\ 11)} \circ u_x \circ (c\ 2) } 
 \mathtt{\eqs\eqs  u_x \circ (c\ 14) }.
 \end{equation}
%
Since the looping pre-condition $\mathtt{(t \  \overset{?}{<}) \circ \pair{(c\ 2), (c\ 11)}}$
evaluates into $(\mathtt{c}\ true)$, and due to (imp$_3$), we have
 \begin{equation}
 \mathtt{\big[}   \mathtt{(lpi \ b \ f)}   \circ   \mathtt{f}   \mathtt{\big|}   \mathtt{id_{\unit}}   \mathtt{\big]}   \circ   \mathtt{pbl}  
 \circ   (\mathtt{c}\ true)   \circ   \mathtt{u_x \circ (c\ 2) }   \mathtt{\eqs\eqs  u_x \circ (c\ 14) } .
 \end{equation}
%
By rewriting the Lemma~\ref{pblt}, we get
\begin{equation}
 \mathtt{\big[\texttt{(lpi b f)} \ \circ \ \mathtt{f} \big| \mathtt{id_{\unit}}\big] \circ} 
 \mathtt{in_1 \circ u_x \circ (c\ 2) } 
 \mathtt{\eqs\eqs  u_x \circ (c\ 14) } .
 \end{equation}
Here, we first convert $\eqs\eqs$ into $\eqs\eqw$ then rewrite (w$\_$lcopair$\_$eq), and end up with
\begin{equation}
 \mathtt{(lpi\ b\  f) \circ f \circ  u_x \circ (c\ 2) } 
 \eqs\eqw   \mathtt{u_x}   \circ   \mathtt{(c\ 14)} 
 \end{equation}
in which the second appearance of \texttt{f} unfolds into
\begin{equation}
 \mathtt{(lpi}  \ \mathtt{b\  f)}   \circ   \mathtt{u_x}   \circ   \mathtt{(t}\   \mathtt{+)} 
 \mathtt{\circ \ \pair{l_x, c\ 4} \circ  u_x \circ (c\ 2) \eqs\eqw  u_x \circ (c\ 14) } . 
 \end{equation}
Since, there is no exceptional case, we are freely back to $\eqs\eqs$. 
By rewriting $\mathtt{commutation}$ $\mathtt{-lookup}$ $\mathtt{-constant}$ $\mathtt{-update}$,
we obtain
\begin{equation}
 \mathtt{(lpi}  \ \mathtt{b}  \  \mathtt{ f)}   \circ   \mathtt{u_x}   \circ   \mathtt{(t} \   \mathtt{+)} 
 \circ   \mathtt{\pair{c\ 2, c\ 4}}   \circ   \mathtt{u_x}   \circ   \mathtt{(c\ 2)}   \eqs\eqs   \mathtt{u_x}   \circ   \mathtt{(c\ 14) } .
 \end{equation}
The rule (imp$_1$) gives
\begin{equation}
 \mathtt{(lpi}  \ \mathtt{b}  \  \mathtt{f)}   \circ   \mathtt{u_x} 
 \circ   \mathtt{(c\ 6)}   \circ    \mathtt{u_x}   \circ   \mathtt{(c\ 2)}   \eqs\eqs    \mathtt{u_x}   \circ   \mathtt{(c\ 14) } .
 \end{equation}
Now, we
rewrite the lemma \texttt{interaction-update-update} (see Figure~\ref{IMP-pp-st}) and get
\begin{equation}
 \mathtt{(lpi\ b\  f)}   \circ   \mathtt{u_x} 
 \circ   \mathtt{(c\ 6)}   \eqs\eqs    \mathtt{u_x}   \circ   \mathtt{(c\ 14) } . 
\end{equation}
\item For the second loop iteration, rewriting (imp-li) gives
\begin{equation}
 \mathtt{\big[}  \texttt{(lpi b f)}  \circ   \mathtt{f}   \mathtt{\big|}   \mathtt{id_{\unit}}   \mathtt{\big]} 
 \circ   \mathtt{pbl}   \circ   \mathtt{(t\ \overset{?}{<})}   \mathtt{\circ} 
 \mathtt{\pair{l_x, (c\ 11)}}   \circ   \mathtt{u_x}   \circ   \mathtt{(c\ 6)}   \eqs\eqs   \mathtt{u_x}   \circ   \mathtt{(c\ 14) } .
 \end{equation}
where looping pre-condition evaluates into ($\mathtt{c} \ true$).
Therefore, we iterate the above procedure, given in the step 1, once again and derive
\begin{equation}
 \mathtt{(lpi\ b\  f)}   \circ   \mathtt{u_x} 
 \circ   \mathtt{(c\ 10)}   \eqs\eqs    \mathtt{u_x}   \circ   \mathtt{(c\ 14) } . 
 \end{equation}
%
\item In the third iteration, rewriting the (imp-li) gives
\begin{equation}
 \mathtt{\big[}  \texttt{(lpi b f)}  \circ   \mathtt{f}   \mathtt{\big|}   \mathtt{id_{\unit}}   \mathtt{\big]} 
 \circ   \mathtt{pbl}   \circ   \mathtt{(t\ \overset{?}{<})}   \mathtt{\circ} 
 \mathtt{\pair{l_x, (c\ 11)}}   \circ   \mathtt{u_x}   \circ   \mathtt{(c\ 10)}   \eqs\eqs   \mathtt{u_x}   \circ   \mathtt{(c\ 14) } .
 \end{equation}
As in step 2, the  looping pre-condition evaluates into ($\mathtt{c} \ true$)
forcing us to reiterate the above procedure, given in the step 1, which results in
\begin{equation}
 \mathtt{(lpi\ b\  f)}   \circ   \mathtt{u_x} 
 \circ   \mathtt{(c\ 14)}   \eqs\eqs    \mathtt{u_x}   \circ   \mathtt{(c\ 14) } . 
 \end{equation}

\item In the fourth step,  rewriting the (imp-li) gives
\begin{equation}
 \mathtt{\big[}  \texttt{(lpi b f)}  \circ   \mathtt{f}   \mathtt{\big|}   \mathtt{id_{\unit}}   \mathtt{\big]} 
 \circ   \mathtt{pbl}   \circ   \mathtt{(t\ \overset{?}{<})}   \mathtt{\circ} 
 \mathtt{\pair{l_x, (c\ 11)}}   \circ   \mathtt{u_x}   \circ   \mathtt{(c\ 14)}   \eqs\eqs   \mathtt{u_x}   \circ   \mathtt{(c\ 14) } .
 \end{equation}
 By rewriting $\mathtt{commutation}$ $\mathtt{-lookup}$ $\mathtt{-constant}$ $\mathtt{-update}$,
we obtain
\begin{equation}
 \mathtt{\big[}  \texttt{(lpi b f)}  \circ   \mathtt{f}   \mathtt{\big|}   \mathtt{id_{\unit}}   \mathtt{\big]} 
 \circ   \mathtt{pbl}   \circ   \mathtt{(t\ \overset{?}{<})}   \mathtt{\circ} 
 \mathtt{\pair{(c\ 14), (c\ 11)}}   \circ   \mathtt{u_x}   \circ   \mathtt{(c\ 14)}   \eqs\eqs   \mathtt{u_x}   \circ   \mathtt{(c\ 14) } .
 \end{equation}
Finally here, the looping pre-condition $\mathtt{(t\ \overset{?}{<})}   \mathtt{\circ} 
 \mathtt{\pair{(c\ 14), (c\ 11)}}$ evaluates into $\mathtt{(c} \ false \mathtt{)}$
yielding
%
\begin{equation}
 \mathtt{\big[}  \texttt{(lpi b f)}  \circ   \mathtt{f}   \mathtt{\big|}   \mathtt{id_{\unit}}   \mathtt{\big]} 
 \circ   \mathtt{pbl}   \circ   (\mathtt{c}\ false)   \circ   \mathtt{u_x} 
 \circ   \mathtt{(c\ 14)}   \eqs\eqs   \mathtt{u_x}   \circ   \mathtt{(c\ 14) } .
 \end{equation}
We rewrite the Lemma~\ref{pblf}, and get
\begin{equation}
 \mathtt{\big[}  \texttt{(lpi b f)}  \circ   \mathtt{f}   \mathtt{\big|}   \mathtt{id_{\unit}}   \mathtt{\big]} 
 \circ   \mathtt{in_2}   \circ   \mathtt{u_x}   \circ   \mathtt{(c\ 14)}   \eqs\eqs   \mathtt{u_x}   \circ   \mathtt{(c\ 14) } .
 \end{equation}
Now we rewrite (s$\_$lcopair$\_$eq) and handle
\begin{equation}
 \mathtt{id_\unit} 
 \circ   \mathtt{u_x}   \circ   \mathtt{(c\ 14)}   \eqs\eqs   \mathtt{u_x}   \circ   \mathtt{(c\ 14)} 
 \end{equation}
which is trivial, since the identity term disappears when to compose and the equality sort $\eqs\eqs$ is reflexive. \qed
\end{enumerate}

\begin{lemma}
For each $\mathtt{x\ y: Loc}$, $\mathtt{e: EName}$, if program pieces
$\mathtt{prog5}$ and $\mathtt{prog6}$ are given as in the following listings, then
$\mathtt{dCmd \ (prog5)}$ $\mathtt{\eqs\eqs\ dCmd \ (prog6)}$.

\noindent\begin{minipage}{.45\textwidth}
\begin{lstlisting}[caption=prog5,frame=tlrb]{Name}
/* prog5 */
x := 1;
y := 20;
TRY(
	while (true)
		do 
		(	if (x <= 0) then (THROW e)
			else x := x - 1
		)
) CATCH e => (y := 7);
\end{lstlisting}
\end{minipage}
\qquad\qquad
\begin{minipage}{.40\textwidth}
\begin{lstlisting}[caption=prog6,frame=tlrb]{Name}
/* prog6 */
x := 0;
y := 7;
\end{lstlisting}
\end{minipage}

\end{lemma}

\proof
In the proof structure,
we first tackle with the $\mathtt{downcast}$ operator. The second task is to deal with the first loop 
iteration which has the state but no exception effect. In the third, 
we study the second iteration of the loop where an exception is thrown which is
followed by the abrupt loop termination.
Finally, in the fourth step, we explain
the exception recovery and the program termination.
Below is the sketch of $\mathtt{prog5}$:
%
$$\begin{array}{ll}
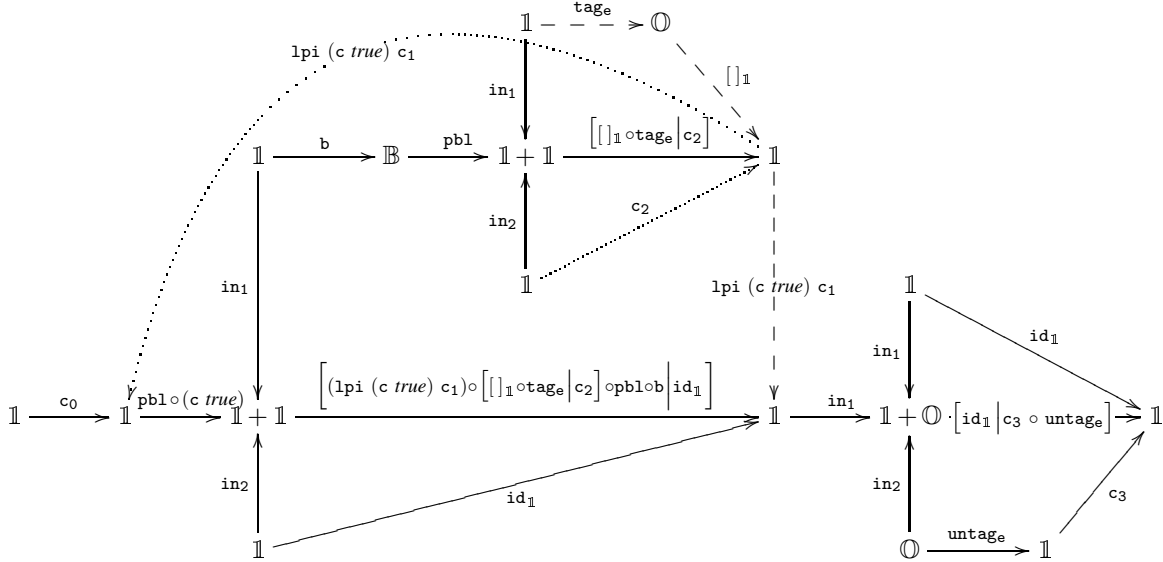
\xymatrix@R=3pc@C=2.5pc{
&&&&\unit\ar[d]_{\mathtt{in_1}} \ar@{-->}[r]^{\mathtt{tag_e}}&\empt  \ar@{-->}[rd]^{\mathtt{\copa_\unit}}&&&\\
 &&\unit \ar[dd]_{\mathtt{in_1}} \ar[r]^{\mathtt{b}} & \mathbb{B} \ar[r]^{\mathtt{pbl}} 
 & \unit+\unit \ar[rr]^{\mathtt{\big[ \copa_\unit \circ tag_e \big|c_2\big]}}& &\unit
 \ar@{-->}[dd]|{\mathtt{lpi\ (c \ \emph{true}) \ c_1}}
  \ar@/^-8pc/@{..>}[llllldd]|{\mathtt{lpi\ (c \ \emph{true}) \ c_1}} &&  \\
&&&&\unit\ar[u]^{\mathtt{in_2}} \ar@{..>}[urr]^{\mathtt{c_2}} & & &\unit  \ar[d]_{\mathtt{in_1}} \ar[rrd]^{\mathtt{id_{\unit}}} &\\
\unit \ar[r]^{\mathtt{c_0}} & \unit  \ar[r]^{\mathtt{pbl \,\circ\, (c\ \emph{true})}} & \unit+\unit \ar[rrrr]^{\mathtt{\Big[ (lpi\ (c \ \emph{true}) \ c_1) \circ  \big[ \copa_\unit \circ tag_e \big|c_2\big] \circ pbl  \circ b \Big|id_\unit \Big]}}
 & &  & & \unit \ar[r]^{\mathtt{in_1}} & 
 \unit + \empt  \ar[rr]|{\big[\mathtt{id_\unit} \big| \mathtt{c_3} \ \circ \ \mathtt{untag_e} \big]} & & \unit    \\
&& \unit \ar[u]^{\mathtt{in_2}} \ar[urrrr]_{\mathtt{id_\unit}} & & && & \empt \ar[u]^{\mathtt{in_2}} \ar[r]^{\mathtt{untag_e}} & \mathtt{\unit} \ar[ru]_{\mathtt{c_3}}&   \\ 
} 
\end{array}$$
\noindent
where 
$\mathtt{b = (x \overset{?}{\leq} 0)}$,
$\mathtt{c_0 = (x \triangleq 1; y \triangleq 20)}$,
$\mathtt{c_1 =}$ $\mathtt{(if\ (x \overset{?}{\leq} 0)}$ \texttt{then} $\mathtt{(THROW\ e)}$  $\mathtt{else}$
$\mathtt{ (x \triangleq x-1))}$,  $\mathtt{c_2 =}$ $\mathtt{(x \triangleq x - 1)}$, 
$\mathtt{c_3 = (y \triangleq 7)}$. The dotted arrows depict the
normal loop iterations while dashed ones are to identify the program behavior after the exception
of name \texttt{e} is raised.
Using the rules of the logic $\Log_{st+exc}$,
we can reduce the above diagram into the one given below with respect to the
equality sort $\eqs\eqs$:
$$\begin{array}{ll}
\xymatrix@R=2pc@C=3pc{
\unit \ar[r]^{\mathtt{c \ 0}} & \mathbb{Z} \ar[r]^{\mathtt{u_x}} & \unit
 \ar[r]^{\mathtt{c \ 7}} & \mathbb{Z} \ar[r]^{\mathtt{u_y}} & \unit
} 
\end{array}$$
which is actually the \texttt{prog6} when sketched.

\begin{enumerate}
\item
Initially, we have
\begin{multline}
 \downarrow   \mathtt{\Big(}   \mathtt{\big[}   \mathtt{id_\unit}   \mathtt{\big|}   \mathtt{\mathtt{c_3}}  \circ   \mathtt{untag}_ \mathtt{e}   \mathtt{\big]} 
 \circ   \mathtt{in_1}   \circ   \mathtt{\Big[}   \mathtt{(lpi\ (c \ \emph{true})\ c_1)}   \circ   \mathtt{\big[}   \mathtt{\copa_\unit}   \circ   \mathtt{tag}_\mathtt{e}   \mathtt{\big|}   \mathtt{c_2}   \mathtt{\big]}   \circ   \mathtt{pbl}   \circ   \mathtt{b}   \mathtt{\Big|}   \mathtt{id_\unit}   \mathtt{\Big]}  \circ \mathtt{pbl} \circ  \mathtt{(c \ \emph{true})}   \mathtt{\Big)} \\  \circ   \mathtt{u_y}   \circ   \mathtt{(c\ 20)}   \circ   \mathtt{u_x}   \circ 
 \mathtt{(c\ 1)}   \eqs\eqs   \mathtt{u_y}   \circ   \mathtt{(c\ 7)}   \circ   \mathtt{u_x}   \circ   \mathtt{(c\ 0)} .
 \end{multline}
We convert
$\eqs\eqs$ into $\eqs\eqw$, then rewrite the (w$\_$downcast) rule and get
\begin{multline}
 \mathtt{\big[}   \mathtt{id_\unit}   \mathtt{\big|}   \mathtt{c_3}   \circ   \mathtt{untag}_\mathtt{e}  \  \mathtt{\big]} 
 \circ   \mathtt{in_1}   \circ   \mathtt{\Big[}   \mathtt{(lpi\ (c \ \emph{true})\ c_1)}   \circ   \mathtt{\big[}   \copa_\unit   \circ   \mathtt{tag}_\mathtt{e}   \mathtt{\big|}   \mathtt{c_2}   \mathtt{\big]} 
 \circ   \mathtt{pbl}   \circ   \mathtt{b}   \mathtt{\Big|}  \mathtt{id_\unit}   \mathtt{\Big]} \\  \circ \mathtt{pbl} \circ  \mathtt{(c \ \emph{true})}   \circ   \mathtt{u_y}   \circ   \mathtt{(c\ 20)}   \circ   \mathtt{u_x}   \circ   \mathtt{(c\ 1)}   \eqs\eqw   \mathtt{u_y}   \circ   \mathtt{(c\ 7)}   \circ   \mathtt{u_x}   \circ   \mathtt{(c\ 0)} .
 \end{multline}
Rewriting
\texttt{commutation-update-update}, on both sides, gives
\begin{multline}
 \mathtt{\big[\mathtt{id_\unit} \big| \mathtt{c_3} \circ \mathtt{untag_e} \big]} 
 \mathtt{\circ in_1 \circ \Big[ (lpi\ (c \ \emph{true})\ c_1) \circ  \big[ \copa_\unit} 
 \circ   \mathtt{tag} _\mathtt{e}   \mathtt{\big|}   \mathtt{c_2}   \mathtt{\big]}   
 \circ   \mathtt{pbl}   \circ   \mathtt{b}   \mathtt{\Big|}   \mathtt{id_\unit}   \mathtt{\Big]} \\  \circ \mathtt{pbl} \circ   \mathtt{(c \ \emph{true})}   \circ   \mathtt{u_x}   \circ   \mathtt{(c\ 1)}   \circ   \mathtt{u_y}   \circ   \mathtt{(c\ 20)}   \eqs\eqw   \mathtt{u_x}   \circ   \mathtt{(c\ 0)}   \circ   \mathtt{u_y}   \circ   \mathtt{(c\ 7)} . 
 \end{multline}
 
Rewriting Lemma~\ref{pblt} yields
\begin{multline}
 \mathtt{\big[\mathtt{id_\unit} \big| \mathtt{c_3} \circ \mathtt{untag_e} \big]} 
 \mathtt{\circ in_1 \circ \Big[ (lpi\ (c \ \emph{true})\ c_1) \circ  \big[ \copa_\unit} 
 \circ   \mathtt{tag} _\mathtt{e}   \mathtt{\big|}   \mathtt{c_2}   \mathtt{\big]}   
 \circ   \mathtt{pbl}   \circ   \mathtt{b}   \mathtt{\Big|}   \mathtt{id_\unit}   \mathtt{\Big]} \\  \circ \mathtt{in_1}   \circ   \mathtt{u_x}   \circ   \mathtt{(c\ 1)}   \circ   \mathtt{u_y}   \circ   \mathtt{(c\ 20)}   \eqs\eqw   \mathtt{u_x}   \circ   \mathtt{(c\ 0)}   \circ   \mathtt{u_y}   \circ   \mathtt{(c\ 7)} . 
 \end{multline}
\item Now we rewrite the rule (w$\_$lcopair$\_$eq), and handle
\begin{multline}
 \mathtt{\big[}   \mathtt{id_\unit}   \mathtt{\big|}   \mathtt{c_3}   \circ   \mathtt{untag}_\mathtt{e}   \mathtt{\big]} 
 \circ   \mathtt{in_1}   \circ   \mathtt{(lpi\ (c \ \emph{true})\ c_1)}   \circ    \mathtt{\big[}   \mathtt{\copa_\unit} 
 \circ   \mathtt{tag}_\mathtt{e}   \mathtt{\big|}   \mathtt{c_2}   \mathtt{\big]} \\   \circ   \mathtt{pbl}    \circ   \mathtt{b}   \circ   \mathtt{u_x}   \circ  
 \mathtt{(c\ 1)}   \circ   \mathtt{u_y}   \circ   \mathtt{(c\ 20)}   \eqs\eqw   \mathtt{u_x}   \circ   \mathtt{(c\ 0)} .
 \end{multline}
By unfolding \texttt{b}, we have
\begin{multline}
 \mathtt{\big[}   \mathtt{id_\unit}   \mathtt{\big|}   \mathtt{c_3}   \circ   \mathtt{untag }_\mathtt{e}   \mathtt{\big]} 
 \circ   \mathtt{in_1}   \circ   \mathtt{(lpi\ (c \ \emph{true})\ c_1)}   \circ    \mathtt{\big[}   \mathtt{\copa_\unit} 
 \circ   \mathtt{tag}_\mathtt{e}   \mathtt{\big|}   \mathtt{c_2}   \mathtt{\big]} \\  \circ   \mathtt{pbl}   \circ   \mathtt{(t\ \overset{?}{\leq})} 
 \circ   \mathtt{\pair{l_x\ (c\ 0)}}   \circ   \mathtt{u_x}   \circ   \mathtt{(c\ 1)}   \circ   \mathtt{u_y}   \circ   \mathtt{(c\ 20)}   \eqs\eqw   \mathtt{u_x}   \circ   \mathtt{(c\ 0)}   \circ   \mathtt{u_y}   \circ   \mathtt{(c\ 7)} .
 \end{multline}
By rewriting the lemma $\mathtt{commutation-}$ $\mathtt{lookup-}$ $\mathtt{constant-}$ $\mathtt{update}$, we obtain
\begin{multline}
 \mathtt{\big[}   \mathtt{id_\unit}   \mathtt{\big|}   \mathtt{c_3}   \circ   \mathtt{untag}_\mathtt{e}   \mathtt{\big]} 
 \mathtt{\circ in_1} 
 \mathtt{\circ}   \mathtt{(lpi\ (c \ \emph{true})\ c_1)}   \circ    \mathtt{\big[}   \mathtt{\copa_\unit}   \circ   \mathtt{tag}_\mathtt{e}   \mathtt{\big|}   \mathtt{c_2}   \mathtt{\big]}   \\ \circ   \mathtt{pbl}   \circ   \mathtt{(t} \   \mathtt{\overset{?}{\leq})} 
 \circ   \mathtt{\pair{(c\ 1), (c\ 0)}} 
  \circ   \mathtt{u_x}   \circ   \mathtt{(c\ 1)} 
 \circ   \mathtt{u_y}   \circ   \mathtt{(c\ 20)}   \eqs\eqw   \mathtt{u_x}   \circ   \mathtt{(c\ 0)}   \circ   \mathtt{u_y}   \circ  ( \mathtt{c\ 7)} .
 \end{multline}
We rewrite the rule (imp$_2$), and get
\begin{multline}
 \mathtt{\big[}   \mathtt{id_\unit}   \mathtt{\big|}   \mathtt{c_3}   \circ   \mathtt{untag}_\mathtt{e}   \mathtt{\big]}   \circ   \mathtt{in_1} 
 \circ   \mathtt{(lpi\ (c \ \emph{true})\ c_1)}   \\  \circ 
 \mathtt{ \big[}  \mathtt{\copa_\unit}   \circ   \mathtt{tag}_\mathtt{e}   \mathtt{\big|}   \mathtt{c_2}   \mathtt{\big]}   \circ   
 \mathtt{pbl}  \circ (\mathtt{c \ \emph{false}})  \circ   \mathtt{u_x}   \circ   \mathtt{(c\ 1)}   \circ   \mathtt{u_y} 
 \circ    \mathtt{(c\ 20)}   \eqs\eqw   \mathtt{u_x}   \circ   \mathtt{(c\ 0)}   \circ   \mathtt{u_y} . 
 \circ   \mathtt{(c\ 7)} .
 \end{multline}
Rewriting the Lemma~\ref{pblf} yields
\begin{multline}
 \mathtt{\big[}   \mathtt{id_\unit}   \mathtt{\big|}   \mathtt{c_3}   \circ   \mathtt{untag}_\mathtt{e}   \mathtt{\big]}   \circ   \mathtt{in_1} 
 \circ   \mathtt{(lpi\ (c \ \emph{true})\ c_1)}   \\  \circ 
 \mathtt{ \big[}  \mathtt{\copa_\unit}   \circ   \mathtt{tag}_\mathtt{e}   \mathtt{\big|}   \mathtt{c_2}   \mathtt{\big]}   \circ   
 \mathtt{in_2}  \circ   \mathtt{u_x}   \circ   \mathtt{(c\ 1)}   \circ   \mathtt{u_y} 
 \circ    \mathtt{(c\ 20)}   \eqs\eqw   \mathtt{u_x}   \circ   \mathtt{(c\ 0)}   \circ   \mathtt{u_y} . 
 \circ   \mathtt{(c\ 7)} .
 \end{multline} 
We now rewrite (s$\_$lcopair$\_$eq) which gives
\begin{multline}
 \mathtt{\big[}   \mathtt{id_\unit}   \mathtt{\big|}   \mathtt{c_3}   \circ   \mathtt{untag}_\mathtt{e}   \mathtt{\big]} 
 \circ   \mathtt{in_1} 
 \mathtt{\circ}   \mathtt{(lpi\ (c \ \emph{true})\ c_1)}   \\ \circ 
 \mathtt{c_2}   \circ   \mathtt{u_x}   \circ 
 \mathtt{(c\ 1)}   \circ    \mathtt{u_y}    \circ   \mathtt{(c\ 20)}   \eqs\eqw   \mathtt{u_x}   \circ   \mathtt{(c\ 0)}   \circ   \mathtt{u_y}   \circ   \mathtt{(c\ 7)} .
 \end{multline}
Here, by unfolding $\mathtt{c_2}$, we have
\begin{multline}
 \mathtt{\big[}   \mathtt{id_\unit}   \mathtt{\big|}   \mathtt{c_3}   \circ   \mathtt{untag}_\mathtt{e}   \mathtt{\big]} 
 \circ   \mathtt{in_1} 
 \circ   \mathtt{(lpi\ (c \ \emph{true})\ c_1)}   \circ 
 \mathtt{u_x}   \circ   \mathtt{(t\ -)} 
 \circ   \mathtt{\pair{l_x, (c\ 1)}}  \\ \circ   \mathtt{u_x}   \circ   \mathtt{(c\ 1)}   \circ   \mathtt{u_y}   \circ   \mathtt{(c\ 20)}   \eqs\eqw 
 \mathtt{u_x}   \circ   \mathtt{(c\ 0)}   \circ   \mathtt{u_y}   \circ   \mathtt{(c\ 7)} .
 \end{multline}
Rewriting the lemma $\mathtt{commutation-lookup-constant-update}$ gives
\begin{multline}
 \mathtt{\big[}   \mathtt{id_\unit}   \mathtt{\big|}   \mathtt{c_3}   \circ   \mathtt{untag}_\mathtt{e}   \mathtt{\big]} 
 \circ   \mathtt{in_1} 
 \circ   \mathtt{(lpi\ (c \ \emph{true})\ c_1)}   \circ 
 \mathtt{u_x}   \circ   \mathtt{(t\ -)} 
 \circ   \mathtt{\pair{(c\ 1), (c\ 1)}} \\  \circ   \mathtt{u_x} 
 \circ   \mathtt{(c\ 1)}   \circ   \mathtt{u_y}   \circ   \mathtt{(c\ 20)}   \eqs\eqw   \mathtt{u_x}   \circ   \mathtt{(c\ 0)}  
 \circ   \mathtt{u_y}   \circ 0 \mathtt{(c\ 7)} .
 \end{multline}
We rewrite (imp$_1$), and get
\begin{multline}
 \mathtt{\big[}   \mathtt{id_\unit}   \mathtt{\big|}   \mathtt{c_3}   \circ   \mathtt{untag}_\mathtt{e}   \mathtt{\big]} 
 \circ   \mathtt{in_1} 
 \circ   \mathtt{(lpi\ (c \ \emph{true})\ c_1)}   \\ \circ 
 \mathtt{u_x}   \circ   \mathtt{(c\ 0)} \circ
 \mathtt{u_x}   \circ   \mathtt{(c\ 1)}   \circ   \mathtt{u_y}    \circ   \mathtt{(c\ 20)}   \eqs\eqw   \mathtt{u_x} 
 \circ   \mathtt{(c\ 0)}   \circ   \mathtt{u_y}   \circ   \mathtt{(c\ 7)} .
 \end{multline} 
We again rewrite the lemma \texttt{commutation-update-update}, and obtain
\begin{multline}
 \mathtt{\big[}   \mathtt{id_\unit}   \mathtt{\big|}   \mathtt{c_3}   \circ   \mathtt{untag}_\mathtt{e}   \mathtt{\big]} 
 \circ   \mathtt{in_1} 
 \circ    \mathtt{(lpi\ (c \ \emph{true})\ c_1)}   \circ 
 \mathtt{u_x}   \circ   \mathtt{(c\ 0)} \\
 \circ   \mathtt{u_y}   \circ   \mathtt{(c\ 20)}   \eqs\eqw    \mathtt{u_x \circ} 
 \mathtt{(c\ 0)}   \circ   \mathtt{u_y}   \circ   \mathtt{(c\ 7)} . 
 \end{multline}
\item We re-iterate the loop via (imp-li), and have
\begin{multline}
 \mathtt{\big[}   \mathtt{id_\unit}   \mathtt{\big|}   \mathtt{c_3}   \circ   \mathtt{untag}_\mathtt{e}   \mathtt{\big]} 
 \circ   \mathtt{in_1} 
 \circ   \mathtt{\big[(lpi \ (c \ \emph{true})\ c_1)}   \circ   \mathtt{c_1}   \mathtt{\big|}   \mathtt{id}   \mathtt{\big]}  \\
\circ \mathtt{pbl}  \circ   \mathtt{(c \ \emph{true})}   \circ 
 \mathtt{u_x}   \circ   \mathtt{(c\ 0)} 
 \circ   \mathtt{u_y} 
 \circ   \mathtt{(c\ 20)}   \eqs\eqw   \mathtt{u_x}   \circ   \mathtt{(c\ 0)}   \circ   \mathtt{u_y}   \mathtt{\circ} 
 \mathtt{(c\ 7)} .
 \end{multline}
We rewrite Lemma~\ref{pblt},  (w$\_$lcopair$\_$eq), then unfold $\mathtt{c_1}$, and get:
\begin{multline}
 \mathtt{\big[}   \mathtt{id_\unit}   \mathtt{\big|}   \mathtt{c_3}   \circ   \mathtt{untag}_\mathtt{e}   \mathtt{\big]} 
 \circ   \mathtt{in_1} 
 \circ   \mathtt{(lpi \ (c \ \emph{true})\ c_1)}   \circ   \mathtt{\big[}   \mathtt{throw}  \ \mathtt{e} \  \mathtt{\unit}   \mathtt{\big|c_2\big]} \\
 \circ   \mathtt{pbl}   \circ   \mathtt{(t\ \overset{?}{\leq})}    \circ 
 \mathtt{\pair{l_x, (c\ 0)}}   \circ 
 \mathtt{u_x}   \circ   \mathtt{(c\ 0)} 
 \circ   \mathtt{u_y} 
 \circ   \mathtt{(c\ 20)}   \eqs\eqw   \mathtt{u_x}   \circ   \mathtt{(c\ 0)}   \circ   \mathtt{u_y}   \circ   \mathtt{(c\ 20)} .
 \end{multline}
By rewriting $\mathtt{commutation-lookup-constant-update}$, (imp$_3$) and Lemma~\ref{pblt},
we have
\begin{multline}
 \mathtt{\big[}   \mathtt{id_\unit}   \mathtt{\big|}   \mathtt{c_3}   \circ   \mathtt{untag}_\mathtt{e}   \mathtt{\big]} 
 \circ   \mathtt{in_1} 
 \circ   \mathtt{(lpi \ (c \ \emph{true})\ c_1)}   \circ   \mathtt{\big[}   \mathtt{throw} \   \mathtt{e} \  \mathtt{\unit}   \mathtt{\big|}   \mathtt{c_2}  \mathtt{\big]}
 \circ   \mathtt{in_1} \\  \circ 
 \mathtt{u_x}   \circ   \mathtt{(c\ 0)} 
 \circ   \mathtt{u_y} 
 \circ   \mathtt{(c\ 20)}   \eqs\eqw    \mathtt{u_x}   \circ   \mathtt{(c\ 0)}   \circ   \mathtt{u_y}   \circ   \mathtt{(c\ 20)} .
 \end{multline}
By (w$\_$lcopair$\_$eq), the exception is raised:
\begin{multline}
 \mathtt{\big[}   \mathtt{id_\unit}   \mathtt{\big|}   \mathtt{c_3}   \circ   \mathtt{untag}_\mathtt{e}   \mathtt{\big]} 
 \circ   \mathtt{in_1} 
 \circ   \mathtt{\big((lpi \ (c \ \emph{true})\ c_1)}   \circ   \mathtt{throw} \  \mathtt{e} \  \mathtt{\unit}   \mathtt{\big)}  \\
 \circ  
 \mathtt{u_x}   \circ   \mathtt{(c\ 0)} 
 \circ   \mathtt{u_y} 
 \circ   \mathtt{(c\ 20)}   \eqs\eqw    \mathtt{u_x}   \circ   \mathtt{(c\ 0)}   \circ   \mathtt{u_y}   \circ   \mathtt{(c\ 20)} .
 \end{multline}
Due to the raised exception, the infinite loop gets abruptly terminated at this step.
Here we unfold the definition of \texttt{THROW} then rewrite \texttt{propagator-propagates} (see Section~\ref{dpotsee}), and get
\begin{multline}
 \mathtt{\big[}   \mathtt{id_\unit}   \mathtt{\big|}   \mathtt{c_3}   \circ   \mathtt{untag}_\mathtt{e}   \mathtt{\big]} 
 \circ   \mathtt{in_1}   \circ   \mathtt{\copa_\unit}   \circ   \mathtt{tag}_\mathtt{e}   \circ  
 \mathtt{u_x}   \circ   \mathtt{(c\ 0)} 
 \circ   \mathtt{u_y} 
 \circ   \mathtt{(c\ 20)}   \eqs\eqw    \mathtt{u_x}   \circ   \mathtt{(c\ 0)}   \circ   \mathtt{u_y}   \circ   \mathtt{(c\ 20)}.
\end{multline}
\item Here, we first cut  $\mathtt{in_1 \circ \copa_\unit \eqs\eqs in_2}$, and rewrite it back in the equation.
Thus, we have
\begin{equation}
 \mathtt{\big[}   \mathtt{id_\unit}   \mathtt{\big|}   \mathtt{c_3}   \circ   \mathtt{untag}_\mathtt{e}   \mathtt{\big]} 
 \circ   \mathtt{in_2} 
 \circ   \mathtt{tag}_\mathtt{e}   \circ   \mathtt{u_x } 
 \circ   \mathtt{(c\ 0)}   \circ   \mathtt{u_y}   \circ   \mathtt{(c\ 20)}   \eqs\eqw    \mathtt{u_x}   \mathtt{\circ} 
 \mathtt{(c\ 0)}   \circ   \mathtt{u_y}   \mathtt{\circ} 
 \mathtt{(c\ 7)} .
 \end{equation}
By rewriting (s$\_$lcopair$\_$eq), we obtain
\begin{equation}
 \mathtt{c_3}   \circ   \mathtt{untag }_\mathtt{e} 
 \circ   \mathtt{tag}_\mathtt{e} 
 \circ   \mathtt{u_x}   \circ   \mathtt{(c\ 0)}   \circ   \mathtt{u_y}   \circ   \mathtt{(c\ 20)}   \eqs\eqw   \mathtt{u_x}   \circ 
 \mathtt{ (c\ 0)}   \circ   \mathtt{u_y}   \circ 
 \mathtt{(c\ 7)} .
 \end{equation}
Since $\mathtt{u_x}$  $\circ$ $\mathtt{(c\ 0)}$ $\circ$  $\mathtt{u_y}$ $\circ$ $\mathtt{(c\ 20)}$
is pure with respect to the exception, we rewrite
(eax$_1$), and get
\begin{equation}
 \mathtt{\mathtt{c_3}} 
 \circ   \mathtt{u_x}   \circ   \mathtt{(c\ 0)}   \circ   \mathtt{u_y}   \circ   \mathtt{(c\ 20)}   \eqs\eqw    \mathtt{u_x}   \circ 
 \mathtt{ (c\ 0)\circ u_y \circ} 
 \mathtt{(c\ 7)} .
 \end{equation}
Unfolding the definition of the command $\mathtt{c_3 = (u_y \circ (c\ 7))}$, we have
\begin{equation}
 \mathtt{u_y}   \circ   \mathtt{(c\ 7)} 
 \mathtt{\circ}   \mathtt{u_x}    \circ   \mathtt{(c\ 0)}   \circ   \mathtt{u_y}   \circ   \mathtt{(c\ 20)}   \eqs\eqw    \mathtt{u_x \circ} 
 \mathtt{ (c\ 0)\circ u_y \circ (c\ 7)} .
 \end{equation}
We now rewrite $\mathtt{commutation-}$ $\mathtt{update-}$ $\mathtt{update}$ on the left, and handle
\begin{equation}
 \mathtt{u_x}   \circ   \mathtt{(c\ 0)} 
 \circ   \mathtt{u_y}   \circ   \mathtt{(c\ 7)}   \circ   \mathtt{u_y}   \circ   \mathtt{(c\ 20)}   \eqs\eqw    \mathtt{u_x}   \circ 
 \mathtt{ (c\ 0)}   \circ   \mathtt{u_y}   \circ   \mathtt{(c\ 7)} .
 \end{equation}
Finally, it suffices to rewrite $\mathtt{interaction-}$
$\mathtt{update-}$$\mathtt{update}$,
\begin{equation}
 \mathtt{u_x}   \circ   \mathtt{(c\ 0)} 
 \circ   \mathtt{u_y}    \circ   \mathtt{(c\ 7)}   \eqs\eqw    \mathtt{u_x}   \circ 
 \mathtt{(c\ 0)}   \circ   \mathtt{u_y}   \circ   \mathtt{(c\ 7)} .
 \end{equation}
which is trivial since the equality symbol $\eqs\eqw$ is reflexive.
However, it still remains to prove the previous cut $\mathtt{in_1 \circ \copa_\unit
\eqs\eqs in_2}$: since everything is pure with respect to the exception, we have
\begin{equation}
 \mathtt{in_1 \circ \copa_\unit \eqs\eqw in_2}.
\end{equation}
 Now, rewriting the rule
 (w$\_$empty) gives
 $\mathtt{\copa_{\unit+\unit} \eqs\eqw \copa_{\unit+\unit}}$.
 This is trivial since the equality sort $\eqs\eqw$ is reflexive.\qed
\end{enumerate}

The full \coq proofs of above lemmata can be found here \footnote{\url
{https://github.com/ekiciburak/impex-on-decorated-logic/blob/master/IMPEX_Proofs.v}}, and the entire implementation 
there \footnote{\url{https://github.com/ekiciburak/impex-on-decorated-logic}}.

\noindent
\subsection{Automating decorated proofs}
A rule in a decorated logic only applies if the given term gets decorated as expected by the rule. Therefore,
decoration checks are pretty important and occur pretty often. To automatize this checks,
at the Coq level, we already have tactics \texttt{decorate} and
\texttt{edecorate}. See them here
\footnote{\url
{https://github.com/ekiciburak/impex-on-decorated-logic/blob/master/Decorations.v}}.
Also, we plan to put~\cite{lcck-jar18}'s CoqHammer tool in use to try automatizing such program property proofs,
done within
the scope of decorated
logics, implemented in Coq. 

\subsection{On the completeness of the logic $\Log_{st+exc}$}
\label{sub:tc}
With the logic $\Log_{st+exc}$, no generic program properties
such as
$$\begin{array}{l}
   \xymatrix@R=0.01pc@C=0.5pc{
   \mathtt{dCmd(p_1) \eqs\eqs dCmd(p_2) \implies \forall \,s\, s',\, eval\, p_1 \, s\ s' \implies
eval\, p_2 \,s\ s'}}
\end{array}$$
 can be proven.
 Here, \texttt{eval} denotes the big-step semantics of the commands until reaching \texttt{SKIP}.
Only programs that admit a particular specification can be proven to be equivalent
with respect to the state and exception effects.
The total correctness is based on a syntactic completeness property.
In a way, it is meant to make sure that we are not using too many axioms
to construct a denotational semantics for the \texttt{IMP+Exc} language
using the logic $\Log_{st+exc}$ as  the target language.
This syntactic completeness property is called relative Hilbert-Post Completeness
(rHPC) and elaborately
defined by \cite{DBLP:conf/macis/DumasDEPR15}. Briefly, given 
two logics $\mathtt{L_0}$ and $\mathtt{L}$ such that $\mathtt{L_0 \subseteq L}$
($\mathtt{L_0}$ is a sub-logic of $\mathtt{L}$)
and a theory \texttt{T} of $\mathtt{L}$.
\texttt{T} is relatively Hilbert-Post complete with respect to $\mathtt{L_0}$
if (1) at least one sentence is unprovable in \texttt{T} (not the maximal theory
ensuring consistency),
and (2) every theory containing \texttt{T} can be generated from \texttt{T} and
some sentences from $\mathtt{L_0}$. Here, $\mathtt{L_0}$ can be seen as the
pure logic that governs the denotational semantics of the effect-free subset of
the \texttt{IMP} language where $\mathtt{L}$ is the logic that governs the denotational
semantics of the superset of the \texttt{IMP} language after either the state or
exception effect is added.

We prove, in Theorem 6.8.5 in (\cite{Ekici:2015t}), that the decorated theory of exceptions is
relatively Hilbert-Post complete
with respect to its pure part. However, only the core part of the decorated logic for the state
effect is proven to be rHPC (see Theorem 5.4.9 in \cite{Ekici:2015t}). What we mean by the ``core part'' is
the logic with no categorical pairs. Clearly, when translated to \texttt{IMP} denotational semantics,
it corresponds to the part that governs conditionals and loops.
We can conjecture that the logic is still complete in the presence of categorical pairs. However, the
proof is not yet done. 

In the rHPC proof of the core part for the decorated logic for the state,
we first determine the canonical forms of accessors and modifiers and then show that
both such forms are equivalent to some finite set of equations in the pure sublogic of $\Log_{st}$
with no pairs. In the presence of
categorical pairs, we so far had difficulties to come up with the
canonical forms for accessors and modifiers even though it is clear that such forms exist.
Once we have these forms in hand, it should also be the case that the rules governing pairs
suffice to prove that such forms are equivalent to finite number of equations
made of terms coming from the pure counterpart of the logic  $\Log_{st}$.
We plan to study this in the near future.

It is also proven that if two theories are rHPC with respect to a (pure) logic, then the combination
of these theories remains to be rHPC. Therefore, the logic $\Log_{st+exc}$ without the use of pairs is rHPC.

%

\section{Concluding remarks}
\label{concl}
We have presented frameworks for formalizing the treatment
of the state and the exception effects, first separately, and then combined, using the decorated logic.
Decorations describe what
computational effect evaluation of a term may involve,
and form a bridge between the syntax and its interpretation in 
reasoning about terms by making computational effects explicit in the decorated syntax.
We have designed a denotational semantics for the \texttt{IMP+Exc} language over
the combined decorated logic $\Log_{st+exc}$. This way,
we managed prove some strong equalities between \texttt{IMP+Exc} programs. We have also
encoded the combined logic in the \coq proof assistant and certified related proofs.

\acknowledgements{We wish to thank the two anonymous
reviewers for their careful reading of our manuscript and for    
insightful comments and suggestions. 
We thank Jean-Guillaume Dumas and Dominique Duval for their
support on all aspects of the presented logics. Many thanks also to Damien Pous
for his guidance on \coq related questions. 
This work has been partially supported by
the Austrian Science Fund (FWF) grant P26201 and the European Research Council (ERC)
Grant No. 714034 SMART.}

\bibliographystyle{abbrvnat}
\bibliography{example.bib}


%
%


\end{document}